\documentclass[showpacs, preprintnumbers, nofootinbib, aps, prd, superscriptaddress,10pt, showkeys, notitlepage]{revtex4-1}

\usepackage{graphicx,amssymb,amsmath,amsthm,amsfonts,epsfig}

\usepackage[linktocpage,breaklinks]{hyperref}
\usepackage[usenames,dvipsnames]{color}
\usepackage{epstopdf}
\usepackage{aas_macros}
\usepackage{pifont}
\usepackage[normalem]{ulem}
\definecolor{darkred}{rgb}{0.5,0,0}
\definecolor{darkgreen}{rgb}{0,0.5,0}
\definecolor{darkblue}{rgb}{0,0,0.5}
\definecolor{prussian}{rgb}{0.0, 0.19, 0.33}
\definecolor{richelectricblue}{rgb}{0.03, 0.57, 0.82}
\definecolor{teal}{rgb}{0.0, 0.5, 0.5}
\definecolor{mediumseagreen}{rgb}{0.24, 0.7, 0.44}
\definecolor{lust}{rgb}{0.9, 0.13, 0.13}
\definecolor{ballblue}{rgb}{0.13, 0.67, 0.8}
\definecolor{darkcyan}{rgb}{0.0, 0.55, 0.55}
\definecolor{mountainmeadow}{rgb}{0.19, 0.73, 0.56}
\definecolor{palecarmine}{rgb}{0.69, 0.25, 0.21}
\definecolor{richcarmine}{rgb}{0.84, 0.0, 0.25}
\definecolor{tangelo}{rgb}{0.98, 0.3, 0.0}
\definecolor{venetian}{rgb}{0.784,0.031,0.082}
\definecolor{bdfrance}{rgb}{0.192,0.549,0.906}

\hypersetup{colorlinks=true, citecolor=venetian,
linkcolor=bdfrance, urlcolor=lust}
\usepackage{amsmath,amssymb}
\usepackage{tensor}
\usepackage{mathtools}
\usepackage{amsbsy}
\usepackage{bm}
\usepackage{float}

\newcommand{\be}{\begin{equation}}
\newcommand{\ee}{\end{equation}}
\newcommand{\bea}{\begin{eqnarray}}
\newcommand{\eea}{\end{eqnarray}}
\newcommand{\nn}{\nonumber}

\newcommand{\p}{\prime}
\newcommand{\pp}{\prime\prime}
\newcommand{\Om}{{\Omega}}

\newcommand{\cE}{{\cal E}}
\newcommand{\cC}{{\cal C}}

\newcommand{\cD}{{\cal D}}

\newcommand{\rK}{{\rm K}}
\newcommand{\rJ}{{\rm J}}
\newcommand{\rJP}{{\rm JP}}

\begin{document}

\title{The modification of photon trapping orbits as a diagnostic of non-Kerr spacetimes}

\author{Kostas  Glampedakis}
\email{kostas@um.es}
\affiliation{Departamento de F\'isica, Universidad de Murcia, Murcia E-30100, Spain}
\affiliation{Theoretical Astrophysics, University of T\"ubingen, Auf der Morgenstelle 10, T\"ubingen, D-72076, Germany}

\author{George Pappas}
\email{georgios.pappas@roma1.infn.it}
\affiliation{Dipartimento di Fisica,  ``Sapienza'' Universit\'a di Roma \& Sezione INFN Roma1, 
Piazzale Aldo Moro 5, 00185, Roma, Italy}

\begin{abstract}

Photon circular orbits, an extreme case of light deflection, are among the hallmarks of black holes and are known to play a 
central role in a variety of phenomena related to these extreme objects. The very existence of  such orbits when
motion is not confined in the equatorial plane, i.e. spherical orbits, is indeed a special property of the separable Kerr 
metric and may not occur, for instance, in the spacetime of other more speculative ultracompact objects.
In this paper we consider a general stationary-axisymmetric spacetime  and examine under what circumstances spherical or more 
general, variable-radius, `spheroidal' non-equatorial photon orbits may exist with the ultimate goal of using the modifications 
-- or even loss -- of photon trapping orbits as a telltale of non-Kerr physics. 
In addressing this issue, we first derive a general necessary condition for the existence of spherical/spheroidal orbits and then 
go on to study photon trapping orbits in a variety of known non-Kerr metrics (Johannsen, Johanssen-Psaltis, and Hartle-Thorne).
The first of these is an example of a separable spacetime which supports Kerr-like spherical photon orbits. 
A more detailed analysis reveals a deeper connection between the presence of spherical orbits and the separability of a metric 
(that is, the existence of a third integral of motion). Specifically, a spacetime that does not admit spherical orbits in any coordinates is 
necessarily non-separable. The other two spacetimes considered here exhibit a clear non-Kerr behaviour by having spherical photon orbits 
replaced by spheroidal ones. More importantly, subject to the degree of deviation from Kerr, equatorial photon rings give place to non-equatorial 
ones with an accompanying loss of low-inclination spheroidal orbits. The implications of these results for the electromagnetic and gravitational 
wave signature of non-Kerr objects are briefly discussed. 

\end{abstract} 
  
\maketitle


\section{Introduction}
\label{sec:intro}

The first direct observations of gravitational waves (GWs) by the advanced LIGO/Virgo detector network~\cite{Abbott:2016blz,Abbott:2016nmj,
TheLIGOScientific:2016pea, Abbott:2017vtc, GW170814} saw General Relativity (GR) becoming even more established as the correct theory 
of gravity. However, these observations, spectacular as they may be, have not yet ruled out alternative to GR theories of gravity nor have they 
established `beyond reasonable doubt' the Kerr nature of the compact objects involved in the mergers. Indeed, testing the Kerr hypothesis
even within GR against the possibility of having some other type of exotic horizonless ultracompact object that could be mistaken for
a black hole is by itself a far from easy endeavour, see e.g. \cite{Cardoso:2016rao, Cardoso_Pani:2017, Maselli_etal2017, Glampedakis2018}. 
This  fascinating prospect provides ample motivation for a more detailed study of how non-Kerr compact objects could manifest themselves 
under the scrutinous eyes of electromagnetic and gravitational wave observatories. 

A not often emphasised characteristic of Kerr black holes is the presence of non-equatorial ``circular'' photon (or particle)
orbits. These are in fact spherical (though not closed) trajectories of constant radius $r_0$ and represent the generalisation of the
more familiar concept of the equatorial circular orbit, the so-called  `photon ring'. Spherical/circular photon geodesics leave their mark 
(directly or indirectly) on a variety of phenomena involving black holes.

A black hole illuminated by an external source of light (e.g.~a hot accretion flow) casts a shadow that is fringed by a sharp 
bright ring~\cite{Bardeen1973, Johannsen2010}. The mechanism responsible for the formation of this optical structure is 
the photon circular orbit which acts as a temporary  depository of electromagnetic flux. It is not surprising then that 
photon circular orbits play a key role in the ongoing effort to capture horizon-scale images of the Sgr $\mbox{A}^*$ 
supermassive black hole (and of other black holes in our galactic neighborhood) and use them as an observational test 
of the Kerr metric, see e.g.~\cite{Johannsen2010,Broderick2014}. This program should soon come to fruition with the ongoing 
observations of the Event Horizon Telescope's worldwide constellation of radio telescopes\footnote{https://eventhorizontelescope.org}.

 \begin{figure*}[htb!]
 \includegraphics[width=1\textwidth]{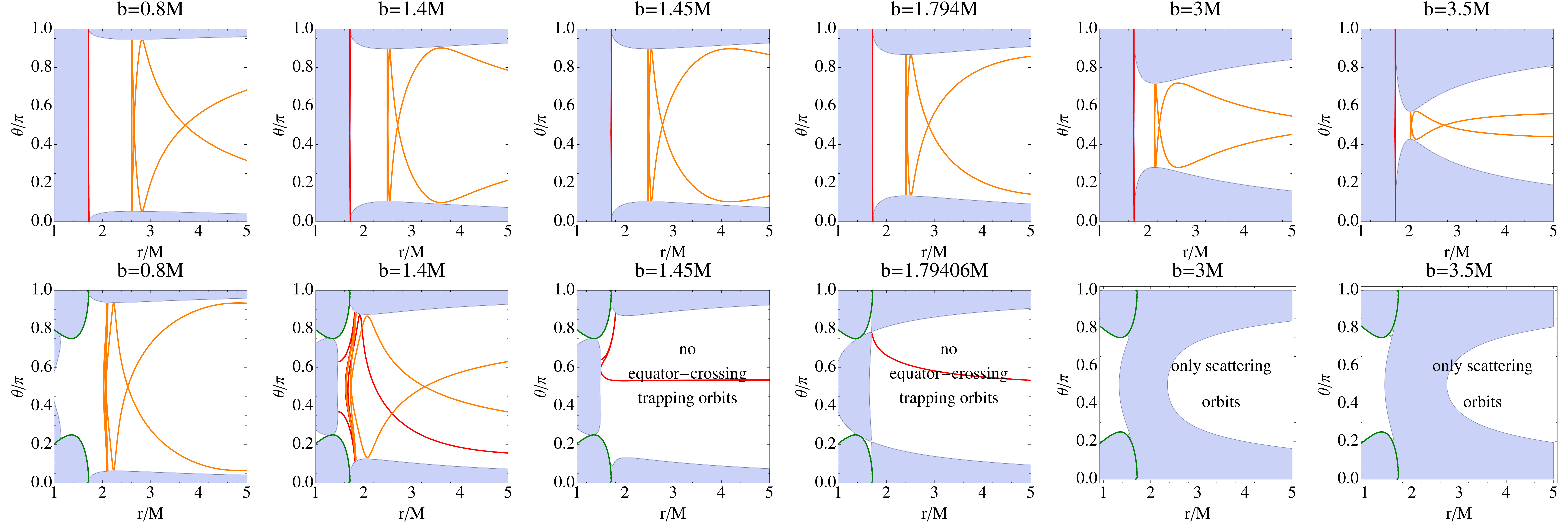}
\caption{\emph{Kerr vs non-Kerr photon orbits}. This figure exemplifies the impact of non-separability on the character of photon trapping orbits.
The shaded area in the $r$-$\theta$ plane represents the region where geodesic motion is not allowed. The green (bottom row) and red (top row) curves mark 
the location of the event horizon. Top row: we show a sequence of quasi-spherical orbits in an $a=0.7M$ Kerr metric with increasing impact parameter $b$; 
in these examples photons are temporarily trapped in the vicinity of a spherical orbits. Bottom row: we show a similar $b$-sequence of orbits in the non-separable 
Johannsen-Psaltis spacetime with deformation parameter $\varepsilon_3=5$ and spin $a=0.7M$. The first two low-$b$ examples, 
although lacking exactly spherical orbits,  display a Kerr-like character in the sense that photons are trapped in spheroidal or nearly spheroidal orbits
that cross the equatorial plane. However, these orbits are lost (together with the equatorial photon ring) as we move towards higher $b$.  The formation
of non-equatorial photon rings (see panel with $b=1.794096M$ ) facilitates the capture of photons in their vicinity in the form of circular-planar or spheroidal 
orbits. In the high-$b$ end of the spectrum, only scattering orbits are present and no photon trapping is possible.}
\label{fig:KerrvsJP}
\end{figure*}

Much of our intuition about wave dynamics in black hole spacetimes is also based on photon spherical orbits.
Quasi-normal mode (QNM) ringdown is intuitively understood in terms of wavepackets temporarily trapped in the vicinity 
of the photon ring, gradually peeling off towards infinity and the event horizon as they circle the black hole.
Indeed, in the eikonal limit of geometric optics the frequency and damping rate of the fundamental QNM are determined,
respectively, by the photon orbit's angular frequency and divergence rate (Lyapunov exponent), 
see~\cite{Mashhoon:1985cya, Ferrari:1984zz, Cardoso:2008bp, Dolan:2010wr, Glampedakis:2017} for more details. 
Similarly, the scattering of plane waves by black holes reveals the presence of a photon ring in the glory  pattern of the scattering 
cross section (see~\cite{Glampedakis2001} and references therein).  
 
Although photon rings are expected to be an ubiquitous orbital feature, present in the spacetime of non-Kerr 
ultracompact objects such as gravastars and bosons stars, the same may not be true for the off-the-equator spherical orbits. 
Excluding the idealised case of spherically symmetric systems, the existence of the latter orbits is not guaranteed unless some 
special conditions are met. Photons moving in non-Kerr spacetimes may be trapped in different kind of orbits or they might not 
be trapped at all, at least for some part of the orbital parameter space. 

This is precisely the issue addressed in this paper.  Specifically, we ask under what circumstances spherical (of constant radius $r_0$) 
or more general `spheroidal'  (of a variable, equatorially-symmetric, radius $r_0 (\theta)$, where $\theta$ is a meridional coordinate) 
photon orbits are allowed when one moves away from Kerr to an arbitrary axisymmetric-stationary and equatorial-symmetric spacetime. 
In this general case we can formulate a necessary `spheroidicity condition' for the existence of the aforementioned orbits.
It is then possible to arrive to the remarkable result that a spacetime that has a photon ring but does not admit 
spherical orbits is necessarily non-separable (and therefore non-Kerr). However, the separability-sphericity connection is not a solid 
one, in the sense that non-separability does not necessarily imply the absence of spherical orbits.

Most non-Kerr spacetime metrics of interest are of course non-separable\footnote{Note that throughout this paper the 
term `separabable' is used as a proxy for the more accurate term `geodesically separable', i.e. the separability associated with 
the Hamilton-Jacobi equation.} and therefore spherical photon orbits could be replaced by spheroidal ones. 
Our results suggest that this may be the generic situation for a large part of the parameter space. For example, considering two of the 
most widely used non-separable metrics in relativistic astrophysics, the deformed Kerr Johannsen-Psaltis metric~\cite{Johannsen:2011dh} 
and the slow rotation Hartle-Thorne metric~\cite{Hartle1967, HT68}, we find that although neither spacetime possesses spherical orbits, orbits of 
the spheroidal type are admitted in both cases (another example is provided by the orbits outside black holes with scalar hair discussed 
in~\cite{Cunha2017b}). However, these orbits too are lost (assuming they cross the equatorial plane) in conjunction with the disappearance 
of the equatorial photon ring when the spin is high and/or the deviation from Kerr is large. The emergence of non-equatorial 
photon rings with their associated new families of photon trapping orbits adds a new layer of non-Kerr phenomenology.

Our results are best summarised if we plot side by side, see Fig.~\ref{fig:KerrvsJP}, Kerr photon trapping orbits against their counterparts in a
strongly deformed Johannsen-Psaltis spacetime assuming the same spin and orbital impact parameter $b$ (the comparison between Kerr and
Hartle-Thorne orbits is similar). One can observe how the two cases are similar in the low $b$ range but gradually deviate as 
we move towards higher $b$.

The deeper astrophysical motivation behind this work lies in the aforementioned role played by circular photon orbits in
creating a black hole shadow and in the GW ringdown produced in the final stage of black hole mergers. 
The key element in both phenomena is the ability of trapping photons/wavepackets in orbit around the black hole for a time 
interval much longer than the system's dynamical timescale. As our results suggest, this ability could be compromised or 
modified if the Kerr metric  were to be replaced by a non-separable metric endowed with strongly non-Kerr spheroidal photon orbits 
or  if  spherical/spheroidal orbits are not  supported at all. The far-reaching consequence of this conclusion is that non-Kerr ultracompact 
objects may look very different compared to Kerr black holes with respect  to their shadow image and QNM ringdown signal.

The remainder of the paper is organised as follows. Sections~\ref{sec:formalism} contains the necessary formalism for
describing photon geodesics in an axisymmetric-stationary metric. In Section~\ref{sec:spherical} we focus on circular motion
and discuss the distinction between spherical and spheroidal non-equatorial orbits. In Section~\ref{sec:circularity} 
we derive the spheroidicity condition describing these orbits. In parallel with our discussion of spherical/spheroidal orbits, in 
Section~\ref{sec:photonring} we study the possibility of having non-equatorial photon rings.
The connection between spherical orbits and the spacetime's separability is the subject of Section~\ref{sec:theorem}. 
In Section~\ref{sec:nonKerr_circ} we search for spherical/spheroidal orbits in three different non-Kerr spacetimes 
(Johannsen-Psaltis, Johannsen and Hartle-Thorne) by means of analytic and numerical solutions of the spheroidicity condition. 
A complementary time-domain study of these orbits is the subject of Section~\ref{sec:tdomain}. Finally, in Section~\ref{sec:conclusions} 
we summarise our results and discuss their implications for the observational signature of non-Kerr objects.


\section{Formalism for general null geodesics}
\label{sec:formalism}

For the general purpose of this paper we consider an arbitrary axisymmetric, stationary and equatorial-symmetric 
spacetime described by a metric $g_{\mu\nu} (r,\theta)$ in a spherical-like coordinate system. 
The spacetime is assumed to satisfy the circularity conditions associated with its two Killing vectors \cite{wald1984}
and by exercising our coordinate choice freedom, the spatial coordinates are all orthogonal. 
The resulting line element takes the form
\be
ds^2 = g_{tt} dt^2 + g_{rr} dr^2 + 2g_{t\varphi} dt d\varphi + g_{\theta\theta} d\theta^2 + g_{\varphi\varphi} d\varphi^2,
\label{metric}
\ee
Geodesics in this spacetime conserve the energy $E$ and the angular momentum component $L$  along the symmetry axis 
(here given per unit mass), $E = -u_t,~L = u_\varphi$, where $u^\mu = dx^\mu/d\lambda$ 
is the four-velocity along the geodesic. Defining the impact parameter $b = L/E$ and rescaling the affine parameter 
$\lambda \to E \lambda$ we can effectively set  $E\to1$ and $ L\to b$ everywhere. These expressions can be inverted to give
\be
u^t = \frac{1}{\cD} \left (\,  g_{t\varphi} b + g_{\varphi\varphi}  \, \right ), \quad
u^\varphi = -\frac{1}{\cD} \left (\,  g_{t\varphi}  + g_{tt} b \, \right ), 
\qquad \cD = g_{t\varphi}^2  - g_{tt} g_{\varphi\varphi}.
\ee
The location of the horizon (if present) is marked by $\cD = 0$; outside the horizon this parameter is positive. 

Assuming null geodesics hereafter, the norm $u^\mu u_\mu = 0$ leads to
\be
 g^{rr} u_r^2 + g^{\theta\theta} u_\theta^2  = \frac{1}{\cD} \left (\, g_{tt} b^2 + 2 g_{t\varphi} b + g_{\varphi\varphi} \, \right )
\equiv   V_{\rm eff} (r,\theta,b),
\label{norm_gen}
\ee
where the effective potential $V_{\rm eff}$ shares the same symmetry properties as the metric. 

In the case of the Kerr metric, the existence of a Carter constant allows the decoupling of the radial 
and meridional motion, with (\ref{norm_gen}) becoming a purely radial equation 
[see Appendix~(\ref{sec:Kerr}) for details].
The `miraculous' property of a third constant is absent in a general axisymmetric-stationary spacetime.  
Instead one is obliged to work with the second-order geodesic equation, which can be written as:
\be
\alpha_\kappa \equiv \frac{d u_\kappa}{d\lambda} = \frac{1}{2} g_{\mu\nu,\kappa} u^\mu u^\nu.
\label{geod}
\ee
The $\theta$-component of this equation is the only one needed here,
\begin{align}
\alpha_\theta &= \frac{1}{2}  \left [\, g_{tt,\theta} (u^t)^2 + g_{rr,\theta} (u^r)^2 + g_{\theta\theta,\theta} (u^\theta)^2
+  g_{\varphi\varphi,\theta} (u^\varphi)^2 + 2  g_{t\varphi,\theta} u^\varphi u^t \, \right ]
\nn \\
& =  \frac{1}{2}  \left (\, \frac{g_{rr,\theta}}{g_{rr}^2} u^2_r + \frac{g_{\theta\theta,\theta}}{g_{\theta\theta}^2} u_\theta^2 \, \right )
+ \frac{1}{2 \cD^2} \Big [\,  g^4_{t\varphi} V_{{\rm eff},\theta} - g^2_{t\varphi} g_{\varphi\varphi} ( g_{tt}  V_{\rm eff} )_{,\theta} 
+ g_{tt} g_{\varphi\varphi} \left \{ (g^2_{t\varphi})_{,\theta} -g_{tt}   g_{\varphi\varphi,\theta}  \right \} V_{\rm eff}  
\nn\\
&\quad + 2 b g_{\varphi\varphi}  ( g_{t\varphi} g_{tt,\theta} - g_{tt} g_{t\varphi ,\theta} ) + g_{\varphi\varphi}^2 g_{tt,\theta} 
-   g_{tt} g_{\varphi\varphi} g_{\varphi\varphi,\theta}  \, \Big ],
%
\label{ath}
\end{align}
For the following discussion of circular orbits we also need to involve the $\lambda$-derivative of (\ref{norm_gen}).
This is,
\be
  \frac{u_r}{g_{rr}} \left ( \,  2 \alpha_r  - \frac{g_{rr,r}}{g_{rr}^2} u_r^2 -  \frac{g_{rr,\theta}}{g_{rr} g_{\theta\theta}} u_r u_\theta \, \right ) 
+ \frac{u_\theta}{g_{\theta\theta}} \left (\,   2 \alpha_\theta  - \frac{g_{\theta\theta,\theta}}{g_{\theta\theta}^2} u_\theta^2
 -  \frac{g_{\theta\theta,r}}{g_{rr} g_{\theta\theta}} u_r u_\theta   \, \right ) 
 =   \frac{u_r}{g_{rr}} V_{{\rm eff},r}  + \frac{u_\theta}{g_{\theta\theta}} V_{{\rm eff},\theta}.
\label{dnorm_gen}
\ee
It can be verified that the insertion of $\alpha_\theta, \alpha_r $ [as computed from (\ref{geod})]  
into (\ref{dnorm_gen}) returns a trivial $0=0$ result.

Returning to Eq.~(\ref{norm_gen}), one can observe that its quadratic form implies that $V_{\rm eff}$ should act 
as a zero-velocity separatrix between allowed and forbidden regions for geodesic motion. This follows from 
\be
g_{rr} (u^r)^2 + g_{\theta\theta} (u^\theta )^2 = 0  \quad \Leftrightarrow \quad  u^r = u^\theta =0 \quad \Leftrightarrow \quad
V_{\rm eff} = 0.
\ee 
Then $ V_{\rm eff} <0 $ ($V_{\rm eff} >0$) marks the allowed (forbidden) region.  Some examples of this 
are given below in Section~\ref{sec:JP}.


\section{Circular, spherical and spheroidal orbits}
\label{sec:spherical}

It is perhaps instinctive to think of circular motion as that associated with a constant radius $r=r_0$.
For non-equatorial motion, a more accurate designation for these orbits would be \emph{spherical} since the trajectory 
is confined on the surface of a sphere of radius $r_0$. This is, for example, how Kerr spherical orbits look like in the familiar 
Boyer-Lindquist coordinates. However, there is a more general way of defining non-equatorial circular orbits, namely, as motion 
confined on a spheroidal-shaped shell $r_0 = r_0 (\theta)$. We shall call this more general type of orbit \emph{spheroidal}. 
From a mathematical point of view we demand the function $r_0 (\theta)$ to be smooth and expandable in even-order Legendre 
polynomials,
\be
r_0 (\theta) = \sum_{\ell}  \beta_\ell P_{2\ell} (\cos\theta),
\label{r0_expansion}
\ee
where $\ell =0,1,2, ...$ and $\beta_\ell$ are constant coefficients. Note that according to this definition $r_0 (\theta)$ is not
required to be single-valued, so that the `spheroidal' shell may actually be torus-shaped (as some of the orbits discussed 
in~\cite{Cunha2017b}). In this case the orbit could intersect the equatorial plane in two distinct radii instead of one.  

Strictly speaking, the distinction between spherical and spheroidal orbits  could be seen as a non-physical gauge degree of 
freedom, in the sense that any spheroidal orbit $r_0 (\theta)$ could be reduced to a spherical one with the help of a suitable coordinate 
transformation (indeed, the new radial coordinate would be given by $u(r,\theta) = r-r_0(\theta)$). Putting aside possible complications 
related to a multivalued and/or non globally defined $r_0 (\theta)$, this procedure of `gauging out' spheroidal orbits 
should be always feasible \emph{provided} we allow for a metric more general than our assumed form (\ref{metric}),  more specifically, 
a metric with an extra mixed $g_{r\theta}$ component (as discussed below, the coordinate transformation 
may not always exist if the metric is assumed to retain the form (\ref{metric})). There is one mathematical and a second 
more practical reason why we do not opt for this approach:  (i) if present, the property of separability requires a metric like (\ref{metric}) 
in order to work, and (ii) all non-Kerr axisymmetric-stationary spacetimes in the literature are of the form (\ref{metric}); transforming one of 
them according to the previous recipe would require a prior knowledge of $r_0 (\theta)$, thus defeating the purpose of the
 entire exercise.

Sticking with our definition of spheroidal orbits we find that $u^r, u^\theta$ are `locked'  to each other,
\be
u^r = r_0^\prime u^\theta \quad \Rightarrow \quad  u_r =  \frac{g_{rr}}{g_{\theta\theta}} r_0^\prime u_\theta,
\label{uruth_circ}
\ee
where a prime stands for a derivative with respect to the argument. 
In this and the following expressions all functions of $r$ are to be evaluated at $r=r_0 (\theta)$.

Taking the $\lambda$-derivative of (\ref{uruth_circ}),
\be
\alpha_r = \frac{g_{rr}}{g_{\theta\theta}} r_0^\p \alpha_\theta + \frac{u_\theta^2}{g^3_{\theta\theta}} 
\Bigg [\, g_{\theta\theta} g_{rr} r_0^{\pp} + r_0^\p \left (\, g_{\theta\theta} g_{rr,\theta} - g_{rr} g_{\theta\theta,\theta}  \, \right ) 
+ (r_0^\p )^2 \left (\,  g_{\theta\theta} g_{rr,r} - g_{rr} g_{\theta\theta,r}   \, \right ) \,\Bigg ].
\label{ar_circ}
\ee
Using (\ref{uruth_circ}) in (\ref{ath}),
\begin{align}
\alpha_\theta &= \frac{ u_\theta^2}{2 g_{\theta \theta}^2}  \left  [ \, g_{rr,\theta} (r_0^\p )^2  +  g_{\theta \theta,\theta} \, \right ]
+ \frac{1}{2 \cD^2} \Big [\,  g^4_{t\varphi} V_{{\rm eff},\theta} - g^2_{t\varphi} g_{\varphi\varphi} ( g_{tt}  V_{\rm eff} )_{,\theta} 
+ g_{tt} g_{\varphi\varphi} \left \{ (g^2_{t\varphi})_{,\theta} -g_{tt}   g_{\varphi\varphi,\theta}  \right \} V_{\rm eff}  
\nn\\
&\quad + 2 b g_{\varphi\varphi}  ( g_{t\varphi} g_{tt,\theta} - g_{tt} g_{t\varphi ,\theta} ) + g_{\varphi\varphi}^2 g_{tt,\theta} 
-   g_{tt} g_{\varphi\varphi} g_{\varphi\varphi,\theta}  \, \Big ].
\label{ath_circ}
\end{align}
Meanwhile, from Eqs. (\ref{norm_gen}) and (\ref{dnorm_gen}) we obtain respectively
\begin{align}
&  \left [ g_{rr} (r^\p_0)^2 + g_{\theta\theta} \right ] u_\theta^2 = g^2_{\theta\theta} V_{\rm eff}, 
\label{norm_circ1}
\\
\nn \\
& r_0^\p \left [ \, 2\alpha_r - \frac{r_0^\p u_\theta^2}{g_{\theta\theta}^2} \left ( \,  g_{rr,\theta} + g_{rr,r} r_0^\p   \, \right )   \, \right ] 
+ 2\alpha_\theta -\frac{u^2_\theta}{g_{\theta\theta}^2} 
\left (\,  g_{\theta\theta,\theta} + g_{\theta\theta,r} r_0^\p \, \right )  =   r^\p_0 V_{\mathrm{eff},r} + V_{\mathrm{eff},\theta}.
\label{dnorm_circ1}
\end{align}


\section{The spheroidicity condition}
\label{sec:circularity}

The previous equations pertaining general non-equatorial spheroidal motion in an arbitrary axisymmetric-stationary
metric can be combined to produce a necessary spheroidicity condition of  the functional form $f(b,r_0,r_0^\p,r_0^{\pp}, \theta)=0$. 

This constraint originates from Eq.~(\ref{dnorm_circ1}) (i.e.  essentially the $\lambda$-derivative of $V_{\rm eff}$) 
after using (\ref{ar_circ})-(\ref{norm_circ1}) to eliminate $\alpha_r$, $\alpha_\theta$ and $u^2_\theta$, respectively.
Once these steps are taken and several terms are combined to form $V_{\rm eff}$ and its derivatives, we arrive at:
\begin{align}
& g_{rr} (r_0^\p)^3 \left ( g_{rr}  V_{\rm eff} \right )_{,\theta} 
+ (r_0^\p)^2 \Bigg [\,    \left (\, g_{\theta\theta} g_{rr,r} -2g_{rr} g_{\theta\theta,r} \,\right ) V_{\rm eff}  - g_{rr} g_{\theta\theta}  V_{{\rm eff},r}  \, \Bigg ]
\nn \\
& +  r_0^\p \Bigg [\,   \left (\,  2 g_{\theta\theta} g_{rr,\theta} -g_{rr} g_{\theta\theta,\theta} \, \right )  V_{\rm eff} 
+  g_{rr} g_{\theta\theta} V_{{\rm eff},\theta} \, \Bigg ] + g_{\theta\theta}  \Bigg [\,    2 g_{rr} V_{\rm eff} r_0^{\pp} 
- \left ( g_{\theta\theta} V_{\rm eff} \right )_{,r}  \, \Bigg ] = 0,
\label{circularity}
\end{align}
where all functions are to be evaluated at $r=r_0 (\theta)$. This equation will become our basic tool for searching for spherical/spheroidal 
orbits in non-Kerr spacetimes (Section~\ref{sec:nonKerr_circ}). It should be noted that (\ref{circularity}) is oblivious to the stability 
of the spheroidal orbit. This extra information is contained in the second derivatives of $V_{\rm eff}$.

As a sanity check of (\ref{circularity}) we consider the Kerr metric in Boyer-Lindquist coordinates with the 
assumption $r_0^\p=0$. The spheroidicity condition reduces to
\be
\cE_\rK (r_0) \equiv r_0^2 (r_0 -3M) + a^2 (M+ r_0) + ab (r_0-M)=0,
\label{eqphKerr}
\ee
which can be identified as one of the two equations that determine non-equatorial Kerr photon orbits
(for $a=0$ this leads to the Schwarzschild photon ring $r_0=3M$). 

This example is indicative of what happens when a spacetime admits $r_0 = \mbox{const.}$ spherical orbits: 
the spheroidicity condition effectively becomes a $\theta$-independent equation for $r_0$.
As we shall see below in Section~\ref{sec:J}, a similar situation arises in the context of the 
separable deformed Kerr metric devised by Johannsen \cite{Johannsen2013PhRvD}.


\section{Non-equatorial photon rings}
\label{sec:photonring}

Apart from the aforementioned 3-D spherical/spheroidal orbits a spacetime may admit 2-D photon rings
where motion takes place along a  trajectory of constant $r$ and $\theta$ on a plane. The equatorial photon ring is the most familiar
example of this family of orbits and is of course a well-known feature of the Kerr spacetime. Interestingly, non-Kerr spacetimes 
may show a richer phenomenology, admitting a symmetric pair of non-equatorial photon rings instead of a single equatorial one. 
This is the subject explored in this section.

We can begin our analysis from Eqs. (\ref{norm_circ1}) and (\ref{dnorm_circ1}) which are still valid for photon rings. 
Specialising to motion with constant $r=r_0$ and $\theta=\theta_0$ (i.e. we need to set $r_0^\p =u_\theta = \alpha_\theta = 0$ 
in the two equations) we arrive to the following three conditions for $V_{\rm eff} (r,\theta,b)$:
\be
V_{\rm eff}  = V_{\mathrm{eff},r}  = V_{\mathrm{eff},\theta} = 0, \qquad \mbox{at} \quad (r,\theta)=(r_0,\theta_0).
\label{ring_conditions}
\ee
These three equations can be solved with respect to $\{r_0,\theta_0, b=b_0\}$;  such a solution implies the existence of a 
non-equatorial photon ring. 

Equatorial photon rings are somewhat simpler to deal with, since  $V_{\mathrm{eff},\theta} (\pi/2)=0$ due to the spacetime's symmetry. 
In this case the problem reduces to the familiar conditions for circular equatorial orbits, i.e. $V_{\rm eff} (r_0,b_0) = V_{{\rm eff},r} (r_0,b_0) =0 $. 
As discussed, for example, in~\cite{Glampedakis:2017}  these two conditions lead, respectively, to equations for the photon ring 
radius and its associated impact parameter (here the upper/lower sign corresponds to prograde/retrograde motion)
\begin{align}
& g_{\varphi\varphi}  \left (g_{tt,r} \right )^2   + 2 g_{tt} \left (g_{t\varphi,r}\right)^2 -g_{tt,r} \left (\, g_{tt} g_{\varphi\varphi,r}
+ 2 g_{t\varphi} g_{t\varphi,r}   \,\right ) 
\mp 2 \sqrt{ \left ( g_{t\varphi,r} \right)^2 - g_{tt,r} g_{\varphi\varphi,r}} \left  (\, g_{t\varphi} g_{tt,r} 
- g_{tt} g_{t\varphi,r} \, \right ) = 0,
\label{eqphring}
\\
\nn \\
& b_0=  \frac{1}{ g_{tt,r}} 
\left [\, \mp \sqrt{(g_{t\varphi,r})^2 - g_{tt,r} g_{\varphi\varphi,r}} - g_{t\varphi,r} \, \right ].
\label{eqbph}
\end{align}

The impact parameter of non-equatorial photon rings is given by the same result (\ref{eqbph}). However, instead of a
single photon ring equation we now have a coupled algebraic system for $\{r_0,\cos\theta_0\}$ coming from 
$V_{\rm eff} = V_{\mathrm{eff},\theta}  = 0  |_{b=b_0}$. The former condition has the same form as Eq. (\ref{eqphring}), 
albeit with a $\theta$-dependent metric. The latter condition leads to
\begin{align}
& g_{\varphi\varphi,\theta}  \left (g_{tt,r} \right )^2   + 2 g_{tt,\theta} \left (g_{t\varphi,r}\right)^2 
-g_{tt,r} \left ( \, g_{tt,\theta} g_{\varphi\varphi,r} + 2 g_{t\varphi,\theta} g_{t\varphi,r} \, \right ) 
\mp 2 \sqrt{ \left ( g_{t\varphi,r} \right)^2 - g_{tt,r} g_{\varphi\varphi,r}} 
\left  ( \, g_{t\varphi,\theta} g_{tt,r} - g_{tt,\theta} g_{t\varphi,r} \, \right ) = 0.
\label{eqphringth}
\end{align}
These equations are solved below in Section~\ref{sec:nonKerr_circ}; there we will see that particular examples of non-Kerr spacetimes 
lead to non-trivial results. The trivial application of these equations is the Kerr spacetime itself where we can demonstrate the impossibility
of non-equatorial photon rings. Indeed, after combining (\ref{eqphring}) and (\ref{eqphringth})  to eliminate the square root term
we obtain $a \cos\theta_0 \left ( a^2\cos 2\theta_0 + a^2  - 6r_0^2 \right ) \left ( a^2\cos 2\theta_0 + a^2  - 2r_0^2 \right )^2 = 0$ which, 
given that $r_0 \geq a$, has $\theta_0 = \pi/2$ as the only acceptable root.


\section{On the connection between spherical orbits \& separability}
\label{sec:theorem}

In the introduction spherical photon orbits were described as a special characteristic of the Kerr spacetime and in particular
of its separable nature. Before embarking on our study of photon trapping orbits in concrete examples of non-Kerr spacetimes it is 
worthwhile to take a detour and examine in more detail the connection between spherical photon orbits  and the separability of a given
spacetime.

The spheroidicity condition (\ref{circularity})  provides the means to establish a remarkable result that can be stated as follows:
if a stationary-axisymmetric spacetime endowed with a photon ring is separable  (in the sense that it admits a third 
integral of motion) then it necessarily admits spherical photon orbits (i.e. orbits with $r_0 =\mbox{const.}$). 
The converse is true in a subset of the solution space of (\ref{circularity}) with $r_0 =\mbox{const.}$ where one needs to further assume 
that the ratio of the metric components $g_{rr}, g_{\theta\theta}$ takes a special factorised form $g_{\theta\theta}/g_{rr} = f(r) h(\theta)$, where 
$h(\theta)$ is a specifically selected function.

A corollary of this proposition is that a spacetime is  \emph{necessarily non-separable} if it possesses a photon ring 
but does not admit spherical orbits in any coordinate system. The rest of this section is devoted to the derivation of these results; 
it should be noted that the sphericity-separability connection is not exclusively about photons but it can be extended to the orbits 
of massive particles (this is discussed in more detail in  Appendix~\ref{sec:particles}). Nor it is exclusively `relativistic'  as it can be shown 
to hold in the context of Newtonian gravity (see Appendix~\ref{sec:Newtcirc}).

For a spacetime $g_{\mu\nu}$ of the general form (\ref{metric}) with  $\{r,\theta\}$ the only non-ignorable coordinates, the 
Hamilton-Jacobi equation for null geodesics becomes \cite{MTW1973}, 
\be 
\frac{(S_{,r})^2}{g_{rr}}  +\frac{(S_{,\theta})^2}{g_{\theta\theta}}  -V_{\textrm{eff}}=0, 
\label{HJ1}
\ee
where $ S(r,\theta)$ is Hamilton's characteristic function. Following the standard separability ansatz~\cite{LLbook} we write 
$S= S_r(r) + S_\theta (\theta)$. Provided the following conditions hold (here $f_1, f_2,h,g$ are arbitrary functions of their argument),
\begin{align}
& g_{\theta\theta} V_{\textrm{eff}}= f_1 (r) h(\theta) + g(\theta), 
\label{sepa_condition1}
\\
\nn \\
& \frac{g_{\theta\theta}}{g_{rr}} = f_2 (r) h(\theta),
\label{sepa_condition2}
\end{align}
we can rearrange (\ref{HJ1}) as, 
\be
f_2(r)  (S_r^\p )^2 - f_1 (r) = \frac{1}{h(\theta)} \left [\, g(\theta) - ( S_{\theta}^\p)^2 \, \right ] = \cC.
\label{HJ2}
\ee
This demonstrates the separability of the system, with $\cC$ playing the role of the third constant (or `Carter constant'). 
On the same issue of separability, Carter~\cite{Carter1968CMaPh} showed that the Hamilton-Jacobi equation as well as
the Schr\"odinger and scalar wave equations are all separable if the metric of a given spacetime can be put in the `canonical' form 
(see \cite{Frolov2017LRR} for a recent review on the subject),
\be 
ds^2= \frac{ Z}{\Delta_r}dr^2 + \frac{Z}{\Delta_{\theta}}d\theta^2  +\frac{\Delta_{\theta}}{Z}  \left (\, P_r d\varphi -Q_r dt \, \right )^2 
+ \frac{\Delta_r}{Z} \left (\, Q_{\theta} dt - P_{\theta} d\varphi \, \right )^2 ,
\label{canonical_metric}
\ee
where $Z=P_r Q_{\theta}-Q_r P_{\theta}$ and $\Delta_\mu = \Delta_\mu (\mu)$ for $\mu = \{r,\theta\}$ (and similarly for the other 
arbitrary functions). One can easily verify that the canonical metric (\ref{canonical_metric}) satisfies the separability 
conditions (\ref{sepa_condition1}), (\ref{sepa_condition2}). As examples of this privileged class of spacetimes we can 
mention the Kerr and Johannsen metrics (although only the former is a solution of the GR field equations).

We now can make contact with the existence of spherical photon orbits. 
According to the spheroidicity condition (\ref{circularity}) an $r_0 = \mbox{const.}$ orbit must satisfy
\be
\left ( g_{\theta\theta} V_{\rm eff} \right )_{,r} |_{r_0} =0 . 
\label{sphericity1}
\ee
The most general solution of this equation takes the form, 
$g_{\theta\theta}V_{\rm eff} = f(r,\theta)(r-r_0)^2 h(\theta) + g(\theta)$,\footnote{This form cannot separate the Hamilton-Jacobi equation.} 
with $f(r,\theta)$ a non-singular function at $r_0$. A special case of this latter form is
\be 
g_{\theta\theta} V_{\rm eff}= f_1(r) h(\theta)+g(\theta),
\label{sphericity2}
\ee
with the additional constraint $f_1^{\p}(r_0)=0$. Now, the expression (\ref{sphericity2}) can be identified as the 
first separability condition (\ref{sepa_condition1}), but separability alone cannot enforce the existence of spherical orbits.
This is the point where we need to invoke the existence of a photon ring in the spacetime under
consideration: for some $\cC=\cC_0$ this requires $V_{r,r} = 0 \Rightarrow  f_1^\p = 0$, with the 
last equation now becoming a relation $r_0 = r_0 (b)$, common for both equatorial and non-equatorial motion as it
depends neither on $\cC$ nor on $\theta$ [in Kerr, this relation is given by Eq.~(\ref{eqphKerr})].

We have thus established the first half of the proposition, namely,  that Hamilton-Jacobi separability 
entails the existence of spherical photon orbits as long as $f_1^{\p}(r)=0$ has roots for some values of $r$. 
As mentioned, the veracity of the converse rests on working within the subclass of solutions of Eq.~(\ref{sphericity1})
that satisfy Eq.~(\ref{sphericity2}) and the assumption that $g_{\theta\theta}/g_{rr}$ is of the form (\ref{sepa_condition2}).

Going beyond our basic result, one can show that  spheroidal orbits \emph{cannot exist} in a separable spacetime 
(and obviously in those coordinates that allow separabilty in the first place). 
Using $ (u_r, u_\theta) = (S_r^\p, S_\theta^\p)$ in (\ref{HJ2}), we obtain a pair of decoupled equations,
\be
 f_2 \left  ( g_{rr} u^r  \right )^2 =  f_1 + \cC \equiv V_r (r,b,\cC),  \qquad  
 \left  ( f_2 g_{rr} u^\theta \right )^2 = \frac{g-h \cC}{h^2} \equiv V_\theta (\theta,b,\cC).
\label{decoupled1}
\ee
Assuming first a spheroidal orbit, $u^r = r_0^\p u^\theta$, these two equations combine to give,
\be
(r_0^\p)^2 V_\theta = f_2 (r_0) V_r (r_0)\equiv \tilde{V}_r (r_0).
\label{spheroeq}
\ee
We can then see that both potentials obey  $V_\theta \geq 0$, $\tilde{V}_r (r_0)\geq 0$. 
At the meridional turning points $\theta_t$ we have $V_\theta (\theta_t) = 0 $ which  means that $\tilde{V}_r [r_0 (\theta_t)]=0$.
A similar argument can be used in the equatorial plane where $ r^\p_0 (\pi/2) =0 $ due to the assumed symmetry of the orbit, 
hence leading to  $\tilde{V}_r [r_0 (\pi/2)]=0$. Given that $\tilde{V}_r (r_0)$ cannot be negative, it can be either zero or
increase and subsequently decrease as $\theta$ moves from $\theta_t$ to the equator. The situation is exactly the same in 
the lower hemisphere and therefore we should have $\tilde{V}^\p_r [r_0(\pi/2)]=0$. Taking the derivative of (\ref{spheroeq}),
\be
2r_0^{\pp} V_\theta+ (r_0^\p) V_\theta^\p = \tilde{V}^\p_r (r_0),
\ee
we can deduce that $r_0^{\pp}(\pi/2)=0$. Using iteration, it is easy to show that the same is true for all higher derivatives $r_0^{(n)} (\pi/2)$.
Combined with $ r^\p_0 (\pi/2) = 0$, this entails that any orbit $r_0 (\theta)$ crossing the equatorial plane can only be an 
$r_0 =r_0 (\pi/2) = r_0 (\theta_t) =\mbox{const.}$ orbit. Therefore, the only possibility is that of spherical orbits in the separability coordinates.

A subtle point of our discussion on spherical orbits  is their inherent coordinate dependence, in the sense
that they occur if one uses the appropriate coordinate system. Spherical orbits should not occur if a different non-separable
coordinate system is employed, but instead one would expect to encounter spheroidal orbits $r_0 (\theta)$, in full agreement with our 
previous result, since the existence of spherical orbits in any other coordinates apart from those that separate the Hamilton-Jacobi equation would 
imply the existence of spheroidal orbits in the first coordinates that allow separability.  An example of this situation is 
provided by the Newtonian Euler potential, see Appendix~\ref{sec:Newtcirc}.

The sphericity-separability interplay is nicely demonstrated by the three non-Kerr spacetimes we study in the following section.
One of them (the Johannsen metric) is separable and, as we are about to see, it admits Kerr-like spherical photon orbits. In contrast, 
the remaining two non-separable spacetimes do not admit such orbits but nevertheless they appear to support the more general spheroidal ones.


\section{Searching for spheroidal orbits in non-Kerr spacetimes}
\label{sec:nonKerr_circ}

\subsection{Strategy}

After having explored the link between the separability of a given spacetime and the existence of 
spherical orbits we go on to consider specific examples of both separable and non-separable metrics.
This case-by-case analysis comprises the deformed Kerr metrics devised by Johannsen-Psaltis 
\cite{Johannsen:2011dh} and Johannsen \cite{Johannsen2013PhRvD}, and the celebrated 
Hartle-Thorne metric \cite{Hartle1967, HT68} which is the `official' GR solution describing the interior and exterior 
spacetime of relativistic stars within a slow-rotation expansion scheme. For brevity, hereafter these three metrics will be 
denoted as `JP', `J' and `HT' respectively.  Amongst these metrics only the J  is separable (by construction) and admits a 
Carter-like constant while the other two acquire this property only in their respective Kerr limits. 

Our overall strategy is based on a two-pronged approach. The first approach is based on the necessary spheroidicity condition
for the existence of spherical/spheroidal orbits. In this section we solve  this condition analytically in the weakly deformed ($\varepsilon_3 \ll 1$) 
JP metric and in the HT metric after being perturbatively expanded with respect to the spin. We show that none of the two spacetimes admits 
spherical orbits in the coordinates that they are given. However, the HT spacetime admits an exact spheroidal solution while in the JP spacetime 
a spheroidal solution can be found in the approximate form of a truncated convergent series. In sharp contrast, we find that spherical orbits are 
allowed in the J metric. 

Solving the spheroidicity condition for the full, unexpanded JP/HT metrics requires a numerical integration; this calculation is performed in this
section. The results obtained suggest the presence of spheroidal orbits in some parts of the parameter space.

This first approach leaves some loose ends in relation with the degree of `decircularization' of the orbits from sphericity. 
This issue is the subject of the second approach and is addressed in Section~\ref{sec:tdomain} with the help of direct numerical integration 
of the geodesic equations. 

Apart from our study of spheroidal orbits, we make contact with Section~\ref{sec:photonring} and apply the results obtained there
to the three aforementioned spacetimes, with the purpose of finding non-equatorial photon rings. To what extent their presence could 
affect the `photon trapping' ability of a non-Kerr spacetime is a topic explored in Section~\ref{sec:tdomain}.

Before embarking on the orbital analysis of the three spacetimes it is worth pausing a moment to 
revisit circular/spherical motion in Kerr. This topic is of course well documented and studied in the literature 
(see e.g. \cite{Bardeen:1972fi,Wilkins1972}) but for the purpose of completeness a brief discussion can be found in 
Appendix~\ref{sec:Kerr}.


\subsection{The Johannsen-Psaltis metric}
\label{sec:JP}

The JP metric belongs to the broader class of the so-called deformed Kerr metrics, the `deformation' in this
instance encoded in the function 
\be
h (r,\theta) =  \varepsilon_3\frac{M^3 r}{\Sigma^2},
\label{JPdeviation}
\ee
where $\varepsilon_3$ is a constant parameter. In terms of the Kerr metric $g_{\mu\nu}^\rK$ 
(listed in Appendix~\ref{sec:Kerr}), the JP metric reads
\begin{align}
& g_{tt}^\rJP = (1+ h) g^\rK_{tt}, \quad g_{t\varphi}^\rJP = (1+ h) g^\rK_{t\varphi}, \qquad 
g_{rr}^\rJP = g_{rr}^\rK (1+h) \left (1+h \frac{a^2 \sin^2\theta}{\Delta} \right )^{-1},
\nn \\
& g_{\theta\theta}^\rJP = g_{\theta\theta}^\rK, \qquad 
g_{\varphi\varphi}^\rJP = g_{\varphi\varphi}^\rK  + h a^2 \left (1 + \frac{2Mr}{\Sigma} \right ) \sin^4\theta,
\end{align}
and it is clear that $\varepsilon_3 \to 0$ corresponds to the Kerr limit. 

The search for spherical/spheroidal orbits is greatly facilitated if we restrict ourselves to a \emph{small deformation} 
$\varepsilon_3$ and work perturbatively with respect to that parameter. We thus consider the ${\cal O} (\varepsilon_3)$ 
`post-Kerr' form of the  JP metric. 

Assuming spherical orbits, the spheroidicity condition (\ref{circularity}) becomes,
\begin{align}
& -16 \left (  r^6_0 +  a^6 \cos^6 \theta \right ) \left(a^2-a b+r^2_0 \right) \cE_\rK (r_0) 
-8 \varepsilon_3 M^3 r^3_0 \left [\, 3 a^4 (4 M+3r_0) \right.
\nn \\
& \left.  -24 a^3 b M+  3 a^2 b^2 (4 M-3 r_0)+2 a^2 r^2_0 (7 r_0-4 M)  + 8 a b M r^2_0 + r^4_0 (5 r_0-12 M)\, \right ] 
\nn \\
&   +\cos^4 \theta  \left [\, 8 a^4 \varepsilon_3 M^3 \left \{\, a^4-a^2 \left(b^2-6 r^2_0 \right)+r^3 (5r_0 -8 M) \, \right \} 
-48 a^4 r^2_0 \left(a^2-a b+r^2_0 \right) \cE_\rK (r_0) \,\right ]
 \nn \\
 &  +\cos^2 \theta \left [\, -48 a^2 r^4_0 \left(a^2-a b+r^2_0 \right) \cE_\rK (r_0)  -32 a^2 \varepsilon_3 M^3 r_0 \left \{\, a^4 (M+2 r_0)-2 a^3 b M 
 \right. \right.
 \nn \\
& \left. \left. +a^2 b^2 (M-2 r_0)+2 a^2 r^2_0 (r_0-3M) +6 a b M r^2_0 -3 M r^4_0 \, \right \} \, \right ] =0 + {\cal O} \left (\varepsilon_3^2 \right ).
\label{circularityJP1}
\end{align}
where the function $  \cE_\rK (r_0)$ was introduced back in Eq. (\ref{eqphKerr}).
The trigonometric functions can be expressed in Legendre polynomials,
\begin{align}
& 2112 a^4 \varepsilon_3 M^3  \left [\, a^4-a^2 \left(b^2-6 r^2_0 \right)+r^3_0 (5r_0 -8M) \, \right ] P_4 (\theta)
\nn \\
& - \left(a^2-a b+r^2_0 \right) \cE_\rK(r_0) \left [\, 1280 a^6  P_6 (\theta) + 1152 a^4  \left(5 a^2+11 r^2_0 \right)  P_4 (\theta) \right.
\nn \\
& \left. + 1760 a^2  \left(5 a^4+18 a^2 r^2_0 +21 r^4_0 \right) P_2 (\theta) + 528 \left(5 a^6+21 a^4 r^2_0 +35 a^2 r^4_0 +35 r^6_0 \right)  \, \right ]
\nn \\  
&   +1760  \varepsilon_3 a^2 M^3  \left [\, 3 a^6 - 3 a^4 b^2 -2 a^4 r_0 (7 M+5 r_0) +28 a^3 b M r_0 + a^2 r^3_0 (60 M-13 r_0) \right.
\nn \\ 
& \left. -14 a^2 b^2 r_0 (M-2 r_0) -84 a b M r^3_0 +42 M r^5_0 \,\right ] P_2 (\theta)  
-616 \varepsilon_3 M^3 \left [\, -3 a^8+a^6 \left(3b^2+20 Mr_0 +22 r^2_0\right) \right.
\nn \\
& \left. -40 a^5 b M r_0 +4 a^4 r_0 \left \{ 5 b^2 (M-2 r_0)+r^2_0 (21 M+40 r_0)\right\}-240 a^3 b M r^3_0
 \right.
\nn \\
& \left. +15 a^2  r^3_0 \left\{ 3 b^2  (4 M-3 r_0)+2 r^2_0 (7 r_0-6 M)\right\}  +120 a b M r^5_0 +15 r^7_0 (5 r_0-12 M)\, \right ] = 0
+ {\cal O} \left (\varepsilon_3^2 \right ).
\label{circularityJP1a}
\end{align}
An $r_0$ solution exists provided the coefficient of each $P_\ell$ term vanishes independently. 
It is straightforward to verify numerically that this is \emph{not} the case for any $r_0>M$. We can thus conclude that
spherical photon orbits do \emph{not} exist in the JP spacetime\footnote{The same conclusion can be reached by means
of a simpler calculation where (\ref{circularityJP1a}) is further expanded with respect to the spin up to ${\cal O}(a^2)$ precision. 
All Legendre polynomials in the resulting expression  share the same non-vanishing coefficient.}.

Having failed to find spherical orbits our next objective is to look for the more general spheroidal $r_0 (\theta)$ orbits. 
In the spirit of our previous post-Kerr approximation we employ an expansion (note that this is equivalent to an 
expansion in the $P_\ell (\theta)$ basis)
\be
r_0 (\theta) = r_\rK + \varepsilon_3 M \sum_{n=0}^{N}  \beta_n \cos^{2n}\theta + {\cal O} \left (\varepsilon_3^2 \right ),
\label{r0expansion}
\ee
where $\beta_n (a,b)$ are constants and $r_\rK (a,b)$ is the Kerr spherical orbit radius i.e. $\cE_\rK (r_\rK)=0$. 
Upon inserting (\ref{r0expansion}) in the spheroidicity condition, the leading order Kerr terms vanish identically leaving
an expression linear in $\varepsilon_3$. After expressing the trigonometric functions in terms of  $P_\ell$
we again arrive to an algebraic equation of the form (\ref{circularityJP1a}), with the maximum $\ell$-order 
depending on the chosen $N$.

To provide a concrete example, we truncate the expansion (\ref{r0expansion}) at $N=3$. The resulting spheroidicity condition 
contains all even-order $P_\ell$ in the range $0 \leq \ell \leq 18$ and therefore leads to an algebraic system of ten equations 
for the four expansion coefficients $\beta_0-\beta_3$. Symbolically, the $N=3$ system takes the following form:
\begin{align}
& P_{18}: \qquad  f_{18}  \beta_{3} = 0,
\\
& P_{16}: \qquad f_{16}^{(1)} \beta_2 +  f_{16}^{(2)} \beta_3 = 0,
\\
&  P_{14}: \qquad f_{14}^{(1)} \beta_1 +  f_{14}^{(2)} \beta_2 + f_{14}^{(3)} \beta_3  = 0, 
\\
&  P_{12}: \qquad f_{12}^{(1)} \beta_1 +  f_{12}^{(2)} \beta_2 + f_{12}^{(3)} \beta_3  = 0, 
\\
& P_{10}: \qquad  f_{10}^{(1)} \beta_0 +  f_{10}^{(2)} \beta_1 +   f_{10}^{(3)} \beta_2 +  f_{10}^{(4)} \beta_3 = 0,
\\
& P_{8}: \qquad   f_{8}^{(1)} \beta_0 +  f_{8}^{(2)} \beta_1 + f_{8}^{(3)} \beta_2 + f_{8}^{(4)} \beta_3 + f_{8}^{(5)} = 0,
\\
& P_{6}: \qquad  f_{6}^{(1)} \beta_0 + f_{6}^{(2)} \beta_1 + f_{6}^{(3)} \beta_2 + f_{6}^{(4)} \beta_3 + f_{6}^{(5)} = 0,
\\
& P_{4}: \qquad  f_{4}^{(1)} \beta_0 + f_{4}^{(2)} \beta_1 + f_{4}^{(3)} \beta_2 + f_{4}^{(4)} \beta_3 + f_{4}^{(5)} = 0,
\\
& P_{2}: \qquad  f_{2}^{(1)} \beta_0 + f_{2}^{(2)} \beta_1 + f_{2}^{(3)} \beta_2 + f_{2}^{(4)} \beta_3 + f_{2}^{(5)} = 0,
\\
&  P_{0}: \qquad  f_{0}^{(1)} \beta_0 + f_{0}^{(2)} \beta_1 + f_{0}^{(3)} \beta_2 + f_{0}^{(4)} \beta_3 + f_{0}^{(5)} = 0,
\end{align}
where $f_i^{(j)}= f_i^{(j)} (r_\rK, b, a)$ are polynomials. An exact spheroidal solution with $N \leq 3$ (assuming it exists) ought to satisfy
all equations in the above system. As we discuss below this is indeed the case in the HT spacetime. In contrast, the JP spacetime
does not admit such an exact truncated solution: the $P_{18}$ equation has $f_{18} \neq 0$ and therefore $\beta_3 =0$.
Moving down one level, we obtain $\beta_2 =0$  from the $P_{16}$ equation and then $\beta_1=\beta_0 =0$ as we move further down. 
We have verified that the situation remains the same even when the $N=4$ expansion is used. 

Without excluding the possibility that an exact solution might exist for some $N >4$ we adopt an alternative approach where
(\ref{r0expansion}) is assumed to be an infinite series. In practise the series has to be truncated at some order $N$ provided
the successive coefficients $\beta_n$ become increasingly smaller. Then for a given $N$, and starting from the lowest order $P_0$, 
one needs to include the necessary number of $P_\ell$ equations so that the system admits a consistent solution. For example for the 
$N=1$ expansion only the $\{P_0, P_1\}$  equations need to be included in the calculation of $\{\beta_0, \beta_1\}$. The $N=3$ 
expansion with its four unknown coefficients requires the simultaneous solution of the $\{P_0, P_2, P_4, P_6\}$ system. 
The successful application of this algorithm would imply that a spheroidal solution does exist at least in the form of an 
approximate truncated infinite series. 

As a case study, we have considered a weakly deformed JP spacetime with $\varepsilon_3=0.1,~a=0.7M$ and an orbital 
impact parameter $b=3.5 M$. The corresponding Kerr spherical radius solves $ \cE (r_\rK) =0$ and we find $r_\rK = 2.02649 M$.
The obtained spheroidal solutions for $ 1 \leq  N \leq 4$ are presented in Table~\ref{tab:LegendreExpansion}. These results are 
strongly suggestive of the convergence of the expansion (\ref{r0expansion}) since $\beta_n \sim 0.1 \beta_{n-1}$ between successive 
coefficients. This also means that the truncation scheme is itself self-consistent. Moreover, we can observe a rapid convergence in the 
value of a given $\beta_n$ as we move to higher $N$ systems, while the residuals of the differential Eq.~(\ref{circularity}) when we substitute 
the approximate solution converge to zero. The resulting $N=4$ solution for $r_0 (\theta)$ is,
\be
r_0  = 1.99884 + 10^{-3} \left (\, 3.11476  \cos^2 \theta - 0.265725 \cos^4 \theta 
+ 0.0248629 \cos^6\theta - 0.0019353 \cos^8\theta \,\right ).
\label{r0N4}
\ee
In the following section we show that this is in excellent agreement with the spheroidal radius extracted from the time-domain analysis 
of JP geodesics.

The main conclusion of the preceding analysis of the spheroidicity condition is that the non-separable JP metric \emph{admits} spheroidal orbits
albeit in the approximate form of a truncated series.

\begin{table*}
\begin{minipage}{135mm}
	\begin{tabular}{|r|c|c|c|c|c|}
	\hline
	  N    &  $\beta_0$ &  $\beta_1$ & $\beta_2$ & $\beta_3$ & $\beta_4$  \\
	\hline
	$~1~$ & -0.275586  & 0.0268995  &  - & - & -\\
	\hline
	$~2~$ & -0.276424  & 0.0308246  & $-2.18738 \times 10^{-3}$  & - &- \\
	\hline
	$~3~$ & -0.276449 & 0.0311300 & $-2.43839  \times 10^{-3}$ & $1.33614 \times 10^{-4} $ & - \\
	\hline
	$~4~$  & -0.276450 & 0.0311476 & $ -2.65725  \times 10^{-3}$ & $2.48629 \times 10^{-4} $ & $-1.9353 \times 10^{-5}$  \\
	\hline
	\end{tabular}
	\caption{Spheroidal orbit solutions of the spheroidicity condition for a JP metric with $\varepsilon_3 =0.1,~ a= 0.7M$ and $b = 3.5M$. 
	 We show the numerical values of the coefficients $\beta_n$ appearing in the expansion (\ref{r0expansion}) when truncated at order $N=\{1,2,3,4\}$. }
	\label{tab:LegendreExpansion}
\end{minipage}
\end{table*}

\begin{figure*}[htb!]
 \includegraphics[width=0.45\textwidth]{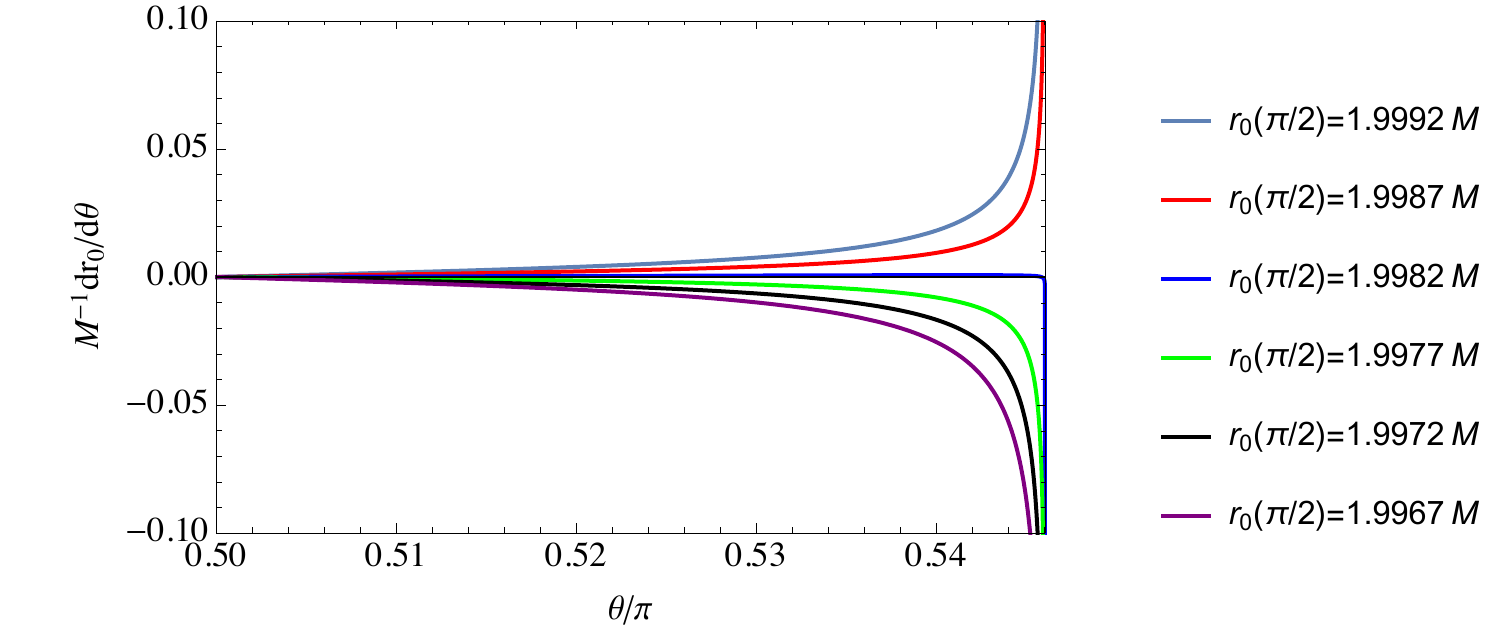}
 \hspace{1cm}
\includegraphics[width=0.45\textwidth]{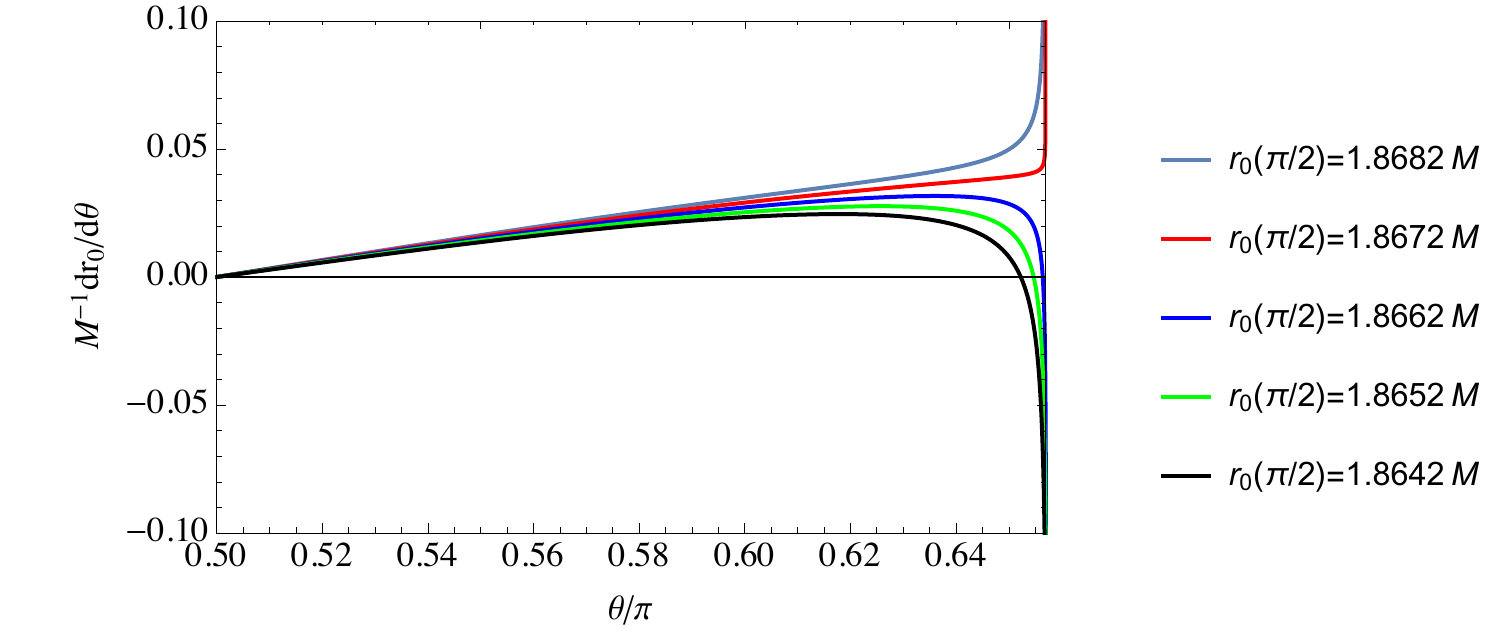}
\caption{\emph{Integration of the JP spheroidicity condition}. We show the numerical solution for $r_0^\p (\theta)$ in two JP metrics with spin 
$a=0.7M$. Left panel: $\varepsilon_3=0.1$ and $b=3.5M$. Right panel: $\varepsilon_3= 1$ and $b=3M$. The initial equatorial value of $r_0$ 
is indicated in the margin. For some value of $r_0$ the integration terminates at the separatrix with $r_0^\p =0$, indicating the presence of a spheroidal orbit.}
\label{fig:dr0dthetaJP}
\end{figure*}

This conclusion is corroborated by the outcome of the direct numerical integration of the spheroidicity condition (\ref{circularity}) 
for the full JP metric (that is, $\varepsilon_3$ is no longer assumed to be small). The integration is initiated at the equatorial plane 
for an initial radius $r_0 (\pi/2)$ which can be arbitrary as long as it lies inside the allowed region for geodesic motion (i.e. $V_{\rm eff}>0$). 
The second necessary initial condition is $r_0^\p (\pi/2)=0$ as dictated by the equatorial symmetry of the problem. The integration proceeds
towards $\theta=\pi$ (or $\theta=0$) and is set to terminate when the $V_{\rm eff}=0$ separatrix is reached. As examples we consider the 
$a=0.7M$ JP spacetime with $\varepsilon_3= \{0.1,1\} $ and respective impact parameter $b=\{3.5 M, 3M \}$ (this latter example is also discussed in 
Section~\ref{sec:tdomainJP}). Figure.~\ref{fig:dr0dthetaJP} displays a typical crop of results. The $r_0^\p$ value at the end of the integration is either a
very large positive or negative number (possibly signalling a divergence), with the integration in many cases terminating well away from the separatrix. 
However, for some particular initial $r_0 (\pi/2)$  the final $r_0^\p$ is zero at the separatrix within some numerical accuracy. 
For the two examples considered here we find that this happens for $ r_0 (\pi/2) \approx \{1.9982M, 1.8671 M \} $. Note that the 
$\varepsilon_3=0.1$ result is in excellent agreement with the analytic solution (\ref{r0N4}). 

As the spin and/or the deformation increases the situation becomes drastically different: the numerical solution for $r_0^\p (\theta)$ 
does not become zero at the separatrix, and therefore no spheroidal orbits should be expected.

\begin{figure*}[htb!]
\includegraphics[width=1\textwidth]{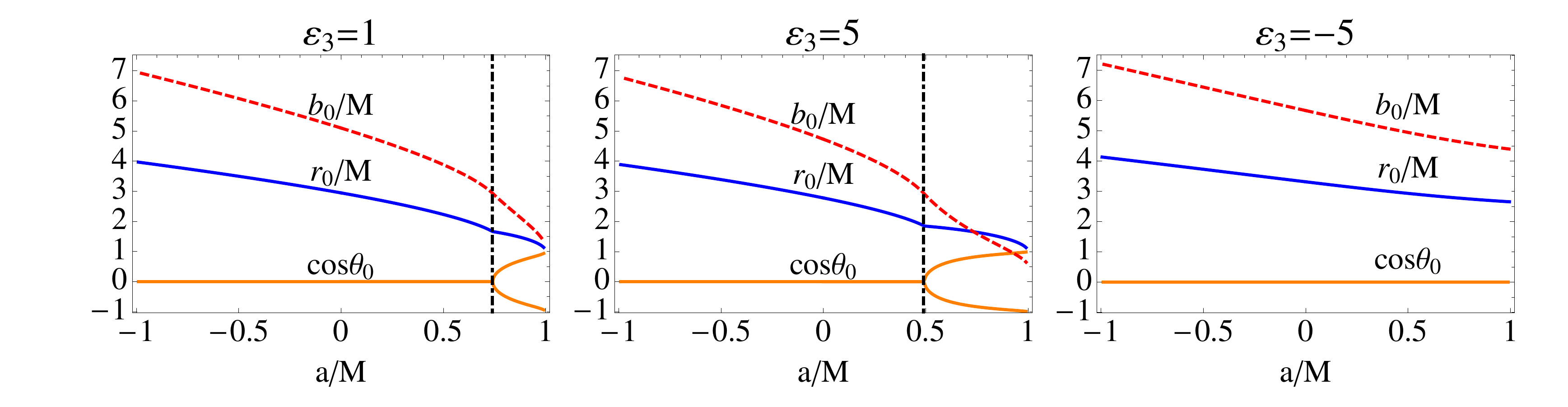}
\caption{\emph{Non-equatorial photon rings in the JP spacetime}. We show the numerical solution $\{r_0,\cos\theta_0,b_0 \}$ of
the photon ring equations (\ref{eqphring})-(\ref{eqphringth}) for the JP spacetime as a function of the spin $a$,
for three values $\varepsilon_3 = \{1,\pm 5\}$ of the deformation parameter. The most striking feature in these plots is the 
photon ring bifurcation at some $a=a_* (\varepsilon_3 > 0)$, marking the emergence of non-equatorial photon rings. 
Note that $a<0 $ ($a>0$) represents retrograde (prograde) orbits. }
\label{fig:JPring}
\end{figure*}

Having completed our investigation of spheroidal orbits we move on to consider the presence of photon rings at given constant 
angle $\theta=\theta_0$ and `radius' $r=r_0$ (this is of course the radial distance from the coordinate origin, not the ring's
actual coordinate radius $r_0 \sin\theta_0$); the previous series expansion does not apply for these circular orbits.
Instead, we can search for photon rings by direct application of Eqs.~(\ref{eqphring}) and (\ref{eqphringth}) of Section~\ref{sec:photonring}.
The numerical solution of the system is shown in Fig.~\ref{fig:JPring} in the form of curves $\{ r_0 (a), \cos\theta_0 (a), b_0 (a)\}$ 
for three values of the deformation, $\varepsilon_3 = \{1,\pm 5\}$.
For most of the spin range the only acceptable solution is that of the familiar equatorial photon ring (see e.g. \cite{Glampedakis:2017} 
for a discussion of the equatorial JP photon ring). Remarkably, and unlike what is known to happen in Kerr,  above a spin threshold $a_* >0$ 
the prograde photon ring in the $\varepsilon_3 >0$ spacetimes \emph{bifurcates} into a pair of symmetrically placed non-equatorial photon rings 
above and below the equator. 
The inclination (radius) of these photon rings increases (decreases) monotonically with the spin until they shrink to a point-like
structure at the north and south poles for $a = r_0 = M$ (in this case the solution can be found analytically).  
As it is easily visible in the figure,  this `photon ring phase transition'  is also reflected in the slope of the $\{r_0 (a), b_0 (a)\}$ curves. 
The bifurcation critical spin is clearly a function $a_* = a_* (\varepsilon_3)$, and lies in a range $0.4 \lesssim a/M \lesssim 0.95$ 
for $ 0.1 \lesssim \varepsilon_3 \lesssim 10$. In stark contrast to what we have just described, $\varepsilon_3 < 0$ JP spacetimes 
only appear to admit  equatorial photon rings, see right panel of Fig~\ref{fig:JPring}.

\begin{figure*}[htb!]
\includegraphics[width=0.45\textwidth]{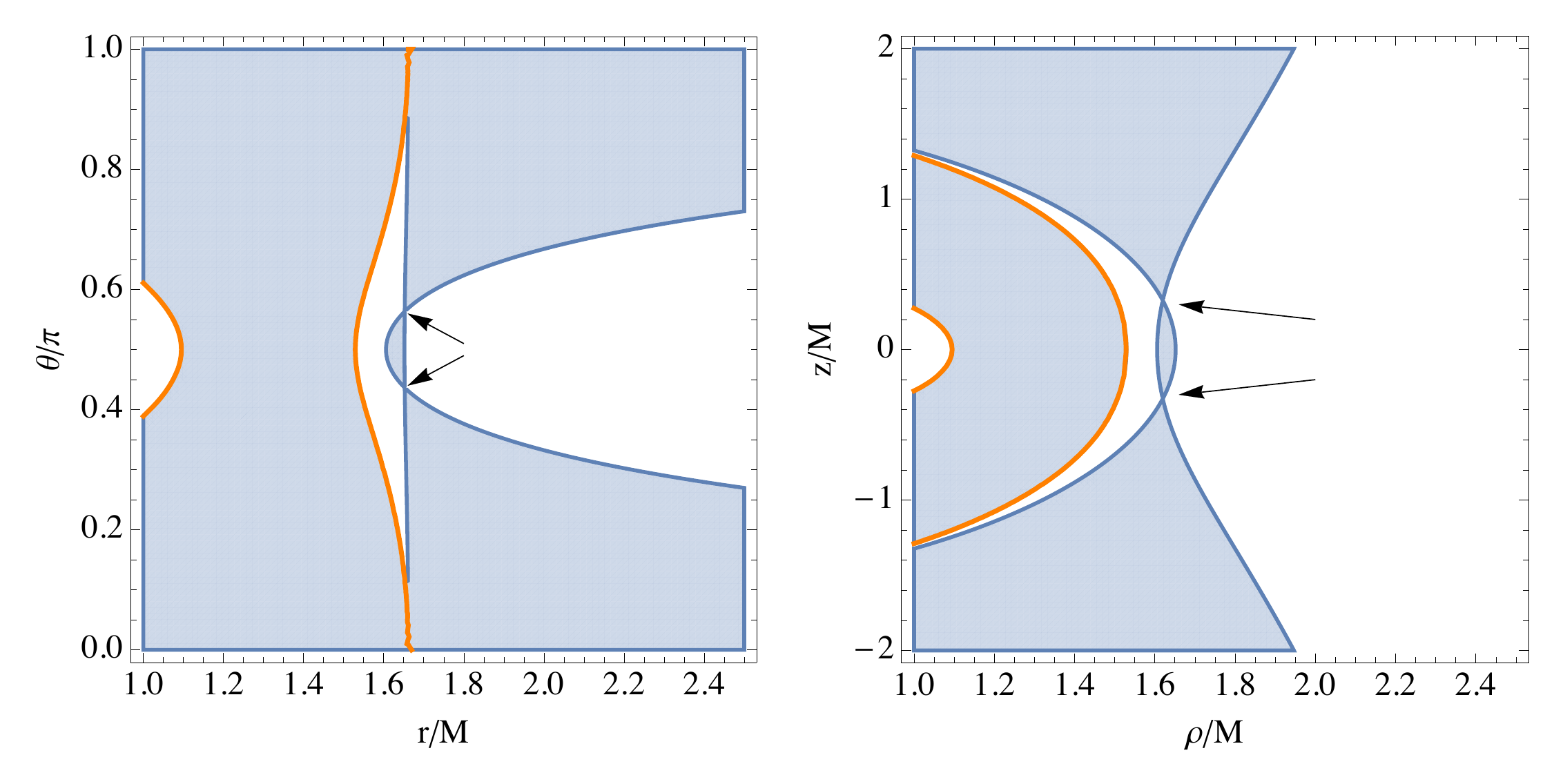}
\hspace{1cm}
\includegraphics[width=0.45\textwidth]{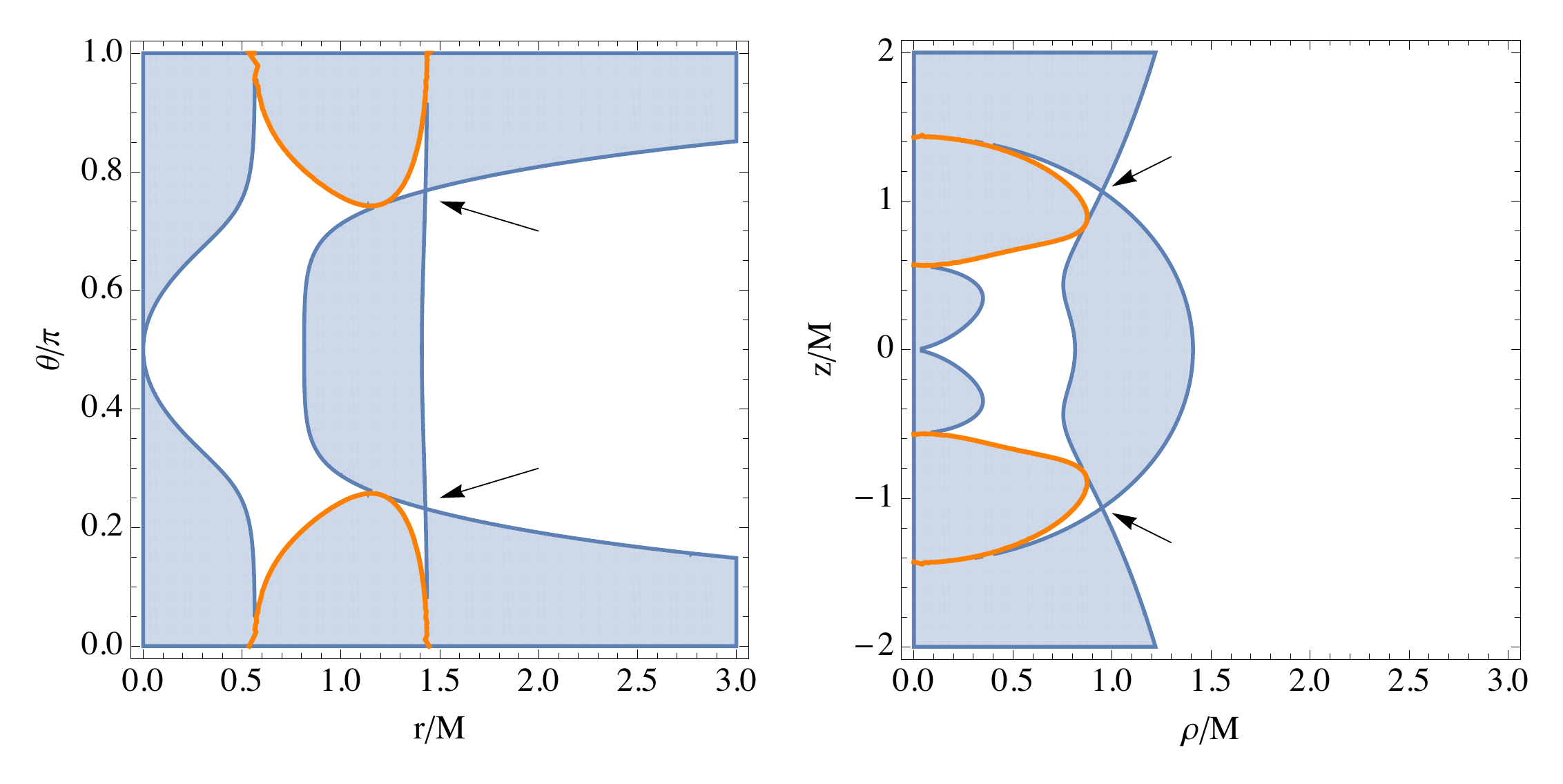}
\caption{\emph{JP separatrix with non-equatorial photon ring structure}. We show the $V_{\rm eff} (r,\theta) =0$ separatrix 
of allowed and forbidden (shaded) regions of photon geodesic motion. As a visualisation aid we use both spherical and cylindrical 
coordinates $\rho=r \sin\theta, z = r \cos\theta$. For these particular examples we have chosen $\varepsilon_3 = 1, a= 0.75M, b_0 = 2.8707 M$ 
(left panel) and $\varepsilon_3 = 1, a= 0.9M, b_0 = 1.9889 M$ (right panel). The arrows indicate the location of the two photon rings while the 
orange curves mark the location of the event horizon.}
\label{fig:JPring_sepa}
\end{figure*}

A complementary view on non-equatorial photon rings is provided by Fig.~\ref{fig:JPring_sepa} which shows 
the separatrix $V_{\rm eff} (r,\theta) =0$ of allowed/forbidden regions of photon geodesic motion. 
The photon ring structure is shaped in the same way as in the equatorial case (see Appendix~\ref{sec:Kerr} for the corresponding
Kerr separatrix figure) but is located off the equatorial plane.

Our results hint at a parameter space correlation between the disappearance of the equatorial photon ring 
(which occurs at high $a/M$ and/or $\varepsilon_3 \gg 1)$ and the absence of the
spheroidal orbits calculated in this section. This correlation is further bolstered by the results of the time-domain analysis, see next Section.

\subsection{The Hartle-Thorne metric}
\label{sec:HT}

We now consider the HT metric~\cite{Hartle1967, HT68}, 
\be
ds^2 = -e^\nu \left ( 1 + 2  h \right ) dt^2 + e^\lambda \left ( 1 + \frac{2 \mu}{r-2m} \right ) dr^2 
+ r^2 \left ( 1 + 2  k \right ) \left \{ \, d\theta^2 + \sin^2\theta  [d\varphi -\left ( \Om -\omega  \right ) dt ]^2  \, \right \} 
+ {\cal O}(\Omega^3),
\label{HTmetric}
\ee
where $\Omega$ denotes the stellar angular velocity. The three metric potentials, $\nu(r), \lambda(r), m(r)$,  are  spherically 
symmetric functions, the latter representing the usual mass function. The rest of the potentials can be expanded in terms
of Legendre polynomials,
\begin{align}
h(r,\theta) & = h_0 (r) + h_2(r) P_2, \quad  \mu(r,\theta)  = \mu_0 (r) + \mu_2 (r) P_2, \quad
k(r,\theta)  = k_2(r) P_2, \quad \omega(r,\theta) = \omega_1 (r) P_1^\p.
\end{align}
In the vacuum exterior of a general relativistic star, the HT metric is most conveniently parametrised in terms of the spin
parameter $\chi = J/M^2 $ (where  $J$ is the ${\cal O} (\Omega)$ angular momentum and $M$ the mass at $\Omega=0$), the quadrupole
moment $ Q = \chi^2 M^3 \left ( 1 - \delta q \right )$ (here expressed in terms of the deviation $\delta q $ from the Kerr quadrupole of the
same mass and spin parameter), the ${\cal O} (\Omega^2)$ shift $\delta m$ in the mass and, finally, the rescaled radial coordinate $x=r/M$:
\begin{align}
& m = M, \quad e^{\nu} = e^{-\lambda} = 1-\frac{2}{x}, \quad  \omega_1  =  \Omega -  \frac{ 2\chi}{ M x^3},
\quad
\frac{\mu_0}{M} = \chi^2  \left ( \delta m - \frac{1}{x^3} \right ), \quad h_0 = \frac{\chi^2 }{x-2} \left (\,  \frac{1}{x^3}   -  \delta m  \, \right ),
\\
\nn \\
h_2 &= \frac{5}{16} \chi^2 \delta q  \left (1- \frac{2}{x} \right )
\left [\, 3 x^2 \log \left (1-\frac{2}{x} \right ) + \frac{2 }{x} \frac{(1-1/x)}{(1-2/x)^2}  (3x^2 -6x -2) \, \right ]
+  \frac{\chi^2}{x^3} \left ( 1 + \frac{1}{x} \right ), 
\\
\nn \\
k_2 & = -\frac{\chi^2}{x^3} \left ( 1 + \frac{2}{x} \right ) -\frac{5}{8} \chi^2 \delta q  \left [\, 3 (1+ x  )
-\frac{2}{x} - 3 \left (1 -\frac{x^2}{2} \right )  \log \left (1-\frac{2}{x} \right )\, \right ],
\\
\nn \\
\frac{\mu_2}{M} & = - \frac{5}{16} \chi^2 \delta q  x  \left ( 1 -\frac{2}{x} \right )^2  \left [\,  3 x^2 \log \left (1-\frac{2}{x} \right )
+ \frac{2}{x} \frac{(1-1/x)}{(1-2/x)^2}  (3x^2 -6x -2) \, \right ]
 -\frac{\chi^2}{x^2} \left ( 1 -\frac{7}{x} + \frac{10}{x^2} \right ).
\end{align}
Hereafter we set $\delta m = 0$ and redefine $M$ as the spin-modified stellar mass.

There are two distinct  ways to proceed when working with an approximate metric like the HT. 
The first one is to use the metric (\ref{HTmetric}) ``as it is" (truncated at a given $\Omega$-order) without making
any further approximation in the geodesic equations.  The second approach is to expand all equations 
to the same perturbative order as the metric.

When the first approach is combined with the spheroidicity condition (\ref{circularity}) we find that spherical orbits are not
admitted. The search for spheroidal solutions can only be done via a numerical integration of the spheroidicity condition. 
The outcome of that calculation is qualitatively similar to the previous JP analysis, and consists of approximate spheroidal orbit solutions
in the low spin portion of the parameter space. The second approach, however, does lead to exact spheroidal orbit solutions. 
The series expansion for $r_0 (\theta)$ is now also an expansion with respect to the spin,
\be
r_0 (\theta) =  3 M - \frac{2 M}{\sqrt{3}} \cos\iota \chi + \chi^2 \sum_{n=0}^N \beta_n \cos^{2n} \theta + {\cal O} (\chi^3),
\label{spheroHT}
\ee 
where the first two terms can be identified with the Kerr (spherical) photon ring radius at leading order with respect
to the spin (the constant inclination parameter $\iota$ is defined in Appendix~\ref{sec:Kerr}).
Assuming $N=3$ as before,  the resulting spheroidicity condition contains the even-order Legendre polynomials
$P_0 - P_{10}$, leading to six independent equations. However, unlike the previous JP system, the present 
overdetermined system of equations does have an acceptable exact solution with $\beta_3 = \beta_2=0$ and
\begin{align}
& \beta_1 = -\frac{M}{144} \left [\, 16 + 27 \delta q \left ( 45 \log 3 -52  \right )   \,\right ],
\label{b1sol}
\\
& \beta_0 =   \frac{M}{27} \left ( 3 \cos 2\iota -4 \right ) + \frac{M}{16} \delta q (3 \cos2\iota +2) (45 \log 3 -52).   
\label{b2sol}
\end{align} 
This solution represents a spheroidal orbit with a $\sim \cos^2\theta$ profile. It should be mentioned that the same result
could have been found via a more direct approach, namely, by postulating a solution 
$r_0 (\theta)= 3M + r_1(\theta) \chi + r_2(\theta) \chi^2$. As expected, the $\chi=0$ spheroidicity condition 
is solved by the Schwarzschild radius $3M$, while $r_1$ and $r_2$ solve the first and second $\chi$-order equations
\begin{align}  
&(b^2  -27 M^2 \sin^2\theta ) r_1^{\pp} -b^2 \cot \theta\,  r_1^\p +  3  M^2 \sin^2 \theta ( 9  r_1  + 2 b  ) =0,
\label{r1HTeq} 
\\
\nn \\
&  \left(b^2  -27 M^2 \sin^2\theta \right)  r_2^{\pp} - b^2 \cot\theta\,  r_2^{\p} + 27 M^2 \sin^2 \theta\, r_2
\nn\\
&+ \frac{M}{32} \sin^2\theta \Big [ \, \frac{128}{9} b^2+15 M^2 \left \{ 27
\delta q (45 \log 3-52)+16\right \} \cos 2\theta + M^2 \left \{ 135 \delta q (45 \log 3-52)+272 \right \} \,  \Bigg ] = 0.
\label{r2HTeq}
\end{align}
When combined with $b=3\sqrt{3}M \cos\iota-2 M \cos^2 \iota \chi$ (this coincides with the Kerr impact parameter of the same spin order) 
these equations lead to the equatorial-symmetric solutions $r_1 = -2M\cos\iota/\sqrt{3} $ and $r_2 = \beta_0 + \beta_1 \cos^2 \theta$. 
Moreover, the equatorial orbit limit of these results agrees with the photon ring radius found in \cite{Glampedakis2018}.

\begin{figure*}[htb!]
\includegraphics[width=0.84\textwidth]{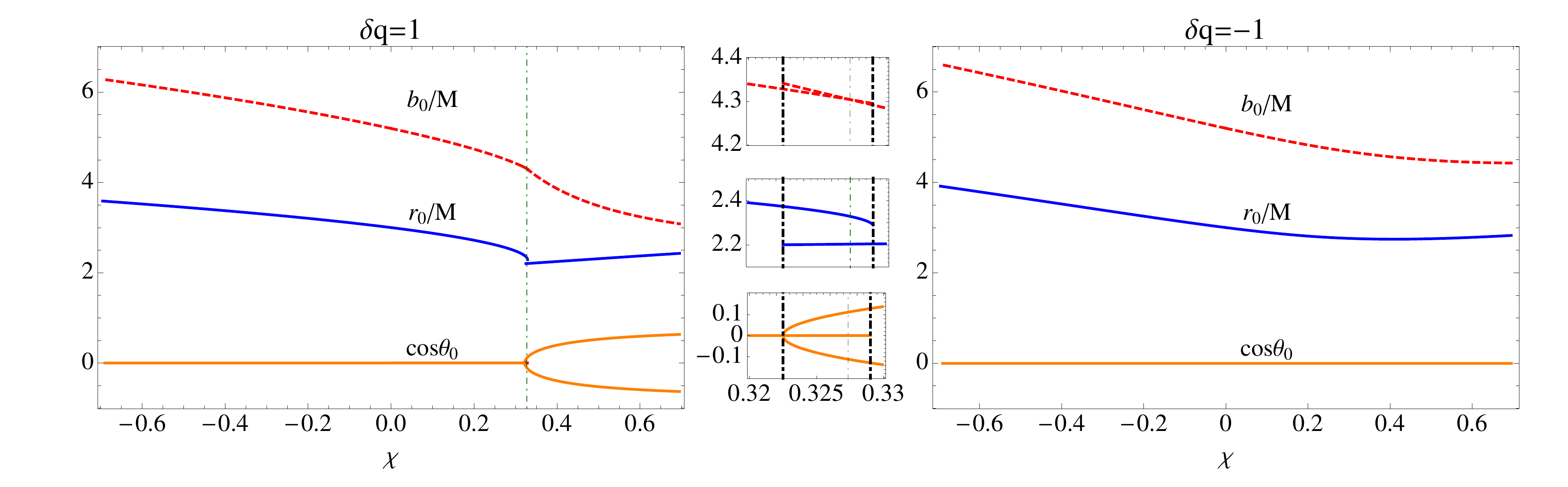}
\caption{\emph{Non-equatorial photon rings in the HT spacetime}. We show the numerical solution $\{r_0,\cos\theta_0,b_0 \}$ 
of the photon ring equations (\ref{eqphring})-(\ref{eqphringth}) for the HT spacetime as a function of the spin $\chi$, for two values 
$\delta q = \pm 1$ of the quadrupolar deviation parameter.  Retrograde (prograde) orbits correspond to $\chi<0 $ ($\chi>0$).
Only the $\delta q >0$ branch admits non-equatorial photon rings, in which case they appear as a bifurcation of the equatorial
photon ring at some critical spin $\chi_*(\delta q)$. The bottom middle panel shows a `pitchfork' structure where an equatorial solution 
is still present for a short spin range past the bifurcation spin $\chi_*$. The corresponding radii and impact parameters are shown in the
other two middle panels. The vertical green line marks the spin  $\chi \approx 0.327 $ at which all three photon ring solutions coexist 
for the same $b_0$.}
\label{fig:HTring}
\end{figure*}

\begin{figure*}[htb!]
\includegraphics[width=0.8\textwidth]{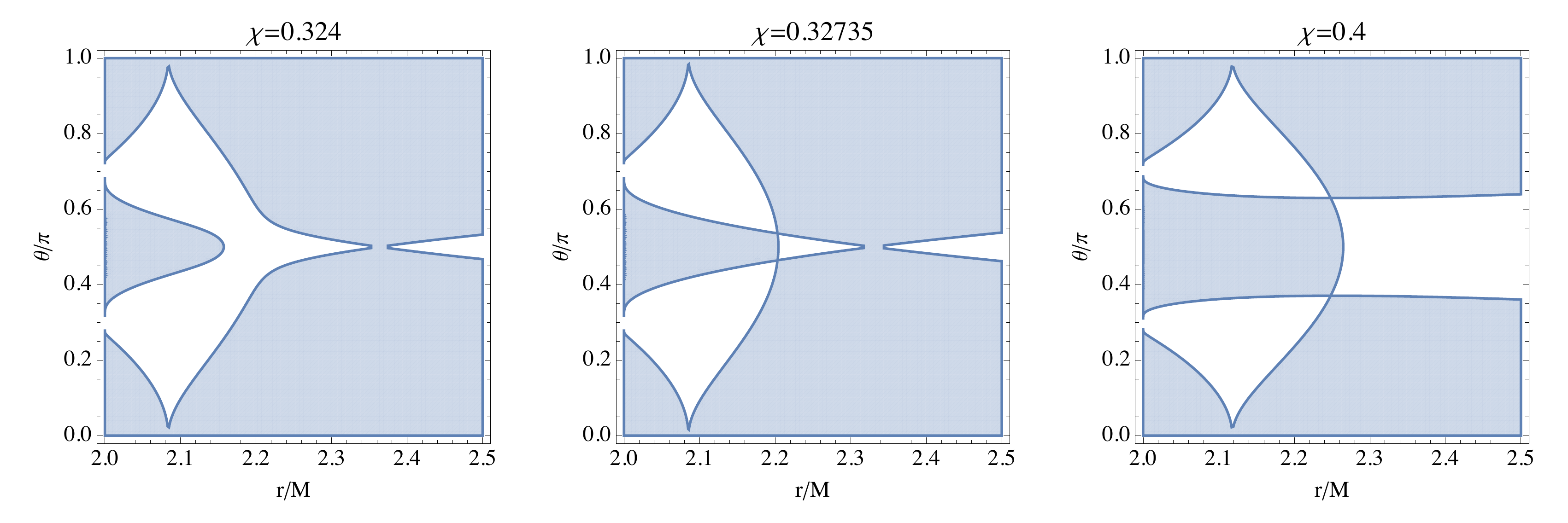}
\caption{\emph{HT separatrices with non-equatorial photon ring structure}. We show a series of $V_{\rm eff} (r,\theta) =0$ separatrices
(shaded area represents the forbidden region for photon geodesic motion) in a $\delta q=1$ HT spacetime. The left panel corresponds to
a situation past the bifurcation spin $\chi_*$ where an equatorial photon ring is still present (for this particular example $\chi=0.324, b_0 = 4.32132 M$). 
The middle panel shows the special coexistence of three photon rings at $\chi = 0.32735, b_0 = 4.30434M$  (vertical green line in
Fig.~\ref{fig:HTring}). For the high-spin example shown in the right panel ($\chi=0.4, b_0 = 3.86143M$), only non-equatorial photon ring
solutions are possible.}
\label{fig:HTring_sepa}
\end{figure*}

We next turn to the study of non-equatorial photon rings in the HT spacetime. In this calculation it is more appropriate 
to use the ``as it is" version of the HT metric; we then numerically solve the photon ring equations.~(\ref{eqphring})-(\ref{eqphringth}).
The results $\{r_0,\cos\theta_0,b_0 \}$ are shown in Fig.~\ref{fig:HTring} as functions of the spin parameter $\chi$. 
To begin with, the $\delta q < 0$ branch of the HT spacetime is Kerr-like, not admitting anything more than an equatorial photon ring.
On the other hand, the situation for $\delta q > 0$ is reminiscent of the previous $\varepsilon_3 >0$ JP results, with the
equatorial photon ring bifurcating to a symmetric pair of non-equatorial ones at some critical spin $\chi_*(\delta q)$. 
A closer inspection of this transition (see the middle plots in Fig.~\ref{fig:HTring}) reveals an interesting triplicity of equatorial and non-equatorial 
photon ring solutions with slightly different values of $r_0$  and $b_0$ in the immediate vicinity of $\chi_*$. The three solutions 
coexist for the same $b_0$ at some particular value of $\chi$ (marked by a vertical green line in Fig.~\ref{fig:HTring}). 
Fig.~\ref{fig:HTring_sepa} describes the corresponding structure of the $V_{\rm eff} (r,\theta) = 0$ separatrix, illustrating 
examples of a generic situation with a pair of non-equatorial photon rings (left plot) and the special case of three coexisting photon 
rings (right plot).
Similarly to what was found in the JP metric, the disappearance of the equatorial photon ring for $\chi > \chi_*$ also 
marks the suppression of spheroidal orbits.

%


\subsection{The Johannsen metric}
\label{sec:J}

Unlike the JP metric, the deformed Kerr metric devised by Johannsen~\cite{Johannsen2013PhRvD} is separable and takes 
the form,
\begin{align}
g_{tt}^\rJ & = - \frac{\tilde{\Sigma}} {N} \left (\, \Delta -a^2 A_2^2  \sin^2\theta \, \right ), \qquad
g_{t\varphi}^\rJ =  - \frac{a \tilde{\Sigma}} {N} \sin^2\theta \left [\, (r^2+a^2) A_1 A_2 -\Delta \, \right ],
\\
g_{\varphi\varphi}^\rJ &=  \frac{\tilde{\Sigma}} {N}  \sin^2\theta \left [\, (r^2+a^2)^2 A_1^2 -a^2 \Delta \sin^2\theta \, \right ],  
\qquad 
g_{rr}^\rJ = \frac{\tilde{\Sigma}}{\Delta A_5}, \qquad g_{\theta\theta}^\rJ = \tilde{\Sigma},
\label{Jmetric}
\end{align}
where
\be
N = \left  [\, (r^2+a^2) A_1 -a^2 A_2 \sin^2\theta \, \right  ]^2, \qquad  \tilde{\Sigma} = \Sigma + f(r). 
\ee
The deformation away from Kerr is encapsulated in the radial functions $\{A_1(r), A_2 (r), A_5 (r), f(r)\}$ ;
in their simplest form these are:
\be
A_1= 1 + \alpha_{13} \left (\frac{M}{r} \right )^3, \qquad  A_2  = 1 + \alpha_{22} \left (\frac{M}{r} \right )^2
\qquad
A_5  = 1 + \alpha_{52} \left (\frac{M}{r} \right )^2, \qquad f  = \varepsilon_3 \frac{M^3}{r}, \qquad
\ee
with  $ \alpha_{13},  \alpha_{22},  \alpha_{52}, \varepsilon_3$ constant parameters. 

Repeating the procedure of the preceding sections, we first look for spherical orbits. 
The spheroidicity condition (\ref{circularity}) for the J metric returns the $\theta$-independent expression,
\begin{align}
&  - r_0^7 \left( a^2 -a b+ r_0^2 \right)  \cE_\rK(r_0)
+ \alpha_{13} \alpha_{22} ab M^5 r_0 \left[\,   r_0^3 (8 M-5 r_0)-5 a^4+2 a^2 r_0 (6 M-5 r_0)   \, \right]
 \nn \\  
 & + \alpha_{13}^2 M^6 \left(a^2+ r_0^2 \right) \left [\, 3 a^4+a^2 r_0 (5 r_0 -7M) + r_0^3 (2 r_0-3 M)\, \right ]
 \nn \\  
 & -2  \alpha_{22} ab M^2 r_0^4 \left [\, a^4-a^3 b+a^2 r_0 (2 r_0 -3 M)+a b r_0 (3 M-2 r_0)+ r_0^3 (r_0-M) \,\right ]
 \nn \\
&  + \alpha_{13} M^3 r_0^3  \left [\, 3 a^6-3 a^5 b+a^4 r_0 (7r_0-8 M)  + 2 a^3 b r_0 (4 M-3 r_0) + a^2 r_0^3 (5r_0 - 8M) \right.
\nn \\
& \left. +a b r_0^3 (4 M-3 r_0)+ r_0^6 \, \right ]  + \alpha_{22}^2 a^2 b^2 M^4 r_0^2 \left [\, 2 a^2+ (3r_0-5 M)   r_0\,\right ]  = 0.
 \end{align}
This is a polynomial with respect to $r_0$ and, when combined with the J metric's Carter constant expression 
(see \cite{Johannsen2013PhRvD}), it leads to a pair of physically relevant roots (for prograde/retrograde motion) 
over a wide range of the deformation parameters. We thus conclude that the J metric admits  Kerr-like spherical photon orbits. 
Furthermore, from the discussion of Section~\ref{sec:theorem} we know that the existence of a separate family of spheroidal 
orbits is ruled out (the same statement is of course true for the Kerr metric itself).  
Finally, a search for photon rings along the lines of the previous two cases reveals the J spacetime to be 
Kerr-like, admitting a single equatorial photon ring.


\section{A time-domain study of spheroidicity}
\label{sec:tdomain}

The spheroidicity condition is an ideal tool for probing the existence of spherical or spheroidal orbits 
in a given stationary-axisymmetric spacetime. However,  in the case where this type of motion is not supported,
it has little to say about the possibility of having `quasi-spherical/spheroidal' orbits, that is, trajectories where a photon 
moves about some mean radius, effectively being trapped for a considerable period of time. Furthermore, one would
like to somehow gauge the degree of `decircularisation' as a function of `departure from separability'. 
To access this kind of information one would have  to rely to direct numerical integration of the geodesic equations and 
consider a spacetime metric that represents a deformation of a known separable metric. The JP and HT metrics are therefore
an ideal choice for this kind of experimentation. These are discussed separately in the following two subsections. 


\subsection{JP orbits}
\label{sec:tdomainJP}

We have performed a series of numerical integrations of the geodesic equations in the JP spacetime; 
in all cases the numerical experiment consists of a photon being launched with $u^r =0$ (i.e. from a radial
turning point) and a suitable $u^\theta \neq 0$ for a given $b$ and black hole parameters $a,\varepsilon_3$. 
Typically (but not exclusively) the photon is initially placed on the equatorial plane and the initial radius is chosen as close as 
possible to the radius below which it would plunge towards the event horizon at $\cD=0$. The orbits are subsequently 
evolved both forward and backward in time. Following this recipe it is straightforward to identify any orbits that could trap 
photons for a considerable time interval (provided such orbits exist in the first place). If present, these orbits are expected to 
appear for $b/M \lesssim {\cal O} (1)$. Photons moving in $b/M \gg 1$ orbits are deflected at a relatively large distance
and as a result they fail to probe the near-horizon region. These orbits are qualitatively similar in both Kerr and non-Kerr 
spacetimes and are of no interest to the present analysis.

\begin{figure*}[htb!]
\includegraphics[width=0.5\textwidth]{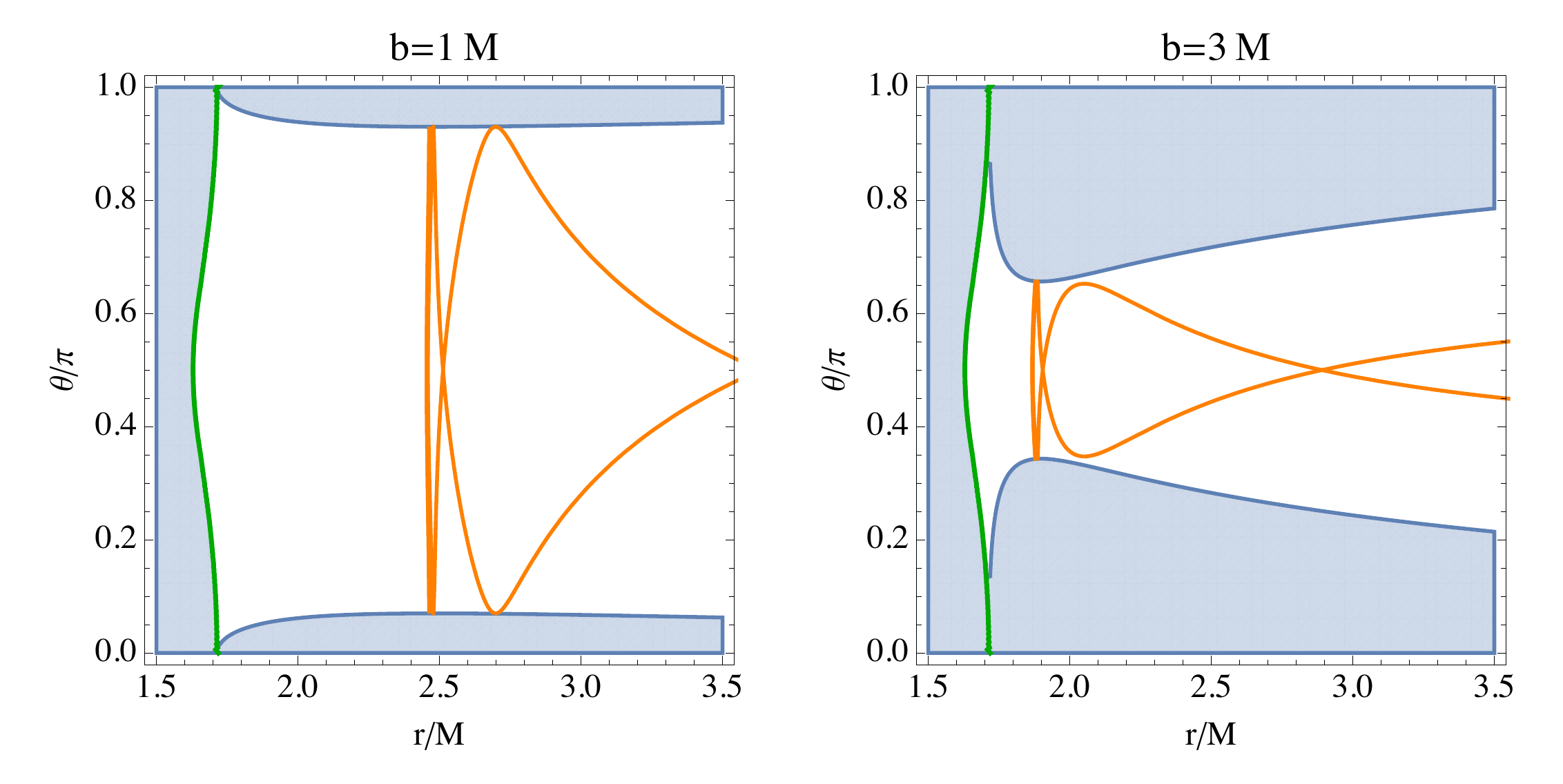}
\includegraphics[width=0.5\textwidth]{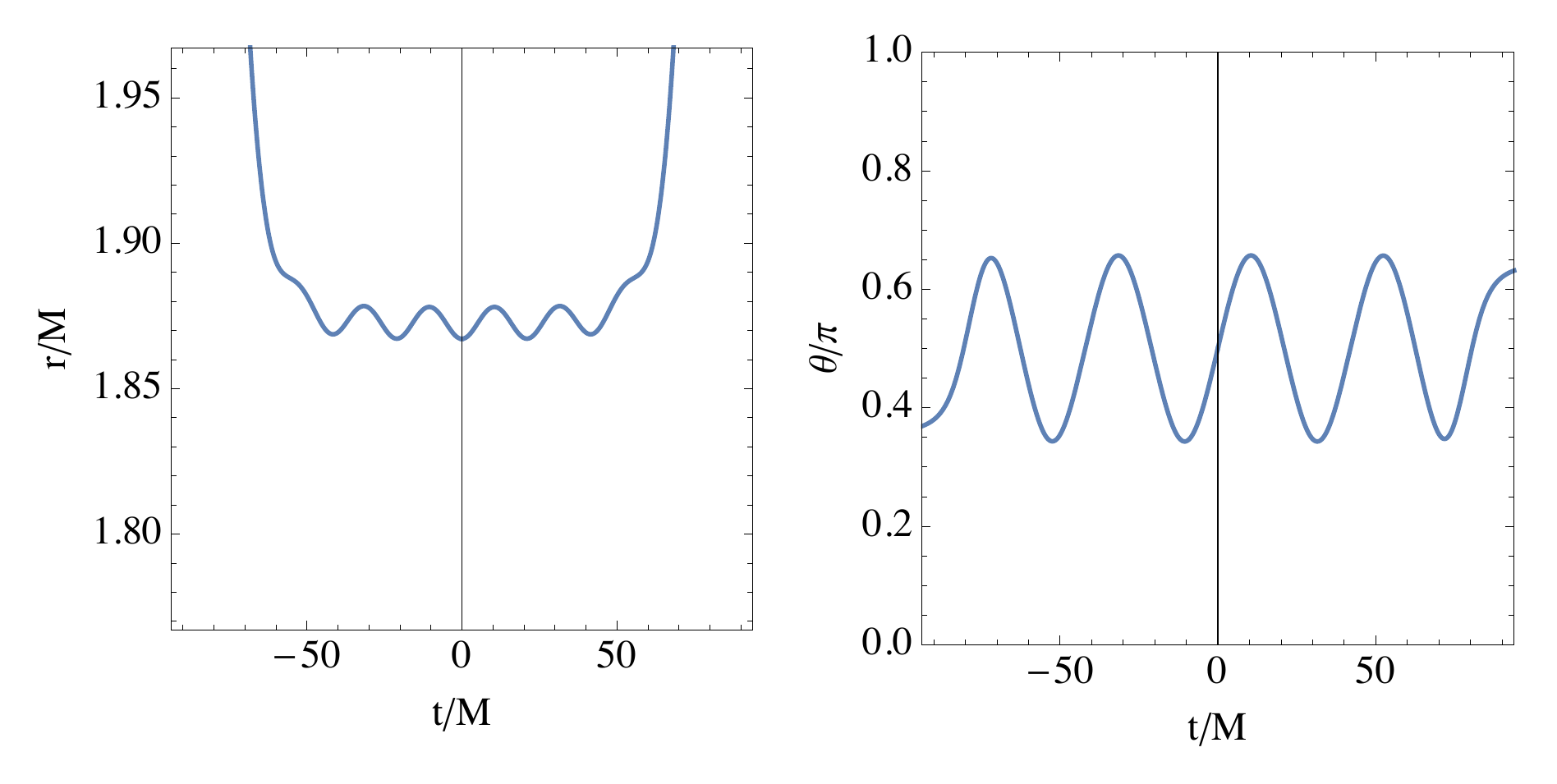}
\includegraphics[width=0.25\textwidth]{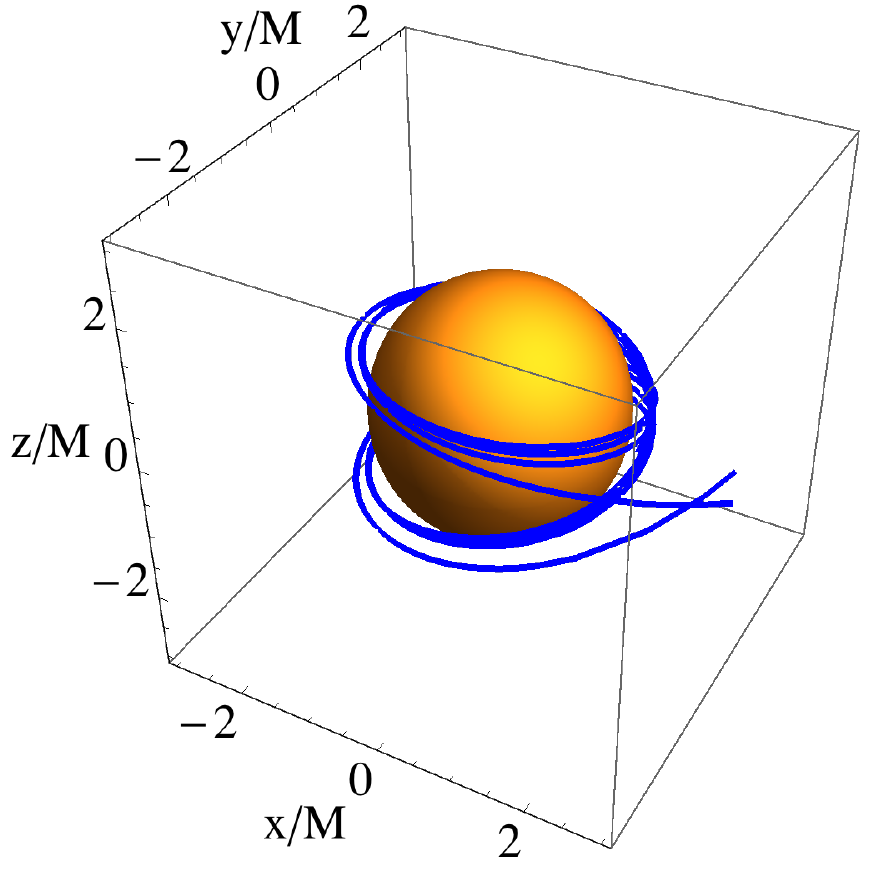}
\caption{\emph{Spheroidal JP orbits}. 
This figure demonstrates the temporary capture of photons in the immediate vicinity of spheroidal orbits in a $\varepsilon_3 =1, a=0.7M$ JP spacetime. 
Top row: orbits for $b=M$ and $b=3M$ superimposed with their respective separatrices (the shaded area
marks the forbidden region $V_{\rm eff} <0$) and event horizons $\cD=0$ (green thick curves). Bottom row: the profiles $\{r(t), \theta(t)\}$ 
of the $b=3M$ orbit and its three-dimensional shape in Cartesian coordinates $ (x,y,z) = r (\sin\theta \cos\varphi, \sin\theta\sin\varphi , \cos\theta) $. 
The quasi-spherical coloured surface represents the JP event horizon.}
\label{fig:JPorbits1}
\end{figure*}

\begin{figure*}[htb!]
\includegraphics[width=0.35\textwidth]{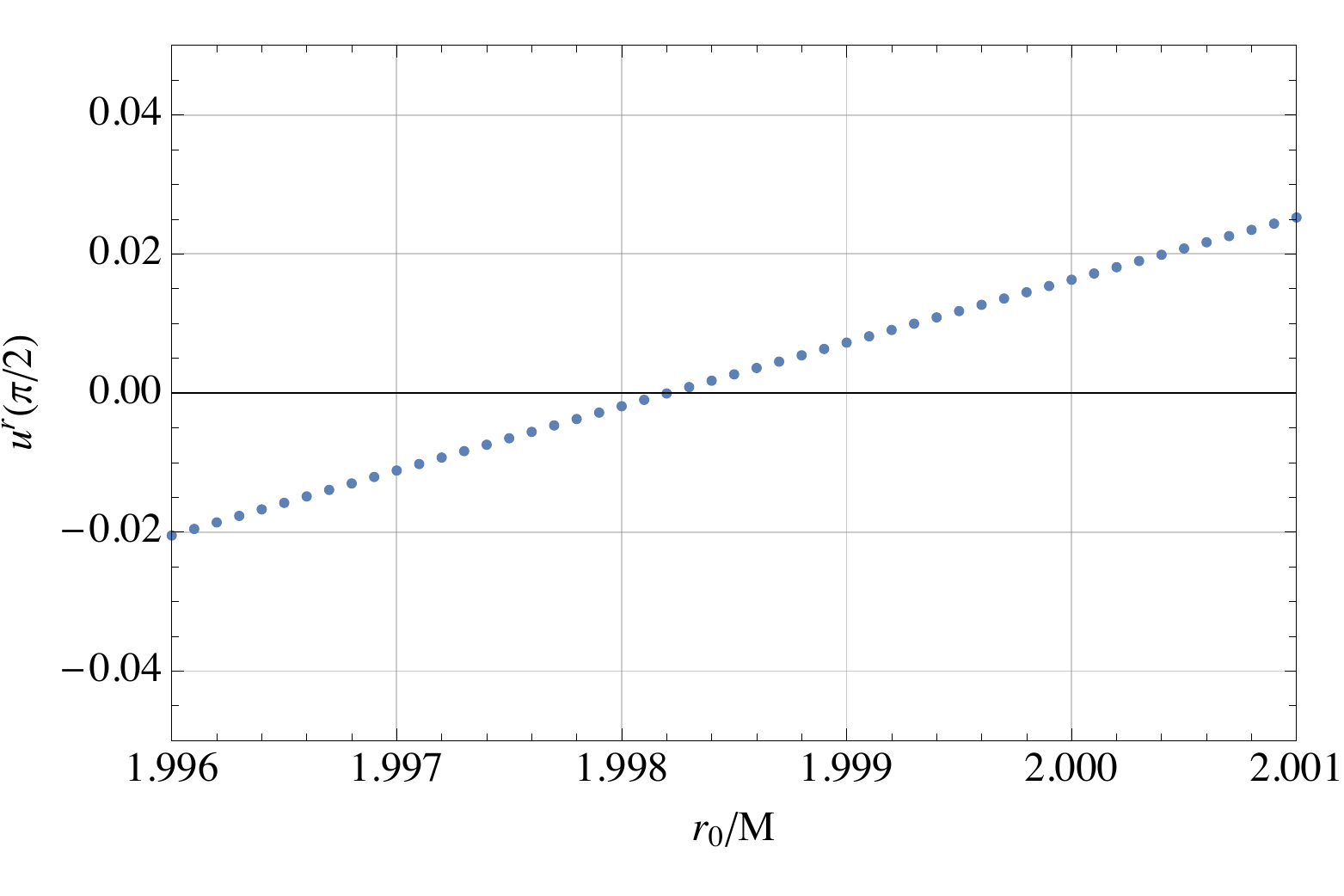}
\hspace{1cm}
\includegraphics[width=0.35\textwidth]{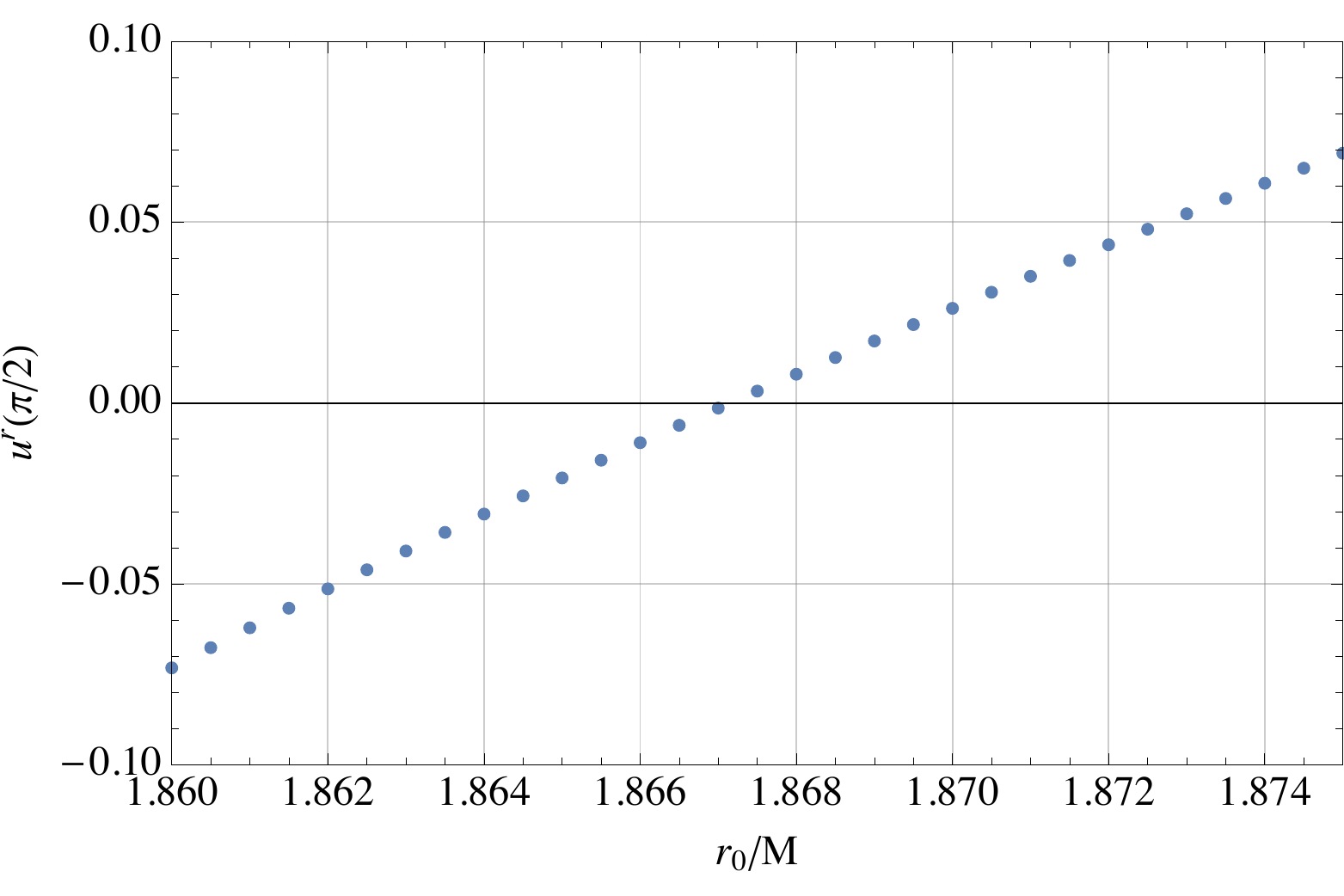}
\caption{\emph{Locating spheroidal orbits in the time-domain}. We show the radial velocity $u^r (r_0)$ as the photon recrosses the equatorial plane 
following its initial launch in the opposite direction with $u^r=0$ and radius $r_0$ in the equatorial plane. A crossing through zero indicates the presence 
of a spheroidal orbit. Although not shown here, the radius at the moment of crossing is virtually identical to the initial $r_0$. 
Left panel: $\varepsilon_3=0.1,~ b=3.5M$. Right panel: $\varepsilon_3=1,~ b=3M$.}
\label{fig:JPureq}
\end{figure*}

The first sample of results correspond to a JP spacetime with $\varepsilon_3 =1$ and $a=0.7M$, see Fig.~\ref{fig:JPorbits1}.
As demonstrated in the previous section, the JP metric admits spheroidal orbits (at least in an approximate sense) for a broad range of
parameters. In the time domain these orbits manifest themselves as trajectories in which the photon spends a considerable amount of time 
($\sim$ few tens of $M$) in the near-horizon strong-field regime, moving about some mean radius. Examples of these orbits are shown in 
Fig.~\ref{fig:JPorbits1} for two choices of the impact  parameter.
The presence of spheroidal orbits can be probed via a complementary time-domain method where a photon is initially launched from the equatorial
plane with $u^r=0$ and, if the orbits is spheroidal, will recross the equatorial plane in the opposite direction with $u^r=0$. The time-inversion and equatorial 
symmetry of the system guarantee that the same motion can be inverted and extended in the opposite hemisphere, thus resulting in a spheroidal orbit. 
Within a given numerical precision such orbits are indeed found; in the examples displayed in Fig.~\ref{fig:JPureq} we plot the `reentry' velocity $u^r$ 
as a function of the initial radius $r_0 = r(0)$. The function clearly passes through zero, and it does so at a radius 
which is in excellent agreement with the results of the previous section (see Eq.~(\ref{r0N4}) and Fig.~\ref{fig:dr0dthetaJP}).

\begin{figure*}[htb!]
\includegraphics[width=0.25\textwidth]{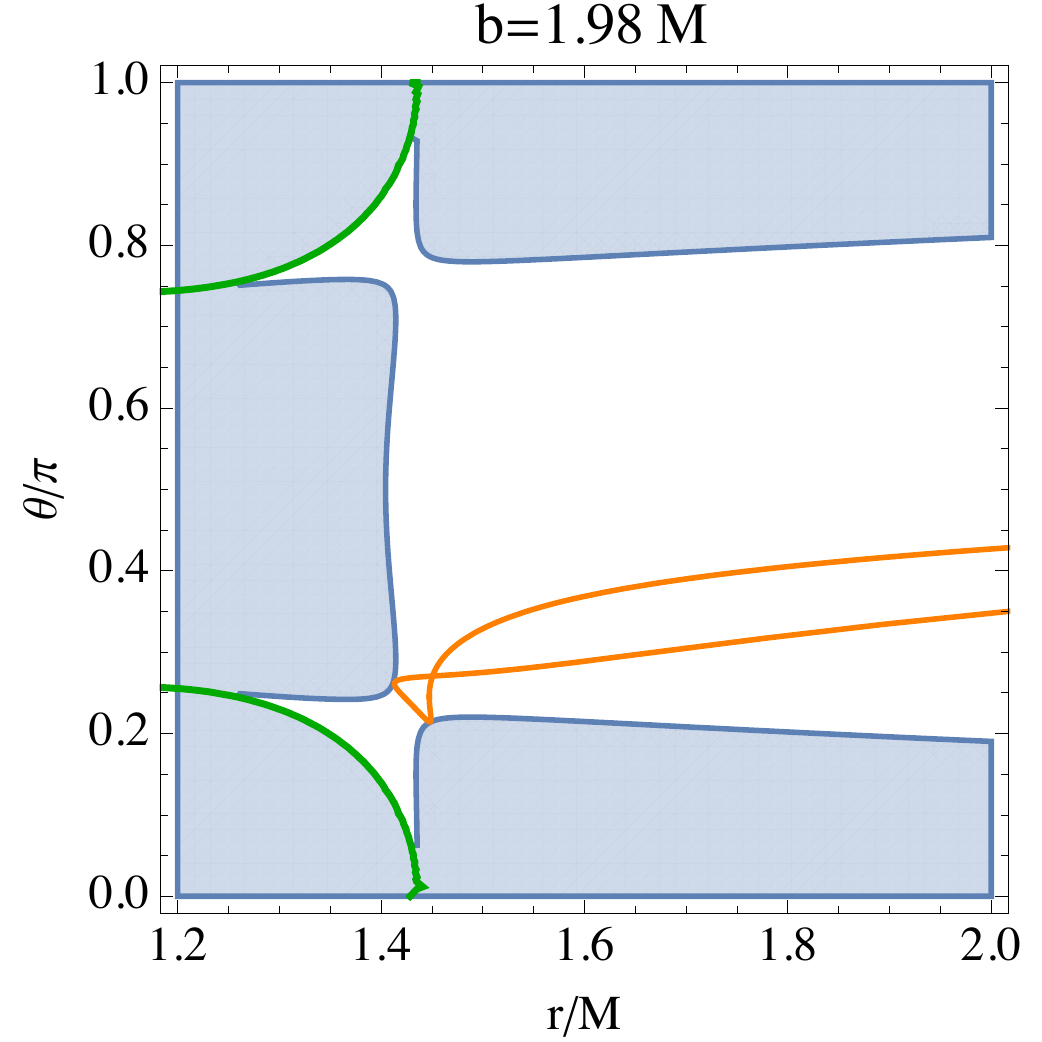}
\hspace{1cm}
\includegraphics[width=0.3\textwidth]{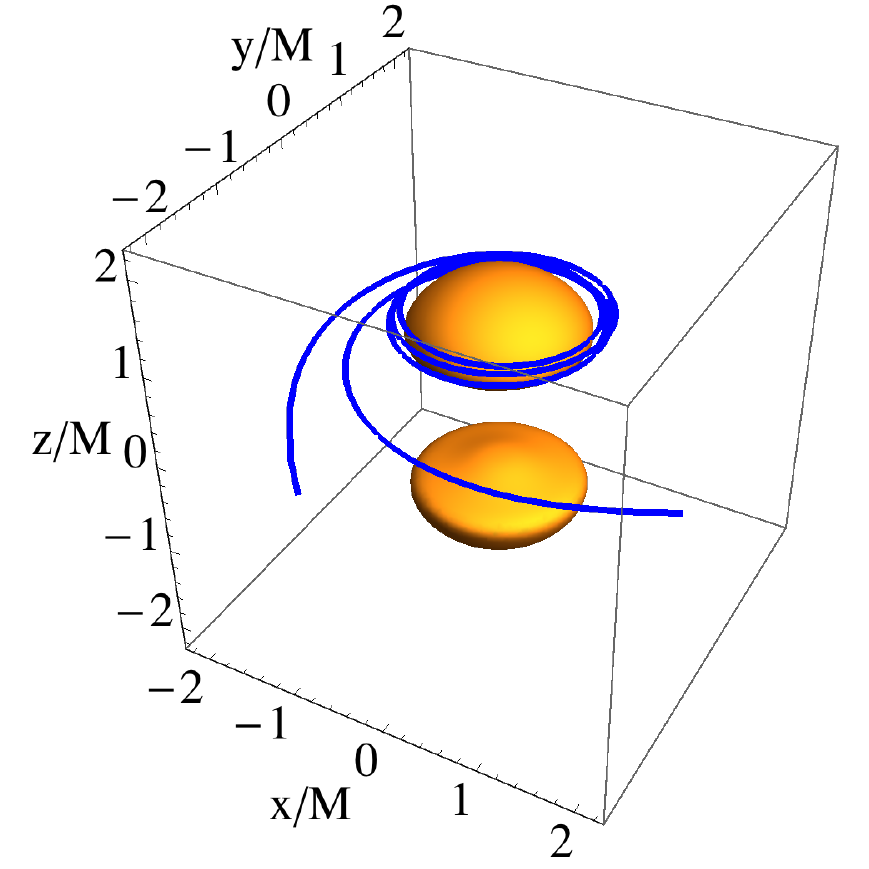}
\caption{\emph{Non-equatorial photon rings and quasi-circular JP orbits}. The $\varepsilon_3 =1, a=0.9M$ JP spacetime considered in 
this figure can support non-equatorial photon rings for a particular value of the impact parameter (see left panel in Fig.~\ref{fig:JPring}). 
For a $b$ close to that value, the potential $V_{\rm eff}$ takes the form shown in the left panel (here we have used $b=1.98M$), featuring 
a pair of `straits'  near the location where the photon rings forms. Thanks to this local behaviour the potential can temporarily trap photon in
quasi-circular/quasi-planar orbits, see orange curve in the left panel. The corresponding three-dimensional orbit is shown in the right panel 
together with the double-lobed event horizon (orange-colored surfaces).}
\label{fig:JPorbits2}
\end{figure*}

As we have seen, non-equatorial photon rings arise in the JP spacetime above some spin threshold (and for a range of $b$). 
This markedly non-Kerr orbital feature may lead to photons being temporary captured in quasi-circular orbits. An example of this is shown 
in Fig.~\ref{fig:JPorbits2} for a  JP spacetime with parameters $\varepsilon_3 =1, a=0.9M$. In this instance, the orbit is 
asymmetric with respect to the equatorial plane and the numerical integration was initiated by placing the photon in the vicinity of
the upper hemisphere photon ring's radius and latitude (a similar orbit can be obtained for the lower hemisphere). A photon in this orbit
is temporarily captured in a quasi-circular/quasi-planar orbit in the vicinity of the upper event horizon. Although we have not examined it in any 
detail, it is likely that this behaviour signals the presence of spheroidal orbits that do not cross the equatorial plane but instead remain 
localised in the vicinity of the non-equatorial photon ring.

\begin{figure*}[htb!]
\includegraphics[width=0.23\textwidth]{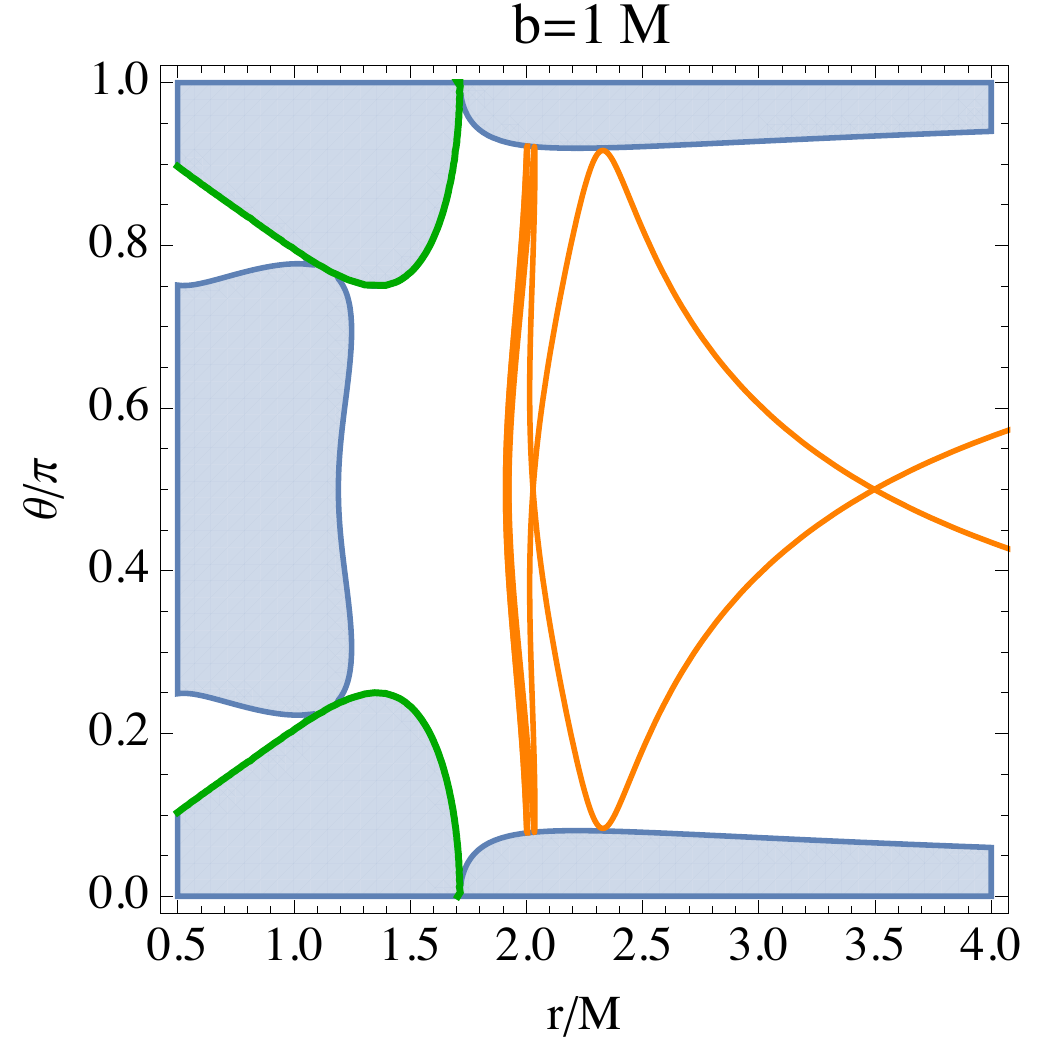}
\includegraphics[width=0.45\textwidth]{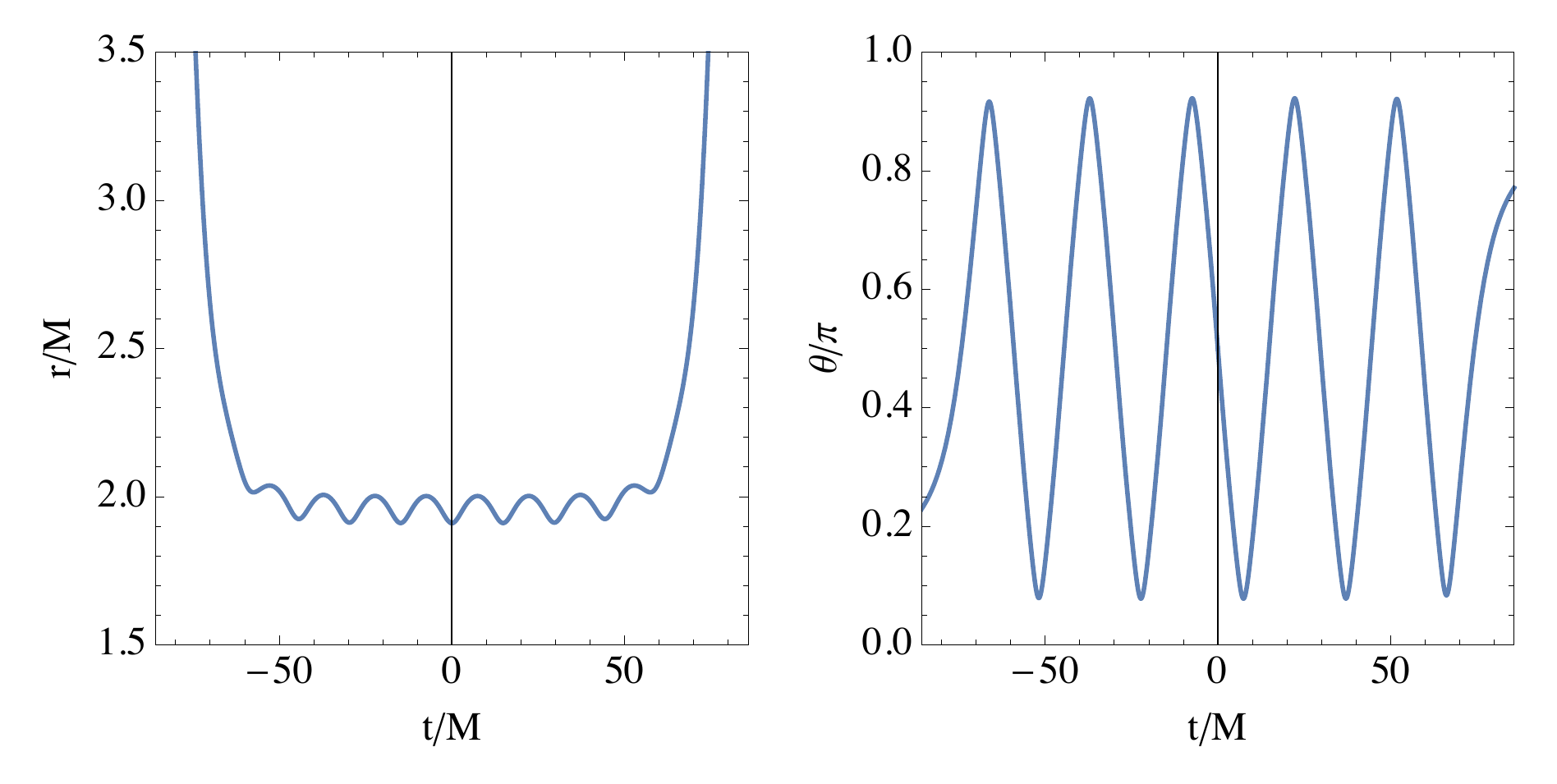}
\includegraphics[width=0.25\textwidth]{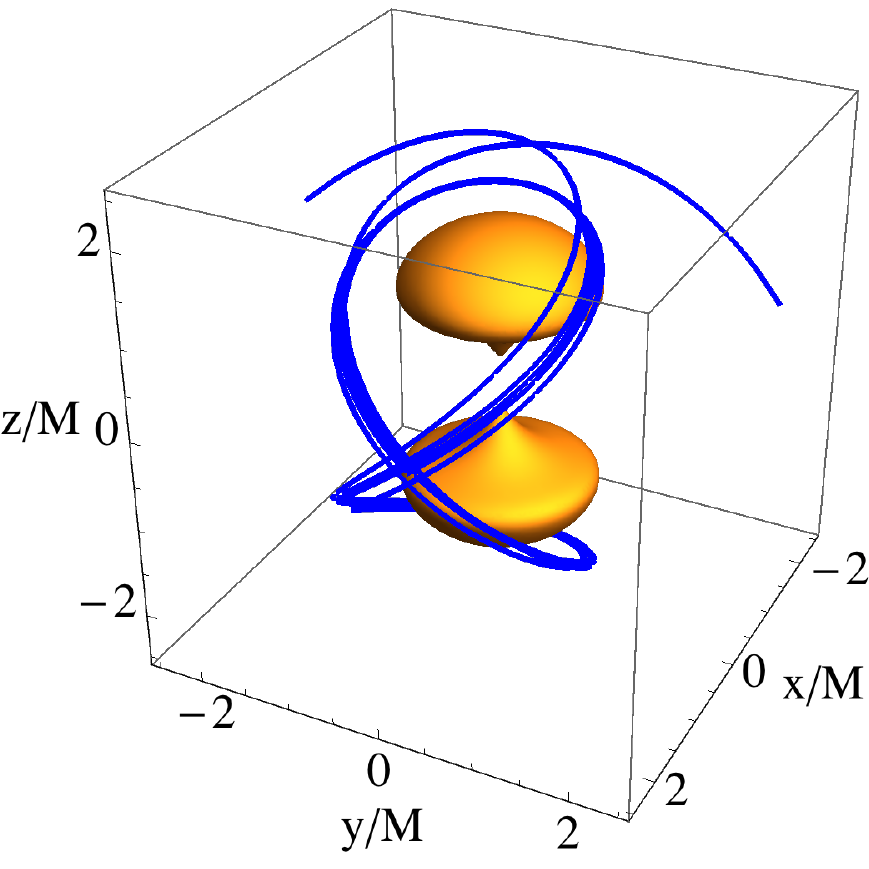}
\caption{\emph{Spheroidal orbit in a strongly deformed JP spacetime}. In this example the JP spacetime has parameters 
$\varepsilon_3 =5, a=0.7M$  and the photon moves in a nearly-spheroidal orbit with $b=1M$. Left: the orbit superimposed with the 
corresponding separatrices. Middle: the orbit's $\{ r(t), \theta(t)\}$ profiles. Right: the orbit's three-dimensional shape in Cartesian coordinates.
The coloured surfaces represent the double-lobed event horizon.}
\label{fig:JPorbits3}
\end{figure*}

Moving on, we consider a highly deformed JP spacetime with $\varepsilon_3 = 5$ and $a=0.7M$.  We first focus on orbits symmetric with respect to 
the equatorial plane and choose the same impact parameter values as in Fig.~\ref{fig:JPorbits1}. The $b=3M$ orbit does not allow the photon
to approach too close to the black hole and therefore is not shown here. For a lower impact parameter such as $b=1M$ the potential/separatrix 
opens up and one finds that spheroidal orbits are supported. An example of such an orbit is shown in Fig.~\ref{fig:JPorbits3}. 
One can clearly see that while being trapped, the photon moves up and down orbiting around both event horizon lobes while the radius 
remains nearly constant. 

\begin{figure*}[htb!]
\includegraphics[width=0.3\textwidth]{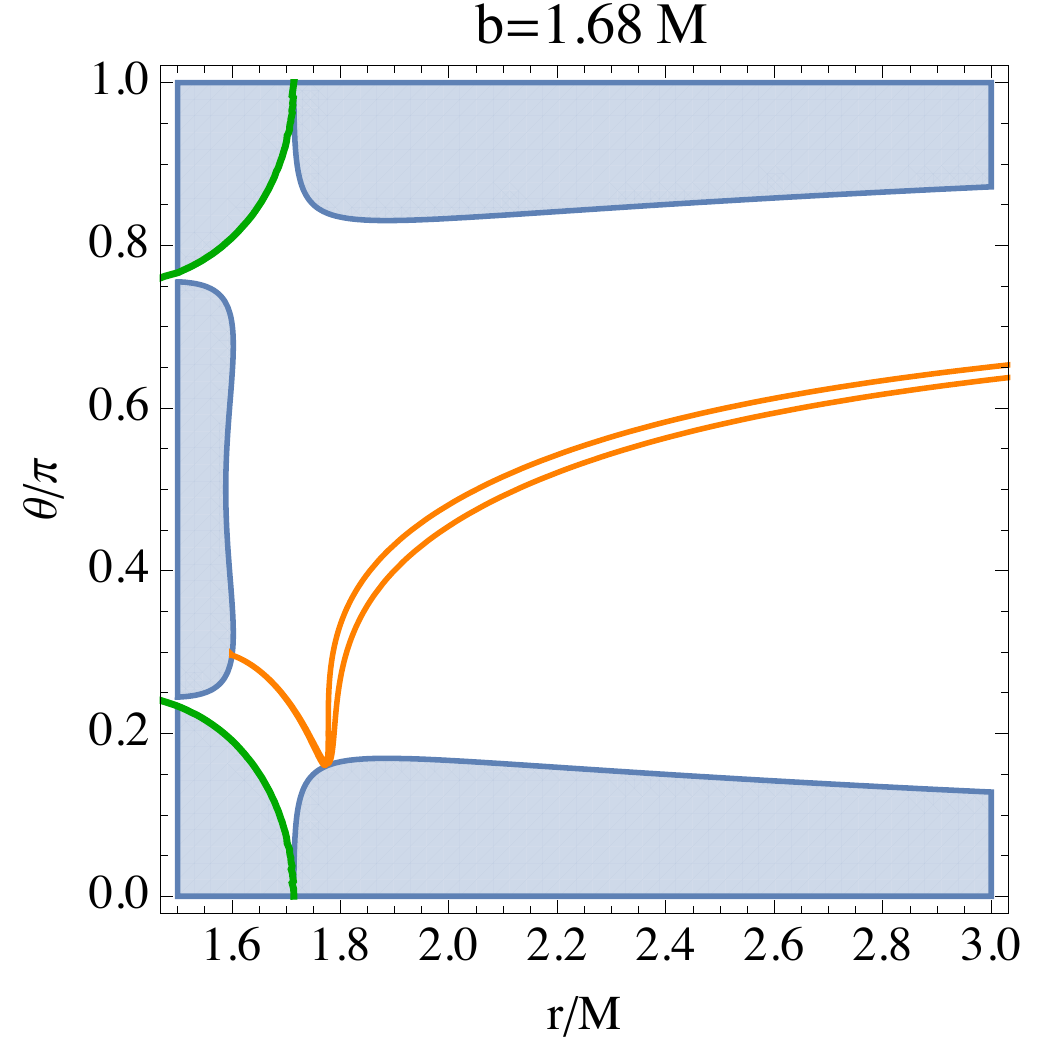}
\includegraphics[width=0.25\textwidth]{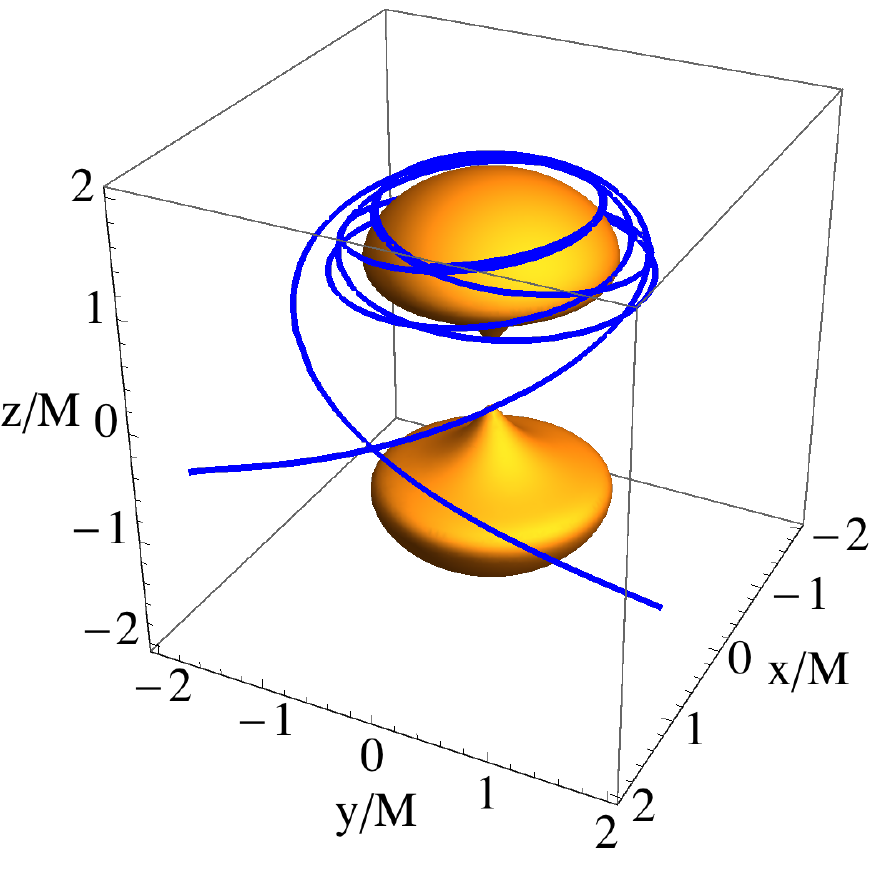}
\caption{\emph{Quasi-circular orbits in a strongly deformed JP spacetime}. We show an example of a quasi-circular orbit in a 
$\varepsilon_3=5, a=0.7M$ JP spacetime. We have chosen an impact parameter $b=1.68M$ close to the value required for the formation of 
non-equatorial photon rings. This orbit is qualitatively similar to the one of Fig.~\ref{fig:JPorbits2}, albeit with a more pronounced $\theta$-motion.
This behaviour suggests the presence of a spheroidal orbit localised well away from the equatorial plane.}
\label{fig:JPorbits4}
\end{figure*}

The $\varepsilon_3=5$ spacetime admits a pair of non-equatorial photon rings for a wide range of $a/M$. As before, it is easy to find
equatorially asymmetric orbits that trap photons in nearly circular trajectories, plausibly near the location of spheroidal orbits that
do not intersect the equatorial plane, see Fig.~\ref{fig:JPorbits4}. 

\begin{figure*}[htb!]
\includegraphics[width=0.3\textwidth]{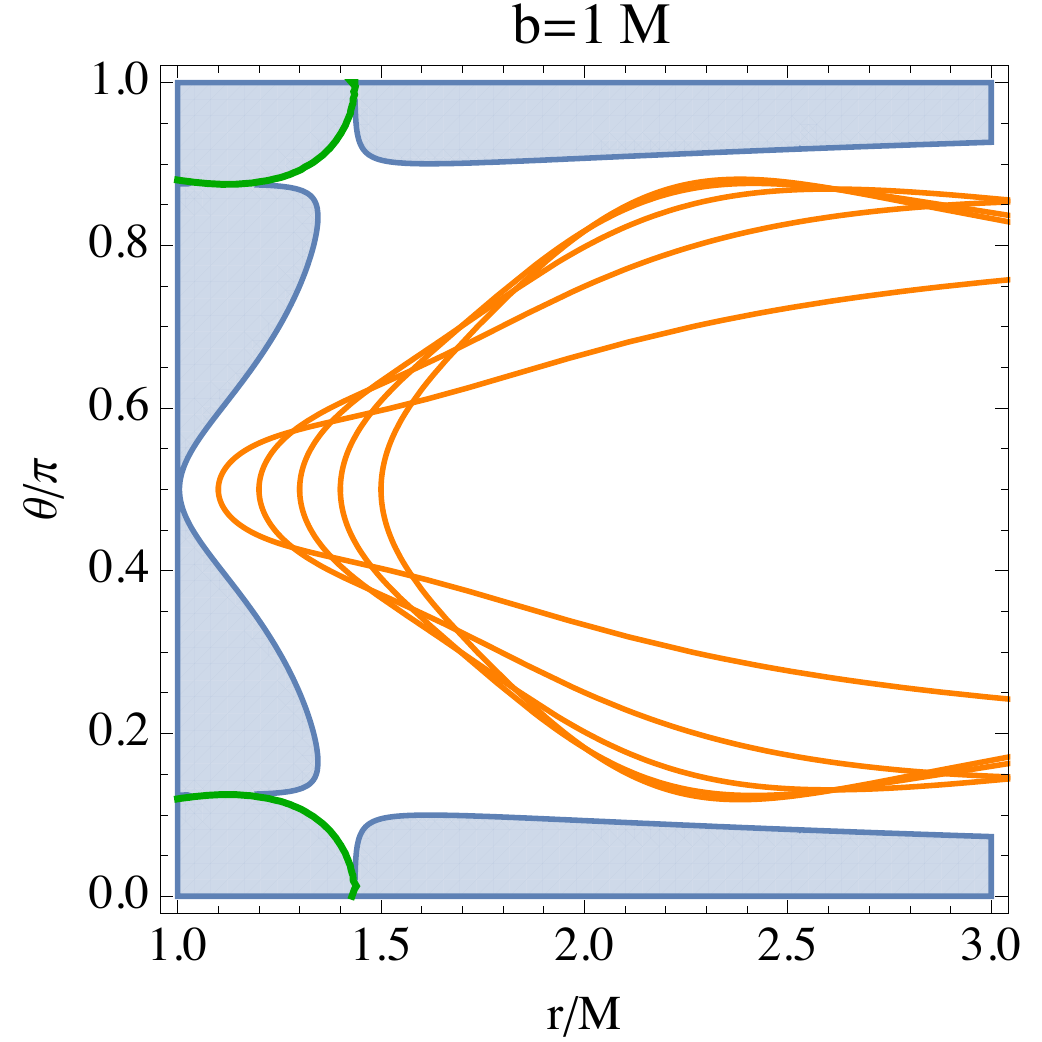}
\hspace{1cm}
\includegraphics[width=0.3\textwidth]{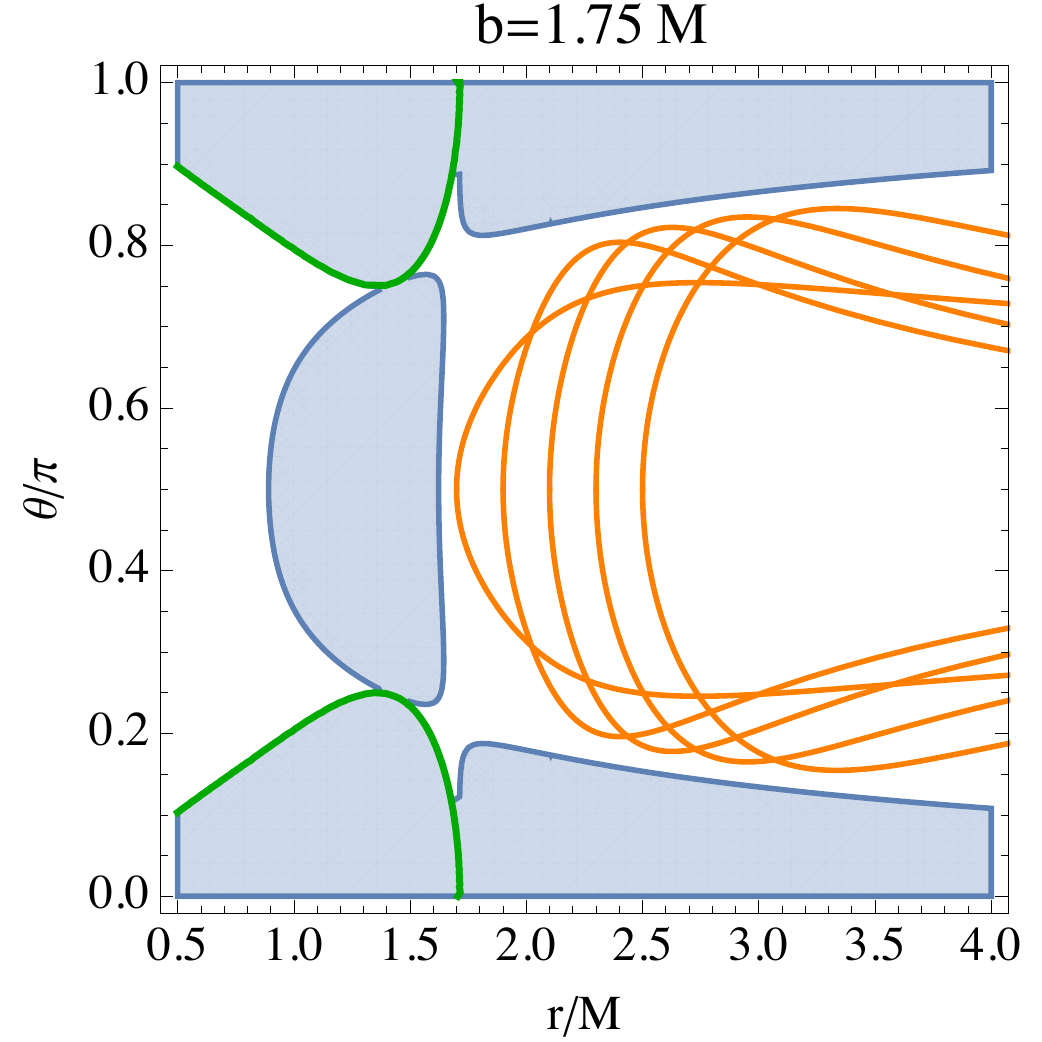}
\caption{\emph{Loss of spheroidal orbits in the JP spacetime}. This plot demonstrates the absence of equatorially-symmetric 
spheroidal orbits that cross the equator in part of the JP parameter space. Left: $\varepsilon_3=1,~a=0.9M,~b=1M$. 
Right: $\varepsilon_3=5,~a=0.7M,~b=1.75M$. In both panels we show a sequence of orbits with varying initial (minimum) radius.}
\label{fig:JPorbits7}
\end{figure*}

The previous examples may suggest that the JP spacetime can always trap photons (in the range of $b$ between the counter-rotating equatorial 
photon ring and the co-rotating non-equatorial photon rings); nevertheless, there exists a portion of the JP parameter space where spheroidal orbits 
that cross the equatorial plane are not allowed. Indeed in that case, the calculation of the equatorial $u^r$ does not return a curve that crosses zero 
as in Fig.~\ref{fig:JPureq}. As already pointed out in Section~\ref{sec:JP}, this parameter space roughly coincides with the one associated with the 
disappearance of the equatorial photon ring. The situation is illustrated in Fig.~\ref{fig:JPorbits7}, for two examples of JP spacetime, 
$\varepsilon_3=1,~a=0.9M$ and $\varepsilon_3=5,~a=0.7M$. 

\begin{figure*}[htb!]
\includegraphics[width=0.3\textwidth]{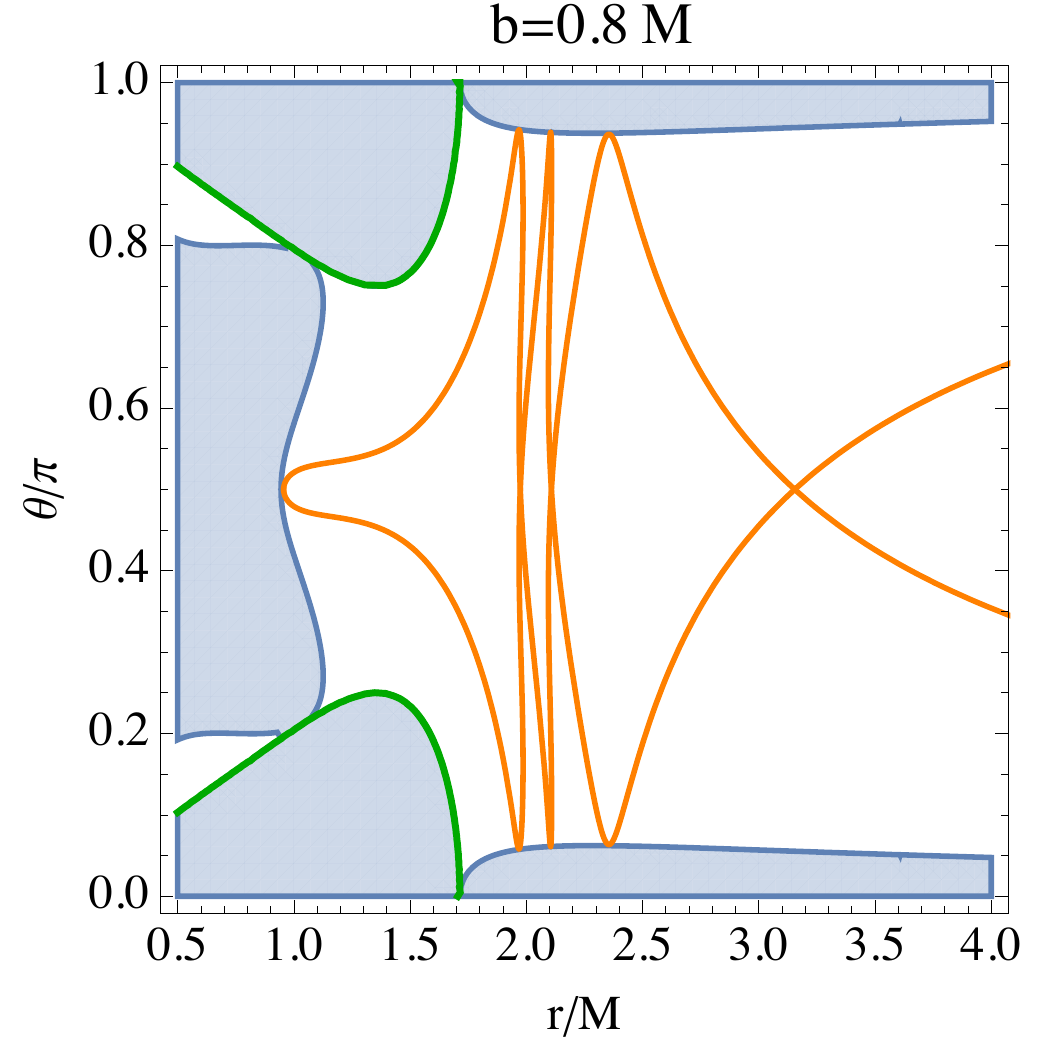}
\includegraphics[width=0.3\textwidth]{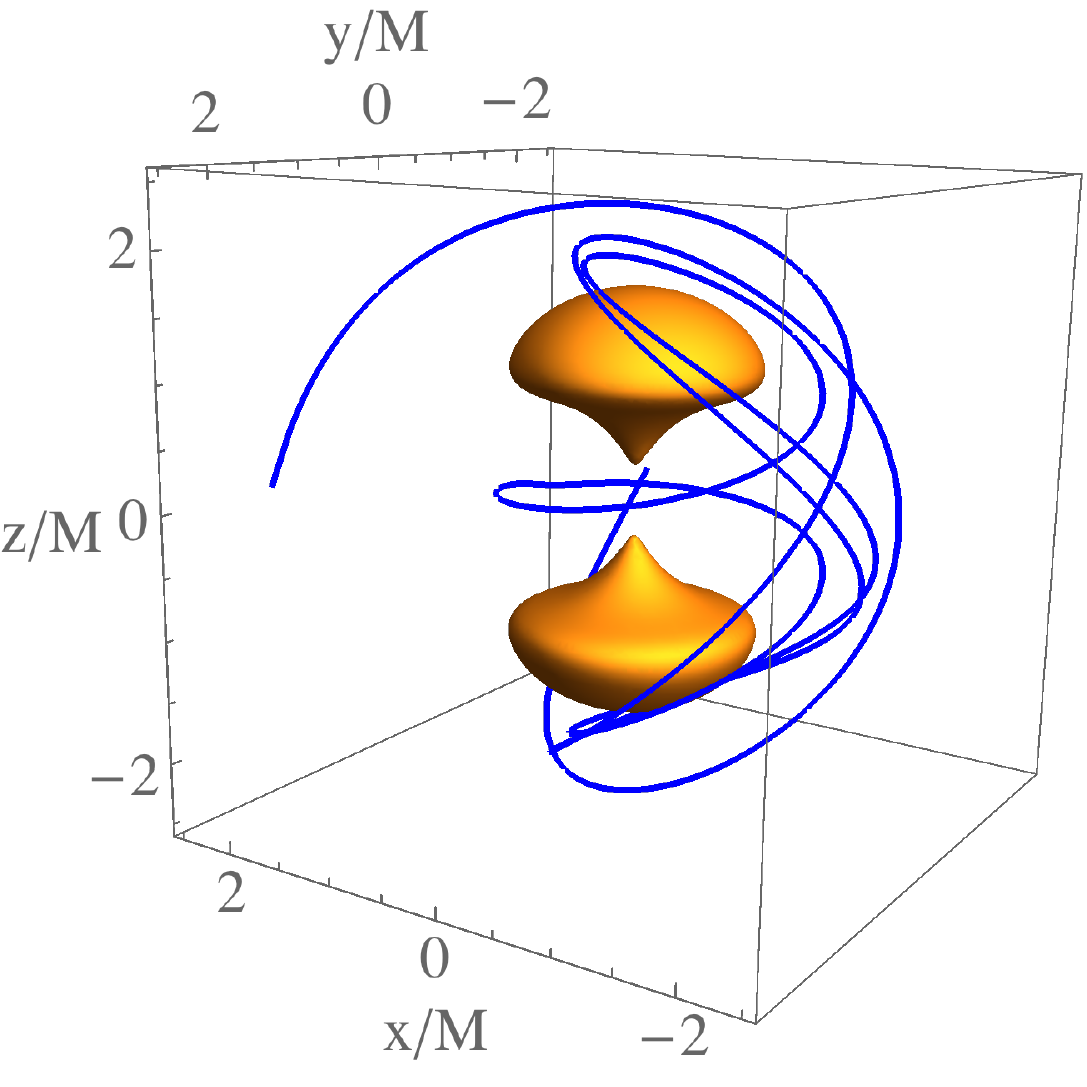}
\\
\includegraphics[width=0.5\textwidth]{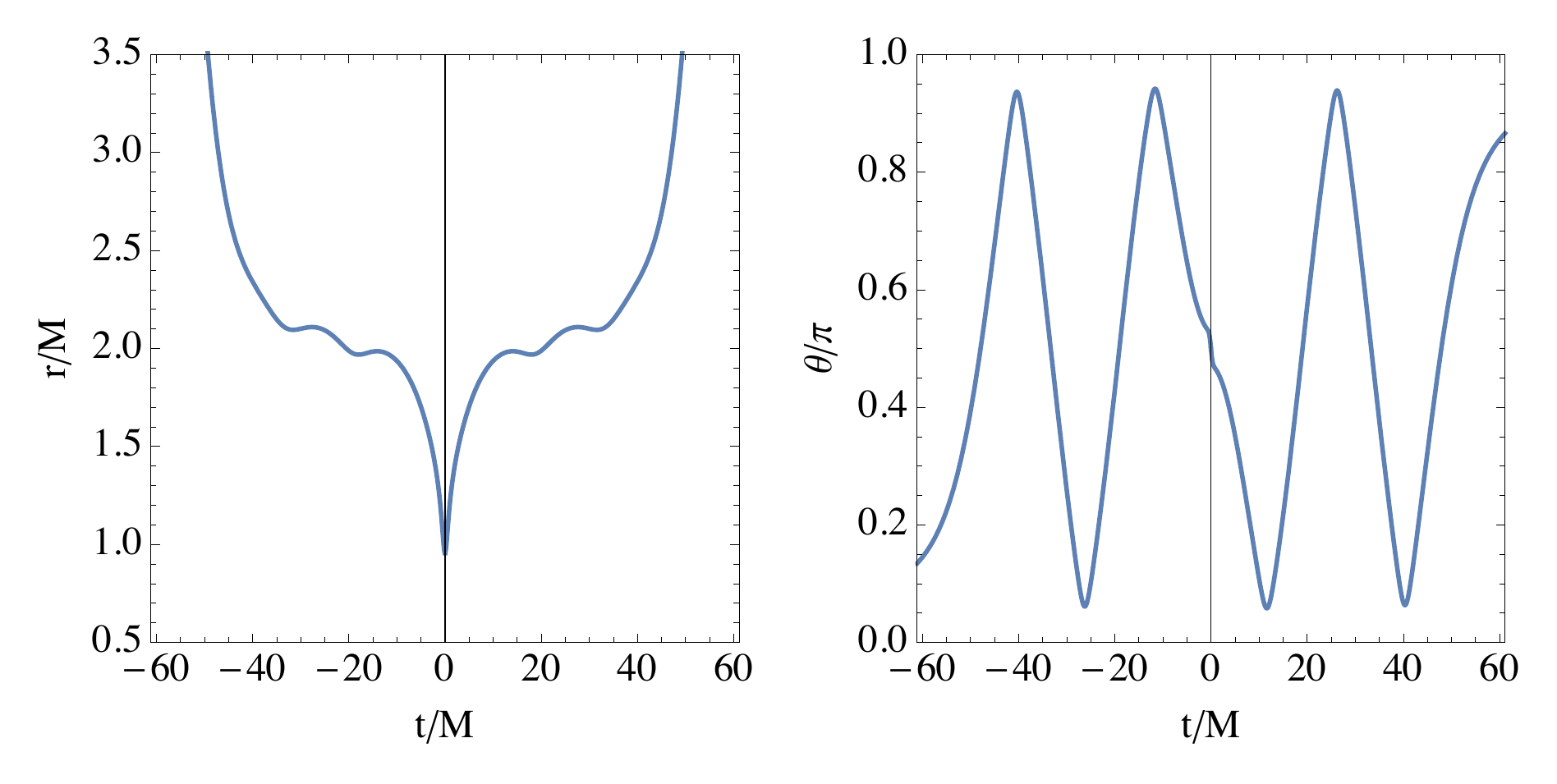}
\caption{\emph{Quasi-spheroidal JP orbit}. A low-$b$ orbit ($b=0.8M)$ in a $\varepsilon_3 =5, a=0.7M$ JP spacetime. This orbit is `exotic' in the 
sense that it allows the photon to travel through the space between the two event horizon lobes. At the same time, this is a `quasi-spheroidal' orbit 
since the photon is temporarily trapped for several revolutions in the vicinity of the black hole near the location of a spheroidal orbit with 
equatorial radius $r \approx 2.022 M$.
Top: the  $r$-$\theta$ projection of the orbit and  the $V_{\rm eff}=0$ separatrix (left panel);  the three-dimensional motion plotted together 
with the event horizon (right panel). Bottom: the $\{ r(t), \theta(t) \}$ profiles of the orbit -- these can be compared against the profiles of the 
orbits of Fig.~\ref{fig:JPorbits3}.}
\label{fig:JPorbits5}
\end{figure*}

\begin{figure*}[htb!]
\includegraphics[width=0.493\textwidth]{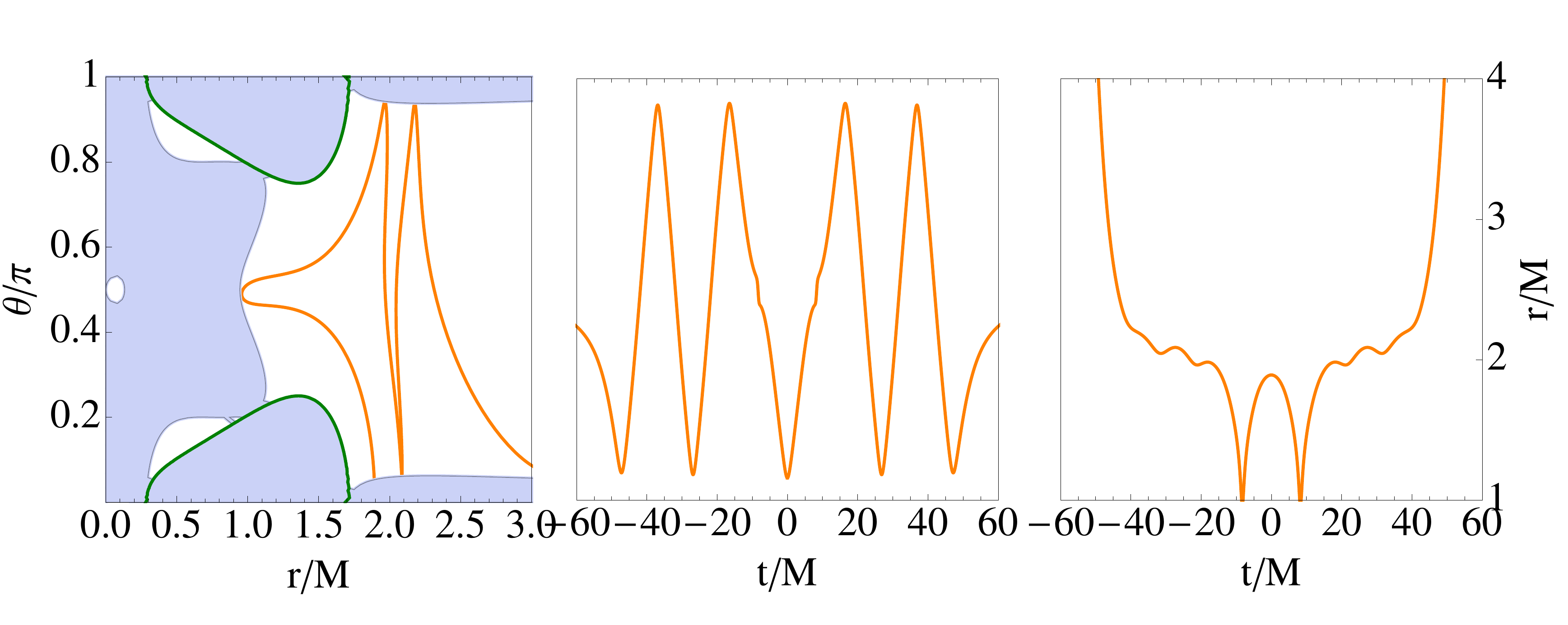}
\includegraphics[width=0.493\textwidth]{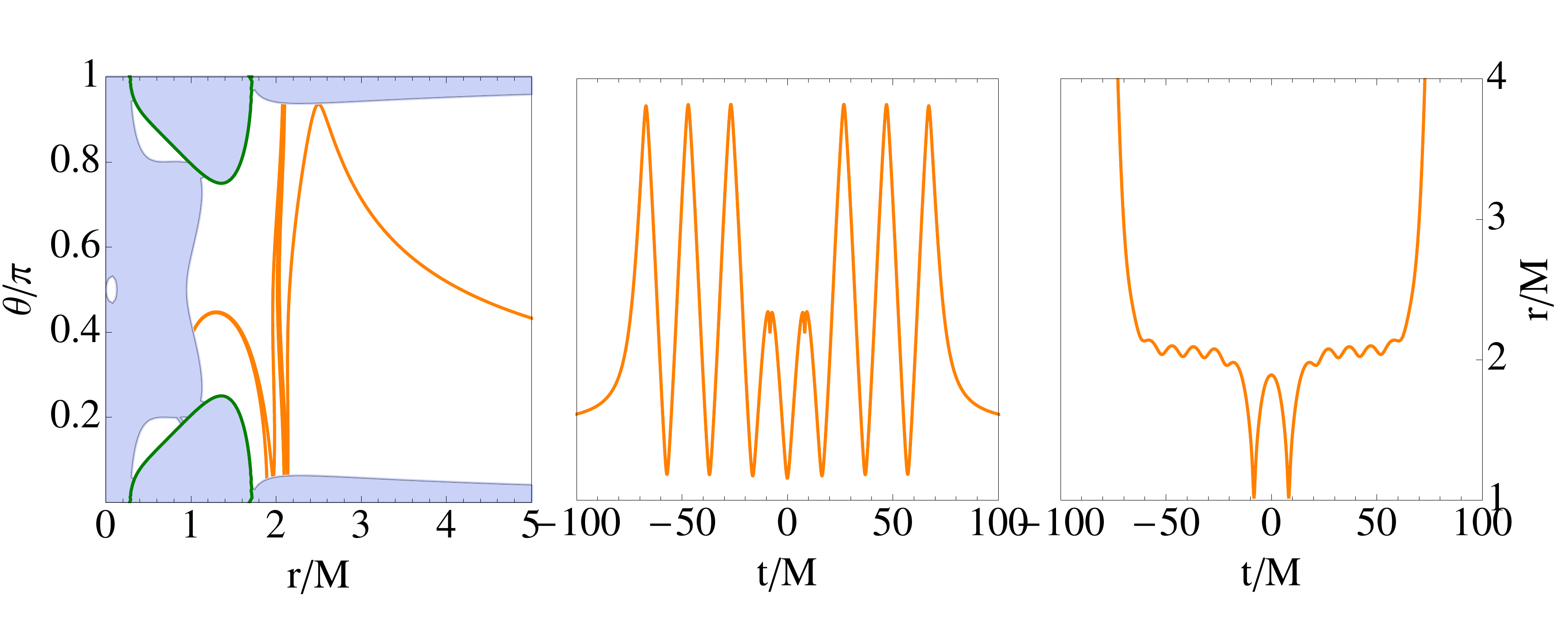}
\caption{\emph{More quasi-spheroidal JP orbits}. For the same parameters as before, i.e., $b=0.8M$, $\varepsilon_3 =5$, and $a=0.7M$, we give 
two examples that are not symmetric with respect to the equatorial plane (unlike the one shown in Fig.~\ref{fig:JPorbits5}). The left plot of each panel 
shows the $r$-$\theta$ projection of the orbit and the $V_{\rm eff}=0$ separatrix, the middle plot shows the the $ \theta(t) $ profile of the orbit, and the 
right plot the $ r(t)$ profile.}
\label{fig:JPorbits6}
\end{figure*}

As a last -- and perhaps most exotic -- example of the rich JP phenomenology, we show in Fig.~\ref{fig:JPorbits5}  an orbit that temporarily 
traps photons in the vicinity of the two event horizon lobes while at the same time forces them to pass through the space between. 
This type of orbit appears in the low-$b$ range and, as a computation of $u^r (\pi/2)$ reveals, 
can be linked to the presence of a spheroidal orbit of the same parameters that crosses the equator at $r \approx 2.022 M$.
The orbit shown in Fig.~\ref{fig:JPorbits5} is completely equatorial-symmetric but one can easily construct asymmetric orbits 
of this type (Fig.~\ref{fig:JPorbits6}). 

The careful reader may have noticed that so far only prograde ($b>0$) orbits have been discussed. Considering retrograde orbits ($b<0$)
one finds that Kerr-like spheroidal orbits and their quasi-spheroidal neighbours (like the ones shown in Fig.~\ref{fig:JPorbits1}) are always possible, 
even for a strongly deformed JP spacetime (at the same time none of the other orbits discussed in this section are present). 
This is not entirely surprising, given that a retrograde-moving photon is kept relatively far away from the black hole, thus being less exposed to 
the non-Kerr metric deviations.

The global conclusion that can be drawn from our numerical study of photon geodesics in the JP spacetime is that
although spherical orbits are formally absent, spheroidal orbits do exist across a wide range of parameters and between the two extremes of 
the counter and co-rotating photon rings, temporarily trapping photons that may happen to move in their vicinity. For a spacetime of 
small/moderate deformation/spin these orbits are essentially quasi-spherical and `Kerr-like'. The exception to the rule is provided by the parameter 
space where the JP spacetime's co-rotating equatorial photon ring is replaced by a pair of non-equatorial ones. In that case spheroidal/quasi-spheroidal 
orbits that cross the equatorial plane are not admitted for a range of $b$-values.  At the same time, the combined emergence of non-equatorial photon rings
and a two-lobed event horizon structure for high spin and/or large deformation opens the possibility of having circular and spheroidal orbits 
with a markedly non-Kerr character. The presence of spheroidal orbits around the off-equatorial photon rings is consistent with \cite{GroverWittig2017PhRvD}.

\begin{figure*}[htb!]
\includegraphics[width=0.6\textwidth]{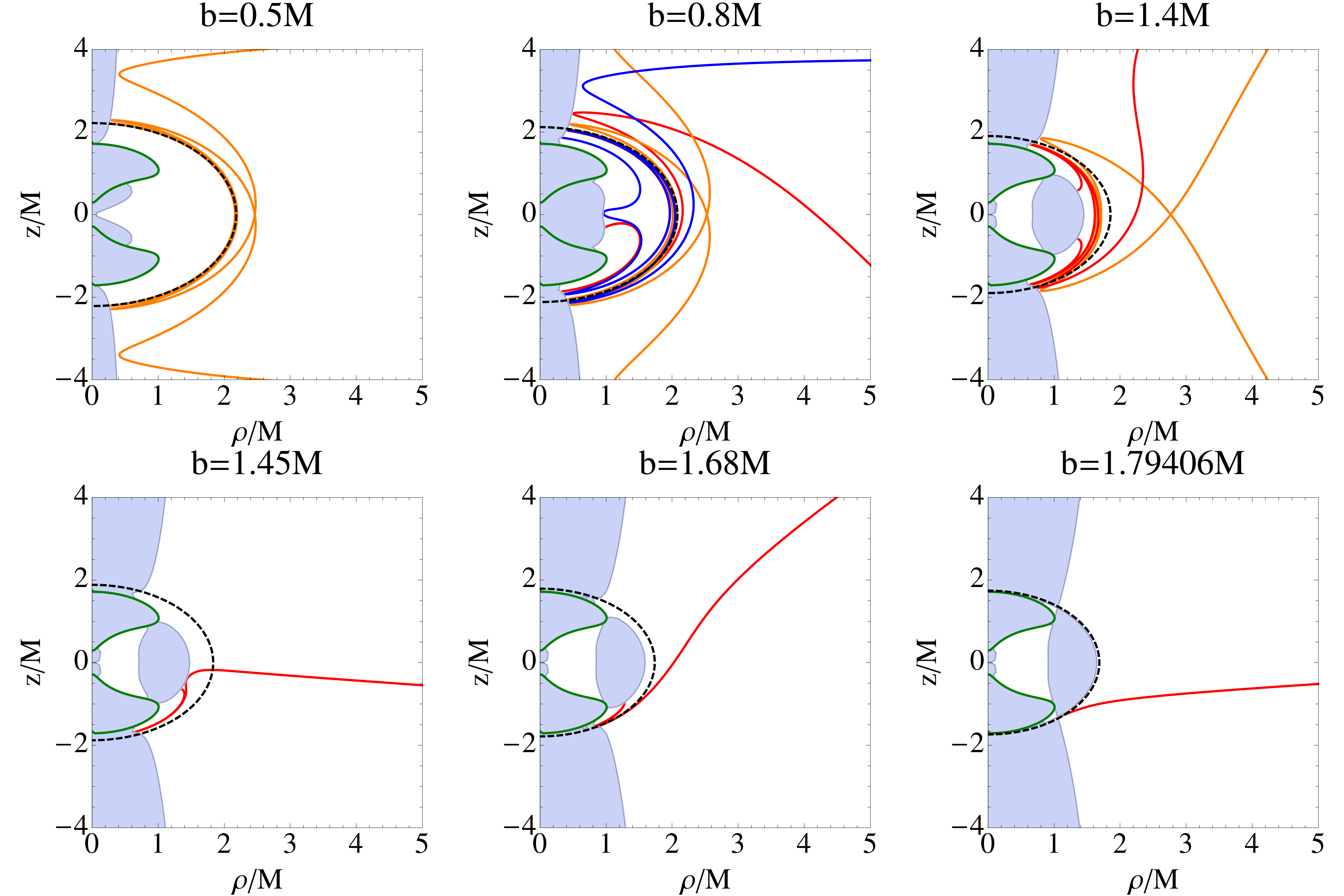}
\caption{\emph{Catalogue of JP photon orbits.} Considering a strongly deformed JP metric with $\varepsilon_3=5$ and $a=0.7M$ 
we show how photon trapping orbits vary as a function of the impact parameter $b$. As an aid for visualising motion in space, 
we use cylindrical coordinates $\{\rho, z \} = \{r \cos\theta, r \sin\theta \}$. In the top row examples, spheroidal orbits that
cross the equatorial plane are present. In the bottom row examples, photons are trapped away from the equator, 
in the region where non-equatorial photon rings form. The dashed curve indicates the approximate spheroidal orbit constructed 
using the methods of Section~\ref{sec:JP}. }
\label{fig:JPe5all}
\end{figure*}

Our results are best summarised with the help of Fig.~\ref{fig:JPe5all} where we provide 
an incomplete but indicative catalogue of orbits present in an $\varepsilon_3=5,~a=0.7M$ JP spacetime as a function of $b$. 
We use cylindrical coordinates to present the various orbits in order to facilitate a more intuitive representation for the reader.

\subsection{HT orbits}
\label{sec:tdomainHT}

Having completed our time domain analysis of the JP photon orbits we move on to a similar study of the HT metric. 
As we have already seen in Section~\ref{sec:HT}, within the slow rotation approximation, the HT metric admits spheroidal 
orbits in the exact sense. These orbits may exist beyond the perturbative regime, as suggested by the results of the 
direct integration of the spheroidicity condition for the `full' HT metric.

The time-domain analysis of this section is also based on the first approach discussed in Section~\ref{sec:HT} where the  
HT metric is used  as it is and the spin parameter $\chi$ and quadrupole $\delta q$ are treated as free parameters. 
The orbital equations are subsequently integrated following the recipe described in the previous section.  
 \begin{figure*}[htb!]
\includegraphics[width=0.8\textwidth]{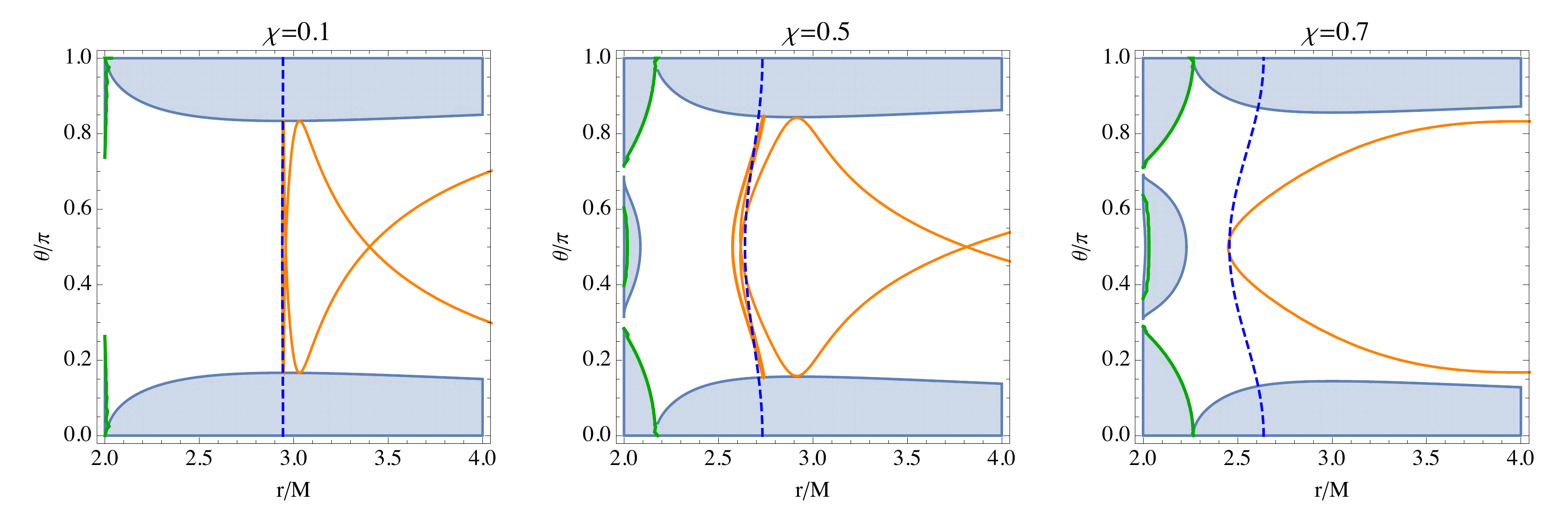}
\caption{\emph{Spheroidal orbits in the HT spacetime}. We show a sample of photon orbits in 
a HT spacetime of quadrupolar deviation $\delta q =1$ and varying spin, together with the corresponding 
separatrix (as always, the shaded area represents the forbidden region $V_{\rm eff} <0$ and the thick green curves represent the event horizon). 
The dashed curve represents the corresponding analytic spheroidal solution $r_0 (\theta)$, see Eqs.~(\ref{spheroHT})-(\ref{r2HTeq}),
for an inclination parameter $\iota = \pi/3$. From left to right: $ (\chi,b/M) = (0.1, 2.54622),~ (0.5, 2.29435 ),~ (0.7, 2.13356)$. 
The first two panels represent cases which admit spheroidal orbits (in both cases these lie very close to the perturbative result). 
Eventually, in the high-spin case of the third panel these spheroidal orbits are lost, thus removing the ability of temporarily trapping photons 
in orbits that cross the equatorial plane (the orbit shown in the right panel is a typical example).}
\label{fig:HTorbits1}
\end{figure*}
 \begin{figure*}[htb!]
\includegraphics[width=0.25\textwidth]{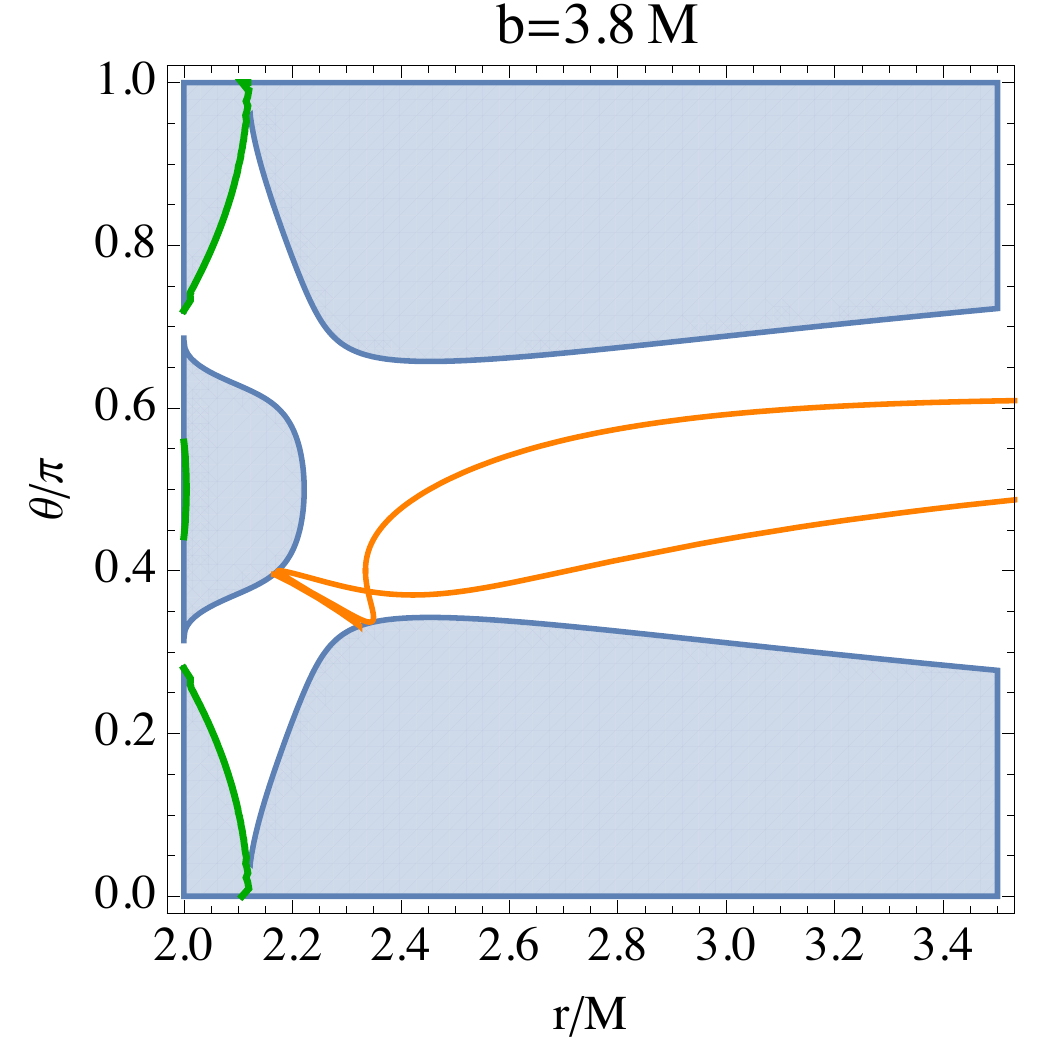}
\hspace{1cm}
\includegraphics[width=0.25\textwidth]{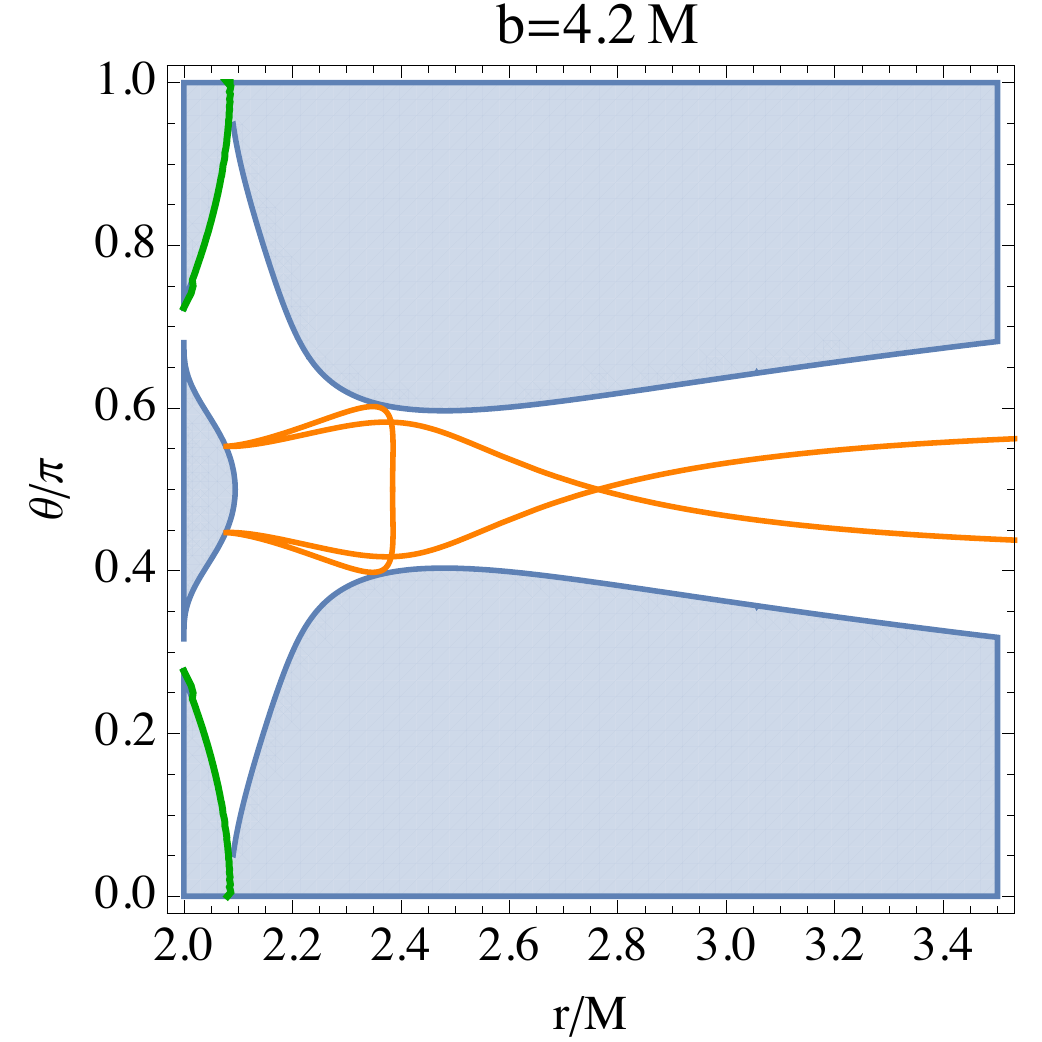}
\caption{\emph{Near-photon ring orbits in the HT spacetime.} We show two examples of orbits (superimposed with the corresponding separatrices)
with parameters $\{\delta q, \chi, b \}$ close to the ones required for the appearance of non-equatorial photon rings. The quadrupolar deformation 
is fixed at $\delta q=1$. Left: this is a quasi-circular/planar orbit with parameters $\chi=0.4,~b=3.8M$. The photon is temporarily trapped near the 
location where a non-equatorial photon ring would appear.  Right: for $\chi=0.32735,~b=4.2M$ the HT spacetime is close to admitting three photon 
rings (e.g. see middle panel in Fig.~\ref{fig:HTring_sepa}). The equatorially-symmetric orbit shown here displays three quasi-circular phases, one for 
each photon ring. In both panels the thick green curves represent the event horizon location.}
\label{fig:HTorbits2}
\end{figure*}
 \begin{figure*}[htb!]
\includegraphics[width=0.25\textwidth]{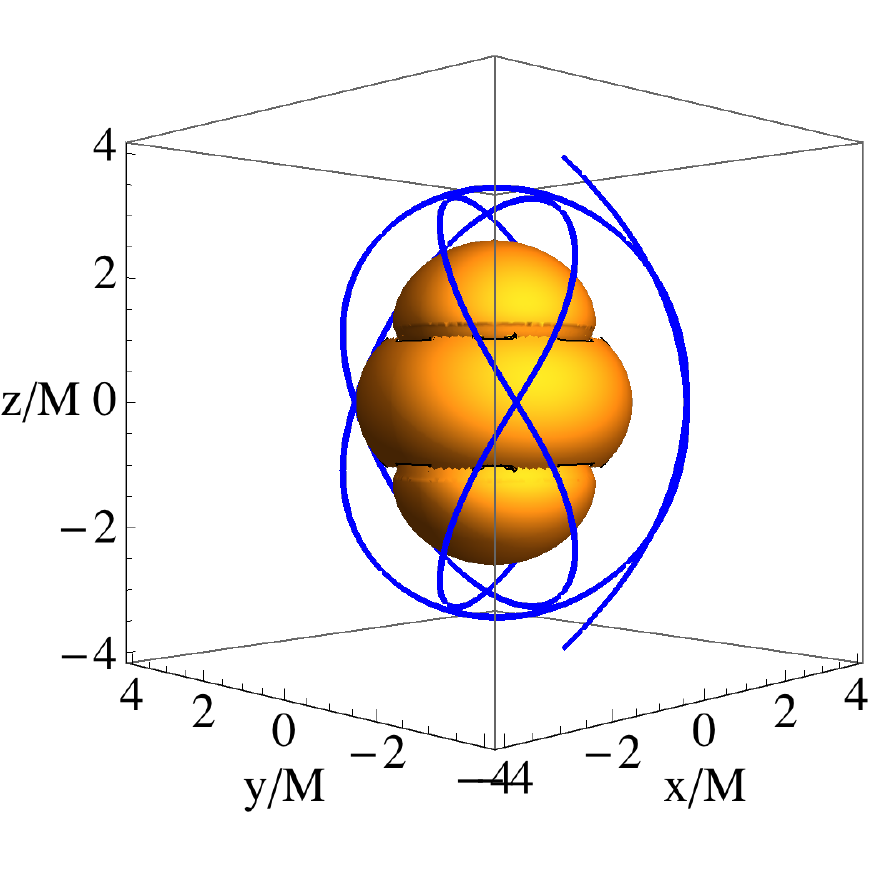}
\hspace{1cm}
\includegraphics[width=0.25\textwidth]{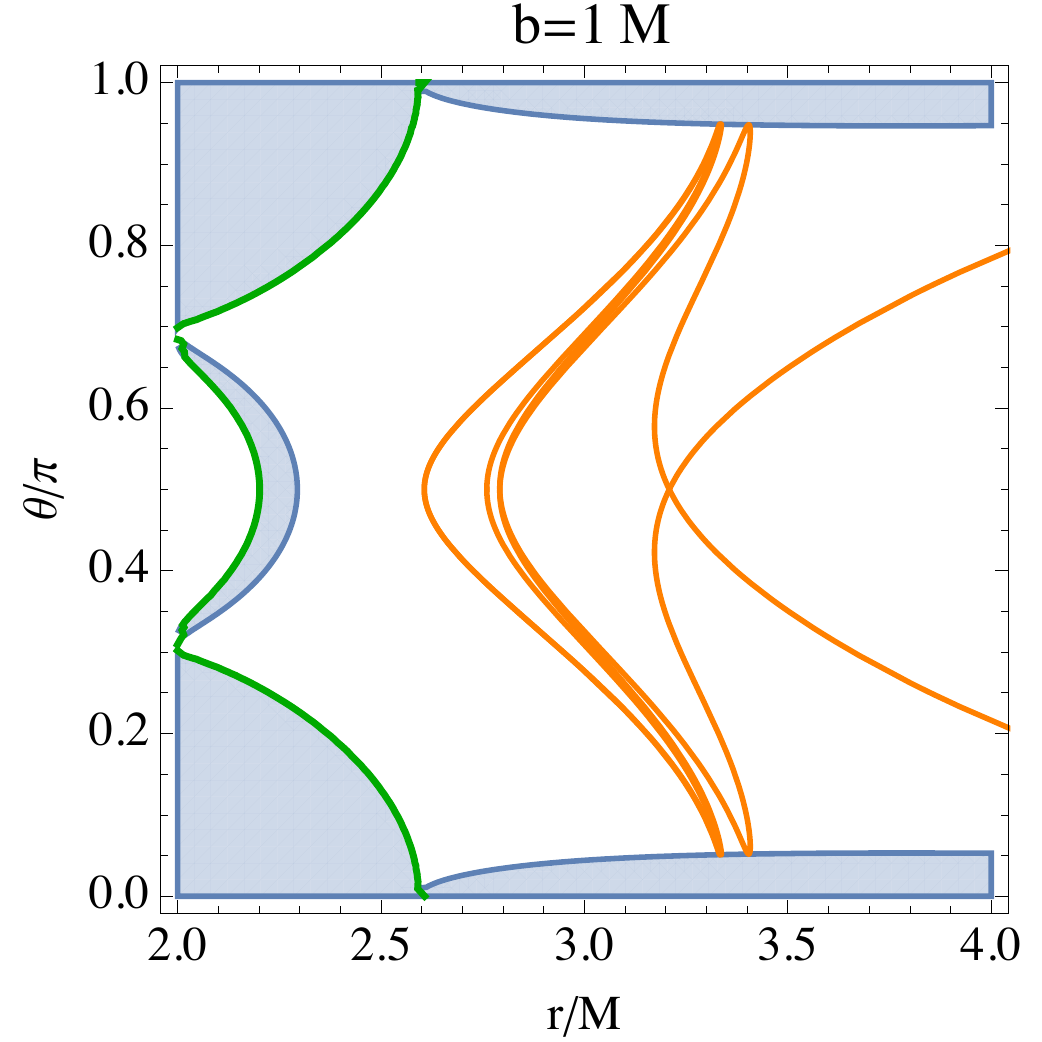}
\caption{\emph{A quasi-spheroidal orbit in a strongly prolate HT spacetime}. In this $\delta q=5, \chi=0.5$ HT spacetime, low-$b$/nearly polar orbits are found
to display a quasi-spheroidal character with a large radial variation. Right: a $b=1M$ orbit superimposed with its corresponding separatrix and HT event horizon 
(thick green curve). Left: the three-dimensional shape of the orbit in Cartesian coordinates, superimposed with the event horizon (orange-colored surface).}
\label{fig:HTorbits3}
\end{figure*}

A calculation of the equatorial $u^r$ along the lines of Fig.~\ref{fig:JPureq} indeed verifies the presence of spheroidal orbits for 
small/moderate $\chi$. For $\chi \gtrsim 0.7$, however, any notion of spheroidal motion through the equatorial plane is washed away 
and the situation qualitatively resembles that of the high-$b$, large-$\varepsilon_3$ JP orbits. This transition is clearly visible in the 
sample of results shown in Fig.~\ref{fig:HTorbits1}.  The first two panels show examples of quasi-spheroidal orbits. 
In the last panel we consider a relatively high spin case where spheroidal orbits are not admitted (this is demonstrated by plotting 
a representative orbit that fails to trap the photon for any considerable amount of time).

As we saw earlier in Section~\ref{sec:HT}, the HT spacetime admits a pair of non-equatorial photon rings above 
a spin threshold $\chi_* (\delta q)$ (and provided the motion is prograde). For example this is the case for the $\delta q =1$ HT 
spacetime considered in Fig.~\ref{fig:HTorbits2}. The chosen spin and impact parameters lie close to the ones required for the 
appearance of photon rings. The equatorially asymmetric $\chi=0.4, ~b=3.8M$ orbit shown in the left panel is quasi-spheroidal 
and suggests that a truly spheroidal orbit may exist in its vicinity (this resembles the situation discussed in the previous JP section, 
see Fig.~\ref{fig:JPorbits3}). In the right panel the HT spacetime is close  to the special `pitchfork' case of three simultaneous photon rings 
(see middle panel in Fig.~\ref{fig:HTring_sepa}). It is then possible to find orbits where photons bounce back and forth/up and down in the 
vicinity of all three photon rings!

Our final case study is concerned with an example of a strongly prolate, $\delta q=5$, HT spacetime. The numerical exploration of the low-$b$ 
regime revealed the existence a new type of  spheroidal orbit, see Fig.~\ref{fig:HTorbits3}. The orbit, which to some extent resembles the JP orbit of 
Fig.~\ref{fig:JPorbits5}, is nearly polar and highly oscillatory with respect to $r$ in its strong-field portion where a photon would 
be temporarily trapped. With increasing $b$ these orbits disappear altogether and the situation is qualitatively similar to the one shown in the right 
panel of Fig.~\ref{fig:HTorbits1}. 

The above results suggest a qualitative similarity between JP and HT photon orbital motion. The latter spacetime can trap photons near 
spheroidal/circular orbits for a large part of the $\{\delta q, \chi, b \}$ parameter space. In many cases however, this is achieved in a strongly 
non-Kerr fashion as exemplified by the orbits shown in Figs.~\ref{fig:HTorbits2} and \ref{fig:HTorbits3}.


\section{Concluding discussion}
\label{sec:conclusions}

In this paper we have explored to what extent photon trapping orbits are modified when one moves away from the Kerr spacetime 
and its spherical orbits. Our main results can be summarised as follows. Motivated by the Kerr spherical photon orbits we have explored the 
connection between such orbits and the spacetime's separability (or, in other words, the existence of a third integral of motion 
like the Carter constant). Considering only those spacetimes that already possess an equatorial photon ring, we have shown that separability 
is compatible with spherical but not spheroidal orbits. Furthermore, a spacetime that does not admit spherical photon orbits in any coordinate system 
is necessarily non-separable. It should be noted, however, that the inverse statement is not necessarily true, that is,
non-separability does not always imply the loss of spherical orbits. Next, we turned our attention to three well-known specific examples of non-Kerr 
spacetimes (Johannsen, Johannsen-Psaltis and Hartle-Thorne) used as proxies of the spacetimes of ultracompact objects, alternative to black holes. 
In accordance with the aforementioned sphericity-separability connection, we have found that the J spacetime, the only separable example among the 
three, admits spherical photon orbits. On the other hand, by means of a general `spheroidicity condition' we have shown that in the JP and HT spacetimes 
spherical orbits are replaced by spheroidal orbits that cross the equatorial plane. This latter type of orbit may be lost when the
deviation away from Kerr is large and/or the spin is relatively high. This presence/absence of spheroidal orbits that cross the equatorial 
plane is strongly correlated with the presence/absence of the equatorial photon ring and is a function of the impact parameter. 
Our numerical time-domain analysis of orbits in these two spacetimes has revealed that photons can be temporarily trapped in 
quasi-spheroidal orbits for a large portion of the parameter space. The emergence of symmetrically-placed non-equatorial photon rings 
above some spin threshold allows photons  to be trapped in quasi-circular/planar orbits in the vicinity of the rings. Although not studied 
rigorously here, localised spheroidal orbits that do not pass through the equator may also appear in the same region.  
In the low-$b$ range one finds quasi-spheroidal orbits with high inclinations and large variations in $r$ in the vicinity of the black hole. 
This highlights one of the key conclusions drawn from the time-domain calculations: even if a photon is temporarily captured in a
quasi-spheroidal orbit, the resulting motion may be significantly non-Kerr. This difference is easily seen in Fig.~\ref{fig:KerrvsJP}'s comparison of Kerr 
against JP photon trapping orbits for the same spin and impact parameter, with the most interesting feature emerging from this being 
that the spacetime  loses the trapping orbits that cross the equatorial plane and do not extend very far from it.

There are at least two ways these results could have a significant impact on the observed electromagnetic and gravitational 
wave signature of black holes (or more precisely, of the putative ultracompact objects that could pass for black holes). 

From the perspective of photon astronomy, the absence of near-equatorial spherical/spheroidal photon trapping orbits in the spacetime of a 
non-Kerr object could cause modifications to the shadow image of a system like a supermassive `black hole' illuminated by a radiating 
accretion flow. At the most basic level, this might translate to a change in the shadow's shape and a suppression/dimming 
of its bright boundary.  Of course, the real situation is likely to be much more complicated than that, with new lensing features 
arising due to the presence, for example, of stable photon orbits and/or non-equatorial photon rings. Significant progress towards 
understanding the rich phenomenology of non-Kerr shadow imaging has been made recently~\cite{Cunha2016, Cunha2017b, Cunha2018}.
 
From the GW side, it is well known that the main black hole QNM ringdown can be understood in terms of gravitational 
wavepackets temporarily trapped in the vicinity of the unstable photon circular orbit before leaking towards infinity and the 
event horizon. This mental picture remains valid for both equatorial $\ell= |m|$ and non-equatorial  $\ell > |m|$ angular 
modes~\cite{Yang:2012he}. As the names suggest, the former (latter) modes are associated with equatorial (non-equatorial) circular
(spherical) photon orbits, where the ratio $|m|/\ell \sim\sin \iota_0$ can be associated to the maximum orbital inclination angle 
$\iota_0$ relative to the equatorial plane, or in other words to the impact parameter $b$. Theoretical modelling (and of course the recent 
GW detections themselves) suggest that among the two families it is the equatorial one (and especially the quadrupolar mode $\ell=m=2)$ 
that typically dominates the ringdown signal.
However, the idea of testing the Kerr hypothesis with QNM `spectroscopy'  is based on the simultaneous observation of several
QNM `lines' \cite{Berti:2007zu, Berti:2016lat, baibhav2018} in which case non-equatorial modes such as the $(\ell,m) = (2,1)$ come to
play an important role. Given that the required ringdown SNR is at least an order of magnitude higher than that of the  strongest 
signal observed so far by LIGO, this kind of test would require a next generation detector. The order of magnitude boost in the SNR 
more or less reflects the relative strength between the quadrupole and other modes. This gap could be shortened for binary systems
with rapidly spinning members and for certain orientations of the spins~\cite{Kamaretsos:2011um, Kamaretsos:2012bs}.

Assume now that instead of Kerr black holes we actually observe  some other type of ultracompact object or a non-GR black hole 
with a non-separable exterior spacetime. The loss of  some classes of spheroidal trapping orbits  would imply a direct impact on the signal 
associated with the prograde non-equatorial $\ell \geq m > 0$ modes. Depending on the actual degree of deformation/non-separability
of the spacetime in question, the rich phenomenology of non-Kerr orbits (e.g. absence of spheroidal orbits, presence of more exotic 
quasi-spheroidal orbits, non-equatorial photon rings)
is likely to lead to a markedly different QNM spectrum: one would expect the dimming of some QNM lines and the appearance of new ones. 
For example, the loss of the equatorial photon ring for some parameters may result in the loss of the $\ell=m$ QNMs.
This could  be the smoking gun signalling the existence of non-Kerr objects; additional evidence might come from the presence of 
late-time `echoes' \cite{Cardoso:2016rao, Maselli_etal2017} and/or a modified early ringdown signal \cite{Glampedakis2018}. 

Apart from the case of exotic UCOs in GR, one could equally well consider black holes in alternative theories of 
gravity, see~\cite{Berti:2015itd} for a comprehensive review. It is known that the Kerr black hole metric is not 
exclusive to GR (although it may lose its uniqueness status in other theories). For such a case our results have no
impact. On the other hand there is a handful of known non-Kerr black hole solutions that are typically `quasi-Kerr'  in the sense that they
deviate slightly from Kerr as a result of being approximate solutions with respect to rotation and/or some coupling constant, 
see for example~\cite{Sopuerta_Yunes2009}. Such non-separable systems are expected to display the same 
phenomenology as the JP/HT orbits discussed here.

Although this work has been focused on photon geodesics, we have seen that the spherical orbits-separability connection 
persists for the case of massive particles. If the behaviour that we have seen for photons carries over to particle spherical/spheroidal
orbits in non-Kerr spacetimes, then it is likely to have a strong impact on the GW waveform of an extreme mass ratio inspiral (EMRI) system, 
where a stellar-mass black hole slowly inspirals in the gravitational field of a supermassive black hole. EMRIs are among the prime targets
for the future LISA space-based GW detector, and are envisaged as the sources that will provide a  
detailed `map' of the Kerr metric.

A detailed study of the implications of our results lies beyond the scope of this paper but becomes the natural objective of 
our future work. As a first instalment, a forthcoming publication will explore the dynamics of scalar waves in non-separable 
spacetimes with the purpose of testing the dependence of the strength of the non-equatorial QNM signal as a function of the 
degree of non-separability. 

\acknowledgements

We thank Theocharis Apostolatos, Emanuele Berti, Georgios Lukes-Gerakopoulos and Thomas Sotiriou for valuable 
comments during the course of this work. We also thank the anonymous referees for their constructive and valuable 
input that has led to a major improvement of this work. We acknowledge networking support by the COST Actions GWverse CA16104 
and PHAROS  CA16214. GP acknowledges financial support provided under the European Union's H2020 ERC, Starting Grant 
agreement no.~DarkGRA--757480.


\appendix

\section{Spherical Kerr orbits}
\label{sec:Kerr}

This appendix provides a compact discussion of spherical photon orbits in the Kerr spacetime. The Kerr metric
in Boyer-Lindquist coordinates is given by the following familiar expressions~\cite{Bardeen:1972fi},
\begin{align}
g_{tt}^\rK &= -\left (1- \frac{2M r}{\Sigma} \right ), \qquad g_{t\varphi}^\rK = -\frac{2Mar}{\Sigma} \sin^2\theta, 
\qquad g_{rr}^\rK = \frac{\Sigma}{\Delta},
\nn \\
g_{\theta\theta}^\rK &= \Sigma, \qquad g_{\varphi\varphi}^\rK = \left (  r^2 + a^2 +  \frac{2Ma^2r}{\Sigma} \sin^2\theta \right ) \sin^2 \theta,
\label{gKerr}
\end{align}
where $\Delta = r^2 -2M r+ a^2$ and $\Sigma=r^2 + a^2 \cos^2\theta$. Making contact with the general formalism 
of Section~\ref{sec:formalism}, we have $\cD \to \Delta, ~ g_{rr} \to \Sigma/\Delta$ plus the Carter constant 
relation~\cite{Bardeen:1972fi}
\be
Q = u_\theta^2 + \cot^2\theta \left ( \, u_\varphi^2  -   \sin^2\theta\,  a^2 u_t^2 \, \right ).
\label{Qeq}
\ee
Using this to eliminate $u_\theta^2$ in the general expression (\ref{norm_gen}) yields a decoupled radial motion 
equation. At the same time (\ref{Qeq}) itself becomes a decoupled equation for the latitudinal motion.  
These two equations take the form (see ~\cite{Bardeen:1972fi} for details)
\be
(\Sigma u^r )^2 = V_r (r,b,Q), \qquad (\Sigma u^\theta)^2 = V_\theta (\theta,b,Q).
\label{KerrEoMs}
\ee
Spherical Kerr orbits are defined as $ u^r = d u^r/d\lambda =0$ at $r= r_\rK$.  It is easy to show that these two requirements 
translate into a pair of conditions for the radial potential,
\be
V_r (r_\rK ,b,\cos^2\iota)  = V^\p_r  (r_\rK,b, \cos^2\iota)= 0.
\label{Vcond_K1}
\ee
Here the constant $\iota$ is a proxy for the orbital inclination, defined as $Q = L^2 \tan^2\iota$. 
One of these equations furnishes an analytic relation  $b=b(r_\rK, \cos^2\iota)$. while the other 
becomes the `photon ring' equation $ \cE_{\rm ph} (r_\rK,\cos^2 \iota) = 0$. This has to be solved 
numerically unless $\iota=0$ (equatorial motion). 
Alternatively, one could solve one of the conditions (\ref{Vcond_K1}) to find $r_\rK = r_\rK (b,\cos^2\iota)$ 
and subsequently use this result to obtain a relation  $f(b,\cos^2\iota)=0$ or equivalently a relation $ Q = Q(b)$ 
between the constants of motion. In both cases the remaining $u^\theta$ equation can be used for calculating 
the orbital period.

 \begin{figure*}[htb!]
\includegraphics[width=0.99\textwidth]{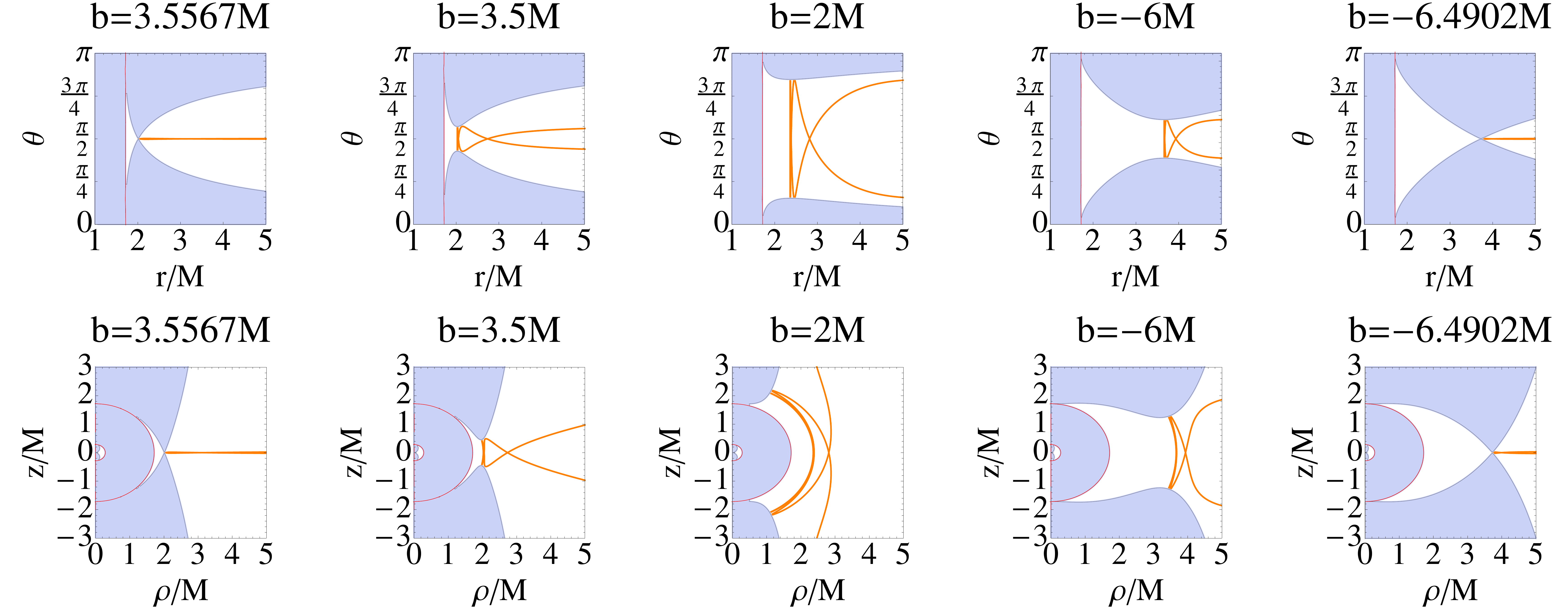}
\caption{\emph{The allowed $r$-$\theta$ region for Kerr spherical photon geodesics}. We show a series of snapshots
of the $V_{\rm eff} = 0$ separatrix between allowed and forbidden (shaded) regions for photon geodesic motion on the $r$-$\theta$ 
(top row) and  $z$-$\rho$ (bottom row) planes. The curves represent marginally capture orbits at the location of the spherical photon orbit, starting
from the prograde equatorial photon ring (leftmost panels) and ending at the retrograde equatorial photon ring (rightmost panels). }
\label{fig:Veff_Kerr}
\end{figure*}

Fig.~\ref{fig:Veff_Kerr} illustrates the allowed $r$-$\theta$ region (i.e. the separatrix $V_{\rm eff} =0$) 
for circular photon geodesics in Kerr. More specifically, we show (i) the unique Schwarzschild 
equatorial circular orbit ($r_\rK = 3M$) as a typical example of equatorial motion, 
and (ii) a non-equatorial orbit  ($r_\rK \approx 2.34M$, $\iota = \pi/3$) in a Kerr black hole with spin $a=0.7M$. 
The qualitative difference between the two cases is evident, with the circular radius acting as a `bottleneck'
in the equatorial case and demonstrating the unstable character of the orbit. 

One last remark concerns the non-existence of spheroidal Kerr orbits. This follows as a a special case of the general
result discussed in Section~\ref{sec:theorem}.


\section{Separability and spherical orbits in Newtonian gravity}
\label{sec:Newtcirc}

The link between separability and the existence of spherical orbits can be firmly established in Newtonian gravity.
Considering point particle motion in an axisymmetric potential $V(r,\theta)$ (here we use standard spherical coordinates
but the following analysis can be extended to include any system of curvilinear coordinates), the total energy is:
\be
\cE = \frac{1}{2} \left (\, \dot{r}^2 +  r^2 \dot{\theta}^2 + r^2 \sin^2\theta \dot{\varphi}^2 \, \right )  + V(r,\theta).
\ee
where a dot denotes a time derivative and the particle mass is taken to be unity.
Apart from the energy, the system conserves its $z$-component of angular momentum
\be
L= r^2 \sin^2\theta \dot{\varphi}.
\ee
Combining these two relations,
\be
2  ( \cE -V ) =  \dot{r}^2 +  r^2 \dot{\theta}^2 + \frac{L^2}{r^2 \sin^2\theta}. 
\label{Newt1}
\ee
The time derivative of this is,
\be
\dot{r} \ddot{r} + r^2 \dot{\theta} \ddot{\theta} + r \dot{r} \dot{\theta}^2  + V_{,r} \dot{r} + V_{,\theta} \dot{\theta}
- \frac{L^2}{ r^3\sin^2\theta} \left ( \dot{r} + r \cot\theta \dot{\theta} \right ) = 0. 
\label{Newt2}
\ee
This equation becomes the Newtonian spheroidicity condition once we eliminate $\dot{\theta}^2$ with the help of 
(\ref{Newt1}), $\ddot{\theta}$ using Newton's second law,
\be
\ddot{\theta} = \frac{1}{r^4} \left (\, \frac{L^2 \cot\theta}{\sin^2\theta} - r^2 V_{,\theta} - 2 r^3 \dot{r} \dot{\theta} \,\right ),
\ee
and finally impose $r=r_0(\theta)$. The end result is,
\begin{align}
& r_0^5 \left [ 2 (V -\cE) + r_0 V_{,r} \right ] + \left ( \frac{ L^2 \cot \theta}{\sin^2\theta} 
- r_0^2  V_{,\theta}  \right )   r_0^2  r_0^\p +  \left [ \frac{L^2}{\sin^2\theta}  + 4 (V-\cE) r_0^2 + r_0^3 V_{,r}  \right ] r_0 (r_0^\p)^2
\nn \\
& \left (  \frac{L^2 \cot\theta}{\sin^2\theta} - r_0^2  V_{,\theta}  \right ) (r_0^\p)^3 
+ \left [ 2 (\cE-V) r_0^2 -  \frac{L^2}{\sin^2\theta}  \right ] r_0^2  r_0^{\pp} = 0, 
\label{Newtcirc}
\end{align}
where all derivatives are to be evaluated at $r_0$. 

For a spherical orbit $r_0 = \mbox{const.}$ the spheroidicity condition reduces to
\be
2 (V-\cE) + r_0 V_{,r} = 0. 
\ee
This expression is satisfied by any central potential, for example an attractive potential $V = -K/r^n$ (with $K>0$) 
leads to $\cE = K(n-2)/2 r_0^n$ which predicts bound circular orbits for any $n<2$. The second family of solutions 
is of the non-central form
\be
V(r,\theta) = \frac{\Theta(\theta)}{r^2},
\ee
with the dipolar field $V \sim \cos\theta /r^2$ being one of the simplest members of this class. As discussed in the 
Landau-Lifshitz textbook~\cite{LLbook} these potentials are precisely the only separable ones in spherical coordinates.  
This short calculation thus demonstrates how the spheroidicity condition singles out the separable potentials in a given 
coordinate system. 

In the case where an otherwise separable potential is written in a different coordinate system spherical orbits become 
spheroidal. An example is provided by the famous `two-centre' Euler potential which, with the exception of the Keplerian one, 
is the only separable axisymmetric and equatorial-symmetric Newtonian potential 
(for more details see e.g.~\cite{GlampedakisApostolatos13}).

This potential is sourced by two point masses $M/2$ placed symmetrically along the $z$-axis at a distance $a$ from the origin.
In fact the Euler potential comes in two flavours, an oblate and a prolate one, in the former case the distance between the centres
being imaginary, $a\to i a$. For the relativist it is the oblate Euler potential that is more interesting since it shares many of the special 
properties of the Kerr metric \cite{GlampedakisApostolatos13, Apostolatos2013JPhCS}. Despite the imaginary distance between the 
two point masses the potential itself is real-valued and takes the form
\be
V_{\rm E} = - \frac{M}{\sqrt{2} R^2} \sqrt{R^2 + r^2 -a^2}, \qquad R^2 = \sqrt{(r^2-a^2)^2 + 4 a^2 r^2  \cos^2 \theta}.
\label{VE}
\ee
The Euler potential is known to be separable and admit spherical orbits (that can be stable or unstable) in an adapted elliptical 
coordinate system~\cite{LLbook}. Here, however,  we will keep working  with standard spherical coordinates and study the 
spheroidicity condition for the potential (\ref{VE}).

The actual calculation is facilitated by a  small-$a$ approximation, effectively treating the Euler potential as a perturbation away 
from a Keplerian potential. This would also mean that this calculation only applies to stable orbits.  
Working to ${\cal O} (a^2)$ precision (which is the leading order deviation from spherical symmetry) 
we use the ansatz
\be
r_0 (\theta) = r_\rK + a^2 r_1 (\theta),
\label{r0a2}
\ee
where $r_\rK$ is the Keplerian circular radius. Upon inserting (\ref{VE}) and (\ref{r0a2}) into (\ref{Newtcirc}) and expanding
we obtain one equation for $r_\rK$ at leading order and another one for $r_1 (\theta)$ at ${\cal O} (a^2)$ order. 
The former equation is simply the Keplerian relation for the energy, $ \cE = - M/2 r_\rK$.
The second equation is somewhat more complicated (here we define  $b = L/\cE$ )
\be
\left ( 4 r_\rK^3 - \frac{b^2 M}{\sin^2\theta}  \right ) r_1^{\pp} + M b^2 \frac{\cot \theta}{\sin^2\theta} r_1^\p + 4 r_\rK^3 r_1 
= ( 1+ 3 \cos 2\theta ) r^2_\rK, 
\label{r1eqE}
\ee 
but can nevertheless be solved exactly. Only the particular solution of this is well behaved in the equatorial plane,  
\be
r_1 (\theta) = -\frac{1}{4 r_\rK^4} \left [ M b^2 - r_\rK^3 (1 -\cos2\theta ) \right ].
\ee
In this expression it is legal to replace $b$ with its value for a Keplerian circular orbit, $ b_\rK^2 = 4 r_\rK^3/M $. Doing so, we find
\be
r_0 (\theta)  =  r_\rK  -\frac{a^2}{2 r_\rK} \left (1  + \cos^2 \theta \right ). 
\label{r1a2}
\ee
We have thus obtained a spheroidal orbit for the ${\cal O} (a^2)$ Euler potential as a result of working in the `wrong' spherical 
coordinates where $V_{\rm E}$ is not separable. 

As already mentioned, the `correct' coordinate system for the Euler problem is the elliptic one, with spherical orbits given 
by $\xi=\xi_0=\textrm{const.}$ The radial elliptic coordinate is defined as  $\xi=(r_1+r_2)/2$, where $r_1, r_2$ are the distances 
from the two centres. We can express $\xi_0$ in spherical coordinates and obtain $r_0 (\theta)$. The result of this exercise  is,
\be 
r_0 (\theta)=\frac{\xi_0\sqrt{a^2+\xi_0^2}}{\sqrt{\xi_0^2+a^2 \cos^2\theta}}=\xi_0+\frac{\sin^2\theta}{2\xi_0} a^2+{\cal O} (a^4).
\label{r0a2full}
\ee
The second $a$-expanded equation should be identical to our previous result. To verify this 
we need to express $\xi_0$ in terms of the Keplerian circular radius $r_\rK$. We find, 
\be 
\xi_0=r_\rK-\frac{b_\rK^2 M}{4 r_\rK^4}a^2+{\cal O} (a^4) =r_\rK-\frac{a^2}{r_\rK}+{\cal O} (a^4),
\ee
which when combined with (\ref{r0a2full}) indeed leads to  (\ref{r1a2}).

One could furthermore ask what happens if the Euler potential is somehow perturbed. A simple way of doing 
this is by adding a small mass $ q M$ at the coordinate origin. The resulting potential,
\be
\tilde{V}_{\rm E} = V_{\rm E}  - \frac{qM}{r},
\ee
is no longer separable in the elliptic coordinates used in the Euler problem. This prompts us to revisit the
issue of the existence of spherical orbits in this new potential. After performing an analysis similar to the one of the 
Euler problem one arrives to an equation for the ${\cal O} (a^2)$ perturbation of the Keplerian circular orbit, 
\be
\left [ 4 (1+q)r_\rK - \frac{4(1+q)r_\rK}{\sin^2\theta}  \right ] r_1^{\pp} + 4(1+q)r_\rK \frac{\cot \theta}{\sin^2\theta} r_1^\p + 4(1+q)r_\rK r_1 
=  1+ 3 \cos2\theta.
\ee 
Given that this equation is almost identical to (\ref{r1eqE}) we expect to find a solution of the same functional form 
as in (\ref{r1a2}). The resulting ${\cal O} (a^2)$ radius of the spheroidal orbit of the $\tilde{V}_{\rm E}$ potential is,
\be  
r_0 (\theta)  = r_\rK -\frac{a^2}{2(1+q)r_\rK} ( 1 + \cos^2\theta ).
\label{r1eqEper}
\ee
The comparison of this result against the same order expansion of the $\xi_0=\mbox{const.}$ expression in 
spherical coordinates reveals a mismatch due to the $(1+q)$  factor. In other words, the spheroidal orbit (\ref{r1eqEper}) 
cannot be mapped onto a spherical orbit in elliptical coordinates.


\section{The spheroidicity condition and separability theorem for particles}
\label{sec:particles}

Most of what we discussed in the main text about photon circular orbits and their connection to the separability
of a given axisymmetric-stationary metric also applies to the case of massive particles. 
The only adjustment one needs to make is to use the appropriate four-velocity normalisation $u_{\mu}u^{\mu}=-1$
and posit the presence of equatorial circular orbits; this results in a modified Eq. (\ref{norm_gen}) with a new effective potential, 
\be   
\tilde{V}_{\rm eff} = E^2 V_{\rm eff} (r,\theta,b)-1. 
\ee
The redefined potential is propagated through all subsequent calculations, leading to the same spheroidicity condition 
as the one derived for photons [Eq. (\ref{circularity})]. For spherical orbits we thus have,
\be 
(g_{\theta\theta}  \tilde{V}_{\rm eff} )_{,r}  |_{r_0}=0. 
\ee
If we assume the same functional form for $g_{\theta\theta}  \tilde{V}_{\rm eff}$ as the one in 
(\ref{sphericity2}) and combine it with the condition (\ref{sepa_condition2}) on the metric, then the two conditions imply the separability of 
the Hamilton-Jacobi equation.

However, photons and particles are found to be on an unequal footing. 
In order for \emph{both} photon and particle spherical orbits to exist, the metric needs to obey the condition, 
\be   g_{\theta\theta}=f_3(r) h(\theta)+\tilde{g}(\theta). 
\label{sphericity3}
\ee
If this requirement is not met, the spacetime can still admit spherical photon orbits but there are no spherical particle orbits. 
One such example is provided by Carter's canonical metric (\ref{canonical_metric}). In its most general form the canonical metric has 
(recall that here a subscript indicates functional dependence),
\be 
g_{\theta\theta}=\frac{P_r Q_{\theta}-Q_r P_{\theta}}{\Delta_{\theta}}, 
\ee
which evidently is not of the required form (\ref{sphericity3}). But there are exceptions to the rule: for example, Kerr 
is a member of the canonical metric family while also being of the special form (\ref{sphericity3}) with $\Delta_{\theta}=1$ and 
$P_r Q_{\theta}-Q_r P_{\theta}=r^2+a^2\cos^2\theta$. This is one more result to be added to the long string of special properties
of the Kerr metric.


\bibliography{biblio.bib}


\end{document}